\documentclass{jfm}
\usepackage{graphicx}
\usepackage{epstopdf, epsfig}
\newcommand{\BibitemShut}[1]{}

\usepackage{graphicx,psfrag,color}
\usepackage{bm}
\usepackage{natbib}
\usepackage{amssymb,amsmath,esint}
\usepackage[mathscr]{eucal}
\usepackage{subfig}
\usepackage{hyperref}
\hypersetup{
    colorlinks=true,       
    linkcolor=blue,          
    citecolor=blue,       
    filecolor=magenta,      
    urlcolor=blue          
}
\title{Comparing non-local granular fluid continuum models for silo discharge: Toward clogging prediction}

\author{Y. Zhou\aff{1}, Y. Wang\aff{1},  M. Li \aff{1} \and P.-Y. Lagr\'ee  \aff{2}
\corresp{\email{pierre-yves.lagree@upmc.fr}}}

\affiliation{\aff{1}Beijing Key Laboratory of Passive Safety Technology for Nuclear Energy, North China Electric Power University, Beijing 102206, China
\aff{2} Sorbonne Universit\'e, CNRS UMR 7190, Institut Jean le Rond $\partial^{\prime}$Alembert, F-75005 Paris, France}

\begin{document}
\pagestyle{empty}
\maketitle
\begin{abstract}
Non-local constitutive theories have received increasing attention in continuum descriptions of granular flows. However, these models have not been systematically compared for silo discharge within a unified numerical framework. We address this gap with two-dimensional finite-volume method (FVM) simulations of silo discharge using the {\usefont{T1}{pzc}{m}{n}Basilisk} platform. We first validate our FVM implementation of the dynamic non-local granular fluidity (NGF) model against the material point method results of \cite{Dunatunga2022}, obtaining quantitative agreement. Second, we relate the discharge rate $Q$ to the outlet-to-particle size ratio $D/d$ and the non-local amplitude $A$. From the simulated $Q$, we then evaluate the clogging probability $J(D/d, A)$ within the probabilistic framework of \cite{Janda2008}. The predicted $J$ decays exponentially with $D/d$, consistent with the experimental trend of \cite{Janda2008}. Rather than directly predicting flow arrest, our approach captures the continuous probabilistic transition. Finally, within the same numerical framework and using identical values of $A$, we compare several non-local constitutive models, including several linearised variants that we derive. Almost all models predict a reduction in the discharge rate with increasing $A$, yet significant quantitative differences are observed among the models. The results are further classified into groups according to their predicted flow behaviour, revealing close correspondences among certain formulations. Notably, using the non-local amplitudes reported in the literature (\cite{Bouzid2013, Henann2013}) yields near-zero discharge rates. Ill-posed issues are discussed. The implementation of all models is open-sourced and computationally efficient.

\end{abstract}


\section{Introduction} \label{sec:Introduction}
The discharge of granular media from a silo through an aperture is a classical problem with many practical and industrial applications. The main scaling relation, generally known as the Hagen-Beverloo relation, predicts that the mass flow rate scales as $(D - kd)^{5/2}$ for a horizontal aperture, where $D$ is the aperture size, $d$ the grain diameter, and $k$ an empirical parameter accounting for the aperture periphery effect (\cite{BEV61, NED92}). However, when the aperture becomes sufficiently small, typically only a few grain diameters wide, the discharge becomes intermittent. Stable arches may then form above the orifice and arrest the flow. Among the various factors influencing hopper flow clogging, the outlet-to-particle size ratio $D/d$ is widely recognized as a key determinant of the clogging probability $J$ (\cite{To2001, Zuriguel2005, Zhou2026}). Many studies have been conducted to quantify the statistics of particle release events in the intermittent range (\cite{Zuriguel2003, Zuriguel2005, Janda2008, Sheldon2010, Zuriguel2014}). Whether a critical orifice size exists beyond which clogging never occurs remains an open question (\cite{Zuriguel2011, Gella2018}). Early studies suggested that the mean avalanche size diverges as the orifice size approaches a critical value, implying a true critical size above which clogging is impossible (\cite{Zuriguel2005}). However, a non-divergent expression in which the average avalanche size increases exponentially with the outlet size raised to the dimensionality of the problem has received increasing support (\cite{To2005, Janda2008, Thomas2013, Gella2018}). This expression implies that clogging can theoretically occur for any finite $D/d$, although its probability becomes practically unobservable for sufficiently large openings. 
Based on the statistics of arch formation, \cite{Janda2008} proposed an exponential relation between $J$ and $D/d$. This relation collapses experimental measurements obtained for different outlet sizes and captures the intermittent nature of clogging.

The study of silo discharge and clogging has been approached through various methodologies. Early investigations relied primarily on experimental measurements, which provided the empirical basis for the Hagen-Beverloo scaling and the statistical characterisation of clogging events (\cite{BEV61, Janda2008}). On the one hand, the discrete element method (DEM) emerged as a powerful tool to resolve particle-scale dynamics and provide detailed information on particle-scale physical quantities (\cite{Arevalo2014, Zhou_15, Arevalo2016}), yet its computational cost remains prohibitive for large-scale studies. On the other hand, continuum models provide a computationally efficient framework for describing granular materials, in particular when they are flowing. Such models can be formulated using the incompressible Navier-Stokes equations supplemented by the visco-plastic $\mu(I)$ rheology. They capture the main features of silo discharge at a substantially lower computational cost than DEM (\cite{STA12, Staron2014, Dunatunga2015, Zhou2017, Fullard2019}). The $\mu(I)$ rheology emerged from a phenomenological effort, partly inspired by the collective work of the GDR MiDi group~(\cite{GDR2004}), which compiled experimental and simulation data from multiple studies. In this framework, the friction coefficient $\mu$ and volume fraction $\phi$ are functions of a single dimensionless number, the inertial number $I$~(\cite{GDR2004, DaCruz2005}). The friction coefficient increases with $I$, while the volume fraction decreases. This local description, later generalised into a tensorial formulation~(\cite{Jop06}), successfully captures many observations across different flow configurations~(\cite{Lagree2011, Staron2014, Dunatunga2015}). 

Despite its practical success, the local $\mu(I)$ rheology has an inherent limitation: its purely local character prevents it from capturing cooperative effects that are intrinsic to granular media, such as creeping  flow in regions below the yield threshold,  finite-size effects, and the long-range influence of distant shear zones~(\cite{Pouliquen1999, GDR2004, Nichol2010, Reddy2011, Andreotti2013, Fullard2019}. These shortcomings have prompted the development of various non-local rheological models over the past decade, and the question of non-locality in granular flows has remained a vibrant and rapidly evolving research topic~(\cite{Kamrin2012, Kamrin2015SM, Kamrin2024, Pouliquen2026}).

 To incorporate non-local effects into a continuum description of dense granular flows, several approaches have been proposed~(\cite{Kamrin2024, Pouliquen2026}). Some rely on an integral formulation to capture the influence of neighbouring regions, inspired by Eyring theory (\cite{Pouliquen2009}) or by introducing averaging kernels in plasticity models (\cite{Dsouza2020}). Other approaches focus on the role of velocity fluctuations, building on modified kinetic theory (\cite{Berzi2024}) or generalized energy fluctuation equations (\cite{Alessio2026}). 

Among the various proposals, four non-local models have gained significant attention within the granular community. The first is the non-local granular fluidity (NGF) model (\cite{Kamrin2012, Henann2013}), which introduces a fluidity field defined as the ratio of shear rate to friction coefficient. The concept of fluidity was introduced in the context of glassy flows by \cite{Goyon2008} and was later extended to dry granular flows. The fluidity satisfies a Helmholtz-type equation that includes a diffusive term, with a cooperative length scale that diverges as the friction coefficient approaches the static yield friction coefficient. The second is the dynamic NGF model~(\cite{Henann2014, Kamrin2015SM, Zhang2017}), which extends the steady-state formulation through a time-dependent evolution equation for the fluidity. This dynamical system was derived from thermomechanical considerations by treating the fluidity and its gradient as kinematic state variables with distinct contributions to the free energy. The dynamic NGF model was subsequently applied by \cite{Dunatunga2022} to predict silo clogging, demonstrating its ability to capture the critical opening size for flow arrest and the formation of stable arches. The third is the $I$-gradient model proposed by~(\cite{Bouzid2013}), which treats the inertial number as a diffusing state variable and expresses the friction coefficient as its local value reduced by a term related to the diffusion of the inertial number. The characteristic diffusion length diverges as the inertial number tends to zero, capturing the increasing spatial correlation near the yield point. Notably, this model shares a similar diffusive mathematical structure with the NGF framework, despite their different physical underpinnings~(\cite{Bouzid2015, Pouliquen2026}). A more recent approach is the $\mu(I,\Theta)$ model~(\cite{Kim2020, Irmer2025}), which introduces granular temperature as an additional state variable representing the fluctuational kinetic energy of particles. Unlike the abstract fluidity field, granular temperature is physically measurable through imaging techniques. A universal scaling relation linking the friction coefficient to the inertial number and granular temperature has been shown to collapse experimental and DEM data across a wide range of flow conditions (\cite{Poon2023, Clarke2024, Hao2023, Breard2024}). This framework has been implemented in a finite-volume solver and validated for steady heap flow, split-bottom Couette flow, and granular-column collapse. It captures non-local phenomena including sustained creep below yield, shear-band broadening, and transient free-surface evolution (\cite{Yuan2026}).

While non-local models have been extensively tested in configurations such as inclined planes, Couette cells, and split-bottom geometries (\cite{Kamrin2019, Pouliquen2026}), their application to silo flow remains relatively limited. Moreover, existing studies on silo flow using non-local rheologies have produced inconsistent results. \cite{Lin2021} used a FVM-based Navier-Stokes solver with the $I$-gradient model and found that adding non-local effects leads to higher velocities compared to the local $\mu(I)$ rheology. In contrast, \cite{Dunatunga2022} employed the material point method (MPM) with the dynamic NGF model and demonstrated that non-local effects slow down the flow and promote clogging. By introducing a separating phase within the rheology, their model was able to capture the formation of stable arches. This discrepancy suggests that the choice of non-local model and numerical method may significantly influence predictions of silo flow, an issue that has not been systematically examined.

Beyond its inability to capture non-local effects, the local $\mu(I)$ rheology also raises concerns regarding its mathematical well-posedness. \cite{Schaeffer1987} showed that, for a constant friction, the Coulomb rheology exhibits a Hadamard-type instability. Furthermore, \cite{Barker2015} demonstrated that the $\mu(I)$ rheology also exhibits a Hadamard-type instability, with unbounded growth rates for short-wavelength perturbations in both the quasi-static ($I \to 0$) and rapid ($I \to \infty$) regimes. This ill-posedness inevitably leads to mesh-dependent numerical solutions and undermines the reliability of simulations based on the local model. Several strategies have been proposed to address this issue, including regularisation of the $\mu(I)$ function~(\cite{Barker2017b, Franci2019}) and the introduction of compressibility effects~(\cite{Barker2017, Heyman2017, Schaeffer2019}). Concerning the non-local models, it is natural to expect that the inclusion of a diffusive Laplacian term in the constitutive laws could serve as a physical means of regularisation, stabilising short-wavelength perturbations and restoring well-posedness. However, \cite{Li2019} showed that while the NGF model successfully suppresses the Hadamard instability, a tensorial generalisation of the $I$-gradient model fails to eliminate the instability and may even exhibit more severe unbounded growth due to the presence of higher-order pressure gradients. It should be acknowledged that, despite the identified ill-posedness, simulations using the local $\mu(I)$ rheology have often produced acceptable results for silo discharge flows~\cite{STA12, Staron2014}. As discussed by \cite{Barker2015}, this may be due in part to the ad hoc regularisations and coarse mesh resolutions that effectively suppress the growth of short-wavelength instabilities. 
Given these theoretical insights, it is of considerable interest to examine whether non-local models can effectively eliminate or at least substantially reduce the numerical instabilities observed in local $\mu(I)$ simulations, and whether different non-local formulations exhibit distinct levels of robustness in practice.

The continuum modelling of granular flows has been pursued using various numerical approaches, each with its own advantages. Lagrangian particle-based methods, such as smoothed particle hydrodynamics (SPH) and the MPM, are particularly attractive for problems involving large deformations, free surfaces and topological changes, as they naturally avoid mesh tangling and remeshing procedures (\cite{Zhu2022, Dunatunga2015, Dunatunga2022}). The lattice Boltzmann method (LBM) has also been explored for non-Newtonian rheology and multiphase interactions in granular systems (\cite{Yang2023, Yang2023b, Shen2026}). Alternatively, Eulerian finite-volume methods (FVM) have been widely used for granular flows with $\mu(I)$ rheology (\cite{Lagree2011, Staron2014}). Given the conflicting predictions reported by \cite{Lin2021} and \cite{Dunatunga2022}, a systematic comparison of different non-local models within a unified numerical framework is needed to separate the effects of the constitutive laws from those of the numerical method.

In the present work, we adopt the {\usefont{T1}{pzc}{m}{n}Basilisk}~platform, an open-source FVM solver, which has been successfully used for granular flows with the local $\mu(I)$ rheology~(\cite{Lagree2011, Staron2014, Zhou2017, Zhou2019, Fullard2019, Zou2022}). Building on these previous implementations, we extend the solver to incorporate four non-local constitutive models: the NGF, the dynamic NGF, the $I$-gradient model, and the $\mu(I,\Theta)$ model. In addition, we test three simplified variants of the NGF-type models: (i) the linearised NGF model; (ii) the linearised constant NGF model, which further replaces the cooperativity length with a constant; and (iii) the constant NGF model, where only the cooperativity length is taken as constant without linearisation. These variants are not intended as physically complete models because the cooperativity length is expected to diverge near the yield point. Instead, they are used as numerical experiments to assess the practical importance of its stress dependence. They also allow us to quantify the effect of linearising the NGF model. All models are systematically compared under identical numerical conditions. 

However, the intermittent nature of clogging poses a challenge for continuum simulations. Individual realisations of silo discharge may exhibit different behaviours, with periods of flow interrupted by clogging events, and the size of individual avalanche events exhibits large fluctuations. This suggests that a deterministic prediction of whether a single realisation clogs is inherently difficult. To circumvent this difficulty, \cite{Dunatunga2022} argued that the continuum solution should predict when the ensemble-averaged flow rate vanishes, which corresponds to the critical opening size below which all realisations clog. Beyond predicting this critical opening size for flow arrest, we further propose that the discharge rate $Q$ obtained from continuum simulations can be used to infer the clogging probability $J$ within a probabilistic framework. This allows us to capture the full range of intermediate clogging states. Our simulations focus on steady-state discharge, where we establish a modified Beverloo relation as a function of $A$ and $D/d$, from which we extract the cutoff size. From the simulated $Q$, we then compute the clogging probability $J(A, D/d)$ within the probabilistic framework of \cite{Janda2008}, and compare the results with experimental observations.

The paper is structured as follows. \S \ref{sec:numericalscheme} describes the numerical method and the implementation of the non-local models. \S \ref{Compare} validates the numerical implementation by comparing the dynamic NGF model with the results of \cite{Dunatunga2022}. \S \ref{DynamicNGFresults} then presents results obtained with the dynamic NGF model using the modified parameter set. We systematically examine how the discharge flow rate $Q$ varies with the orifice size $D$ and the non-local amplitude $A$, and subsequently evaluate the clogging probability $J(A, D/d)$ within the probabilistic framework of \cite{Janda2008}. \S \ref{comparenonlocal} compares the results of the different non-local models under identical numerical conditions.

\section{Constitutive models and their implementation in numerical resolution of Navier-Stokes equations }\label{sec:numericalscheme}
We now introduce the governing equations, followed by the local $\mu(I)$ model and the non-local models considered in this study. The continuum simulations were performed using the {\usefont{T1}{pzc}{m}{n}Basilisk}  solver in two dimensions, which solves the incompressible Navier-Stokes equations for a bi-phasic mixture using a volume-of-fluid (VOF) approach (\cite{Popinet2003, Popinet2009, Staron2014, Popinet2015}): 

\begin{eqnarray}
\nabla \cdot \mathbf{u} &=& 0, \label{eq:conti} \\
\rho \left( \frac{\partial \mathbf{u}}{\partial t} + \mathbf{u} \cdot \nabla \mathbf{u} \right) &=& -\nabla p + \nabla \cdot (2 \eta_{\text{eff}} \mathbf{D}) + \rho \mathbf{G}, \label{eq:mom}
\end{eqnarray}
where $\mathbf{u}$ is the velocity vector, $p$ is the pressure, $\rho$ is the mixture density, $\eta_{\text{eff}}$ is the effective viscosity, $\mathbf{G}$ is the gravitational acceleration vector, and $G$ is its magnitude (to not confuse with $g$ the granular fluidity defined latter). The strain-rate tensor $\mathbf{D}$ is defined as $(\nabla \mathbf{u} + \nabla \mathbf{u}^{\mathrm{T}})/2$. The phase fraction $c$ (with $c=1$ for the granular phase and $c=0$ for air) is advected by the velocity field according to:

\begin{equation}
\frac{\partial c}{\partial t} + \nabla \cdot (c \mathbf{u}) = 0, 
\end{equation}
and the mixture density is obtained as the arithmetic average weighted by the phase fraction:
\begin{equation}
\rho = c \rho_s + (1-c) \rho_f,
\end{equation}
where $\rho_s$ and $\rho_f$ are the densities of the granular and air phases, respectively. 

Before presenting the specific rheological models, we note that a unified regularization technique is employed to avoid the divergence of the viscosity. Specifically, the effective viscosity is bounded by lower and upper limits: $\eta_{\text{eff}}^{\min} = \rho_s \sqrt{ G d^3}$ and $\eta_{\text{eff}}^{\max} = 100$, where $d$ is the mean particle diameter. This treatment, which follows previous continuum simulations~(\cite{Lagree2011, Zhou2017, Zhou2019}), provides a consistent regularisation of the viscosity in both the quasi-static limit ($I \to 0$) and the rapid flow limit ($I \to \infty$), and is applied uniformly across all constitutive models considered in this study.

\subsection{Local $\mu(I)$ model} \label{sec:local}
Following \cite{GDR2004, Jop2005, Jop06}, the granular medium is modelled as a viscoplastic fluid with an effective viscosity $\eta_{\text{eff}}$ that depends on the shear rate $|\dot{\gamma}|$, the local pressure $p$ and the effective friction coefficient $\mu$. The local $\mu(I)$ model is expressed as:
\begin{eqnarray}
\eta_{\text{eff}} = \frac{\mu(I)}{|\dot{\gamma}|} p, \quad
I = \frac{|\dot{\gamma}| d}{\sqrt{p / \rho_s}}, \quad
\mu(I) = \mu_s + \frac{ \mu_2 - \mu_s}{I_0 / I + 1}, \label{eq:muI}
\end{eqnarray}
where $|\dot{\gamma}| = \sqrt{2 D_{ij} D_{ij}}$ is the second invariant of the strain-rate tensor, $d$ is the particle diameter, and $I_0$, $\mu_s$ and $\mu_2$ are material-dependent constants, with $\mu_s$ the static yield friction coefficient and $\mu_2$ the upper limit of the friction coefficient at high shear rates.

\subsection{Non-local granular fluidity (NGF) model}\label{sec:NGF}
For emulsions, a non-local fluidity model was proposed by \cite{Goyon2008} and later generalized to two dimensions by \cite{Bocquet2009}. The fluidity is defined as $f = \dot{\gamma}/\tau$, where $\tau$ is the shear stress, and is governed by the non-local relation:
\begin{equation}
f - f_{\mathrm{loc}}(\tau) = \xi^2 \nabla^2 f,
\end{equation}
where $f_{\mathrm{loc}}(\tau) = \dot{\gamma}_{\mathrm{loc}}(\tau)/\tau$ is the local fluidity, and $\xi$ denotes the cooperativity length. \cite{Bocquet2009} further derived from a kinetic elastoplastic model that the cooperativity length diverges as the stress approaches the yield point (i.e., the critical stress below which the material behaves as a solid).
\cite{Kamrin2012} extended this framework to granular materials by replacing the shear stress $\tau$ with the friction coefficient $\mu = \tau/p$, leading to the definition of granular fluidity:
\begin{equation}
g = \frac{|\dot{\gamma}|}{\mu}. 
\end{equation}
Following the implementation described in \cite{SalvadorVieira2017}, we solve for each time step the following additional set of equations using values obtained from the previous local formulation, in order to obtain a global (non-local) value for the granular viscosity within our FVM framework:

\begin{eqnarray} 
	&&g_{\mathrm{loc}} = \frac{|\dot{\gamma} |}{ \mu_{\mathrm{loc}}(I)},\\
	&& -[\xi(\mu_\mathrm{loc})]^2 \nabla^{2}g  +g = g_{\mathrm{loc}}, \label{gnonlocalequation}\\
	&&\mu =\frac{ |\dot{\gamma} | }{ g},\\ 
	&& \eta_{\mathrm{eff}} = \frac{\mu p}{  |\dot{\gamma} |}.
\end{eqnarray}

Following the theoretical framework of \cite{Bocquet2009}, the cooperativity length $\xi$ is postulated to depend on the distance to the yield point, diverging as the friction coefficient $\mu$ approaches the static yield value $\mu_s$:

\begin{equation}
\label{eq:xi}
\xi (\mu) = A\sqrt{\frac{\mu_2 - \mu}{\Delta\mu(\mu - \mu_s)}} d,
\end{equation}
with $A$ a dimensionless non-local amplitude that quantifies the spatial extent of cooperativity in the flow. The Helmholtz-type equation for $g$, equation~\eqref{gnonlocalequation}, is solved using a multigrid Poisson solver. To avoid division by zero when computing the friction coefficient from the fluidity, we add a small regularisation term $\varepsilon = 10^{-16}$ to the denominator, i.e., $\mu = |\dot{\gamma}| / (g + \varepsilon)$.

The tolerance of the multigrid solver was chosen based on systematic sensitivity tests. We observed that the solver tolerance affects the convergence behaviour of the iterative solver. This effect is particularly pronounced for larger values of the non-local amplitude ($A \geq 1.5$), where the convergence becomes more sensitive. Based on these tests, we selected tolerance values that have negligible influence on the final solution: $10^{-3}$ for $A < 1.5$ and $10^{-4}$ for $A \geq 1.5$. These tolerance settings are applied uniformly across all non-local models considered in this study.

\subsection{Dynamic NGF model}\label{sec:algorithmDymNGF}

To accurately describe size-dependent flow arrest where the steady-state approximation ($g \approx g_{\mathrm{loc}}$) becomes invalid, \cite{Kamrin2015SM} adopted the full dynamic form of the NGF equation:
 \begin{equation}
 t_0\dot{g} = A^{2}d^{2}\nabla^{2}g - \Delta{\mu}\left(\frac{\mu_{s} - \mu}{\mu_{2} - \mu}\right)g - b\sqrt{\frac{\rho_{s}d^{2}}{p}}\mu g^{2}.
    \label{equation1}
\end{equation}

In this equation, $t_0$ is a characteristic relaxation time scale controlling the rate at which the fluidity evolves toward steady state. $b = \Delta\mu / I_0$ is a material constant derived from the local $\mu(I)$ rheology, with $I_0$ a characteristic inertial number.

As in \cite{Dunatunga2022}, a split numerical update is employed to solve the evolution equation. By introducing an intermediate variable $g^*$, the following two expressions are obtained:

 \begin{equation}
 \label{gstar}
\frac{g^* - g^n}{\Delta t} = \frac{1}{t_0} A^2 d^2 \nabla^2 g^*_{i,j} 
\end{equation}

 \begin{equation}
\label{dynamicNGF2}
\frac{g^{n + 1} - g^*}{\Delta t} = \frac{1}{t_0}\left(-\Delta \mu \left(\frac{\mu_s - \mu^{n+1}_\mathrm{loc}}{\mu_2 - \mu^{n+1}_\mathrm{loc}}\right)g^{n + 1} 
- \frac{\Delta\mu}{I_0}\sqrt{\frac{\rho_s d^2}{p^{n + 1}}}\mu^n g^n g^{n + 1}\right). 
\end{equation}

Note that, unlike the explicit treatment in \cite{Dunatunga2022}, the diffusion step in \eqref{gstar} is solved implicitly. In the reaction step \eqref{dynamicNGF2}, we use the local friction coefficient $\mu_{\mathrm{loc}}^{n+1}$, whereas \cite{Dunatunga2022} employs the global friction coefficient $\mu^{n+1}$.

\subsection{$I$-gradient model}
\label{sec:Igradient}

Shortly after the NGF model, an alternative model for non-local flow was proposed by \cite{Bouzid2013}. The proposed form is a more direct non-local expansion of the $\mu(I)$ rheology:
\begin{equation}
\mu(I) = \mu_{\text{loc}}(I)\left(1 - \nu d^{2}\frac{\nabla^{2}I}{I}\right),
\end{equation}
where the non-local effect is accounted for through the Laplacian of the inertial number. To avoid the singularity when $I$ approaches zero, \cite{Lin2021} introduced a Bercovier-Engelman regularisation for the $I$-gradient model, yielding the following local and non-local effective viscosities:
\begin{equation}
\eta_{\text{loc}} = \frac{\mu_{s} p}{\sqrt{|\dot{\gamma}|^{2} + \lambda_{r}^{2}}} +\frac{(\mu_{2} - \mu_{s})p}{(I_{0} / d)\sqrt{p / \rho_{s}} + |\dot{\gamma}|},
\end{equation}
\begin{equation}
\eta_{\text{eff}} = \eta_{\text{loc}}\left(1 - \frac{\nu d \sqrt{p / \rho_{s}}}{\sqrt{|\dot{\gamma}|^{2} + \lambda_{r}^{2}}}\nabla^{2}I\right),
\end{equation}
with $\lambda_{r}$ a small regularisation parameter. Similar regularisation practices have been adopted in other $\mu(I)$-based continuum simulations~(\cite{Franci2019}). Because our viscosity is evaluated only when $p>0$ and $\dot{\gamma}>0$, and we introduce a maximum effective viscosity $\eta_{\text{eff}}^{\max}$, the singularity is naturally avoided. Sensitivity tests confirm that the results are essentially independent of $\lambda_{r}$; we therefore set $\lambda_r = 0$ throughout.

In order to compare with other methods, particularly the linearized constant NGF (see \S\ref{sec:linearisedConstantNGF} for more details), we express:
 \begin{equation}
 \nu = A^2.
\end{equation}

\subsection{$\mu(I,\Theta)$ model}
\label{sec:muItheta}
We also consider a constitutive framework based on granular temperature. Following the work of \cite{Kim2020}, the effective friction coefficient is expressed as a function of both the inertial number $I$ and the dimensionless granular temperature $\Theta = \rho_s T / p$:
\begin{equation}
\label{eq:muITheta}
\mu(I,\Theta) = \mu_{\text{loc}}(I) \left( \frac{\Theta_{\text{loc}}(I)}{\Theta} \right)^P,
\end{equation}
where $\Theta_{\text{loc}}(I)$ is the granular temperature that would be obtained in a purely local homogeneous shear flow at the same $I$, and $P$ is a dimension-dependent exponent. For two-dimensional disks, \cite{Kim2020} found $P \approx 1/8$ from DEM simulations across a wide range of flow geometries. This value is adopted in the present work.

The granular temperature field $\Theta$ is governed by a transport equation proposed by \cite{Irmer2025}:
\begin{equation}
\label{muIthetat0}
t_0 \dot{\Theta} = A^2 d^2 \nabla^2 \Theta - b \Theta + a I^{3/2},
\end{equation}
where $b$ is a dissipation coefficient and $a$ is a shear-induced production coefficient. Following \cite{Irmer2025}, we set $a = 0.15$ and $b = 1$. The diffusion coefficient is written as $A^2 d^2$ to maintain consistency with the notation used in other models, with $A$ denoting the non-local amplitude and $d$ the grain diameter.

The transport equation~\eqref{muIthetat0} is discretised using an implicit backward Euler scheme. The absence of a quadratic nonlinearity in the $\Theta$ equation allows for a fully implicit solution in a single step, in contrast to the split treatment used for the $g$ equation.

The local granular temperature $\Theta_{\text{loc}}$ follows the scaling relation
\begin{equation}
\Theta_{\text{loc}} = \frac{a}{b} I^{3/2}.
\end{equation}
The resulting granular temperature field is then used to compute the friction coefficient $\mu(I,\Theta)$ via equation~\eqref{eq:muITheta}, and the effective viscosity is obtained as
\begin{equation}
\eta_{\text{eff}} = \frac{\mu(I,\Theta) p}{|\dot{\gamma}|}.
\end{equation}

\subsection{Linearised NGF model}

To facilitate a direct comparison with the $I$-gradient model, we introduce a simplified variant of the NGF framework. When the variation of $\dot{\gamma}$ is large compared with the variations of $p$ and $\mu$, the ratio $\nabla^2 g / g$ can be approximated by $\nabla^2 I / I$, which is precisely the form appearing in the $I$-gradient model. For the linearised NGF formulation, we further adopt the approximation $\nabla^2 g / g_{\text{loc}} \approx \nabla^2 I / I$, replacing $g$ by $g_{\text{loc}}$ in the denominator. This replacement is justified when the non-local correction is small ($\xi^2 \nabla^2 g / g_{\text{loc}} \ll 1$), so that its effect is negligible. Under this condition, we obtain the following linearised form of the effective viscosity:
\begin{equation}\label{eq:linearisedNGF}
\eta_{\text{eff}} = \frac{p}{g} = \frac{p}{g_{\text{loc}} + \xi^2 \nabla^2 g} \approx \eta_{\text{loc}} \left( 1 - \xi^2 \frac{\nabla^2 g}{g_{\text{loc}}} \right),
\end{equation}
where $\eta_{\text{loc}} = p / g_{\text{loc}}$ is the local viscosity, and the cooperativity length retains its full state-dependent form:
\begin{equation}
\xi(\mu) = A \sqrt{\frac{\mu_2 - \mu}{\Delta \mu (\mu - \mu_s)}} d.
\end{equation}
The linearised NGF model allows a direct comparison between the NGF and $I$-gradient models.

\subsection{Linearised constant NGF model}
\label{sec:linearisedConstantNGF}

Starting from the linearised form of the NGF viscosity, equation~\eqref{eq:linearisedNGF}, we further simplify the model by taking the cooperativity length as constant:
\begin{equation}
\xi = A d.
\end{equation}
In the $I$-gradient model, the correction term is proportional to $\nabla^2 I / I$ with a constant prefactor $\nu d^2$. Under the approximation $\nabla^2 g / g_{\text{loc}} \approx \nabla^2 I / I$, the linearised constant NGF model becomes structurally identical to the $I$-gradient model. This reduction is primarily a mathematical convenience to establish a structural correspondence between the two frameworks, rather than a physically equivalent alternative. Indeed, by taking $\xi$ constant, the model loses the divergence of $\xi$ as $\mu \to \mu_s$ that is essential for capturing creep in the NGF model.

\subsection{Constant NGF model}
\label{sec:constantNGF}

Finally, we consider a simplified variant in which the state-dependent cooperativity length is replaced by a constant value, while the full Helmholtz-type equation for the fluidity, equation~\eqref{gnonlocalequation}, is retained:
\begin{equation}
\xi = A d.
\end{equation}

In contrast to the linearised constant NGF model introduced above, which is derived from the linearised viscosity equation, this constant NGF model preserves the full Helmholtz-type diffusion of the fluidity without linearisation. By taking $\xi$ constant, the model removes the divergence of $\xi$ as $\mu \to \mu_s$, thereby serving as a baseline to isolate the effects of the state dependence of $\xi$.

In summary, apart from the widely used local $\mu(I)$ rheology, we have introduced seven non-local models: four primary formulations (NGF, dynamic NGF, $I$-gradient, and $\mu(I,\Theta)$), and three simplified NGF variants (linearised NGF, linearised constant NGF, and constant NGF). The non-local amplitude in all models is expressed in terms of the same parameter $A$. Apart from the unified viscosity regularization applied to all models and a small regularisation term $\varepsilon = 10^{-16}$ used in the NGF model to avoid division by zero (see \S\ref{sec:NGF}), no additional regularization is required for the other non-local models. Following \cite{Dunatunga2022}, we initialise each simulation with a short local $\mu(I)$ run to generate a non-zero flow field, which is then used as the initial condition for the non-local simulation. The duration of this initialization phase, $t_{\text{switch}}$, is specified below; we have verified that the steady-state discharge rate is insensitive to its value. A detailed description of the boundary conditions for the physical variables is provided in the subsequent sections.

\section{Convergence test and validation against \cite{Dunatunga2022}}  \label{Compare}

\begin{table}
  \begin{center}
\begin{tabular}{cc}
\textbf{Parameter} & \textbf{Value} \\
$\mu_s$ & 0.3819 \\
$\mu_2$ & 50 \\
$\Delta \mu/I_0$ & 1 \\
$t_0$ (s) & 0.001 \\
$\rho_f$ (kg/ $\text{m}^{3}$) & $1\times10^{-4}$ \\
$\rho_s$ (kg/ $\text{m}^{3}$) & 1 \\
$G$ (m/$\text{s}^2$) & 9.8 \\
$\Delta x$ & $L/2^7$\\
$\Delta t$ (s) & $0.00025$ 
\end{tabular}
 \caption{ \label{tableruns} Dynamic NGF simulation parameters for validation against \cite{Dunatunga2022}, note that in this part, dimension are used to compare with this reference.}
 \end{center}
\end{table}

 \begin{figure}
\begin{centering}

 \hspace{-5.5cm} $(a)$ \hspace{6.cm} $(b)$\\
 \includegraphics[height=5cm]
{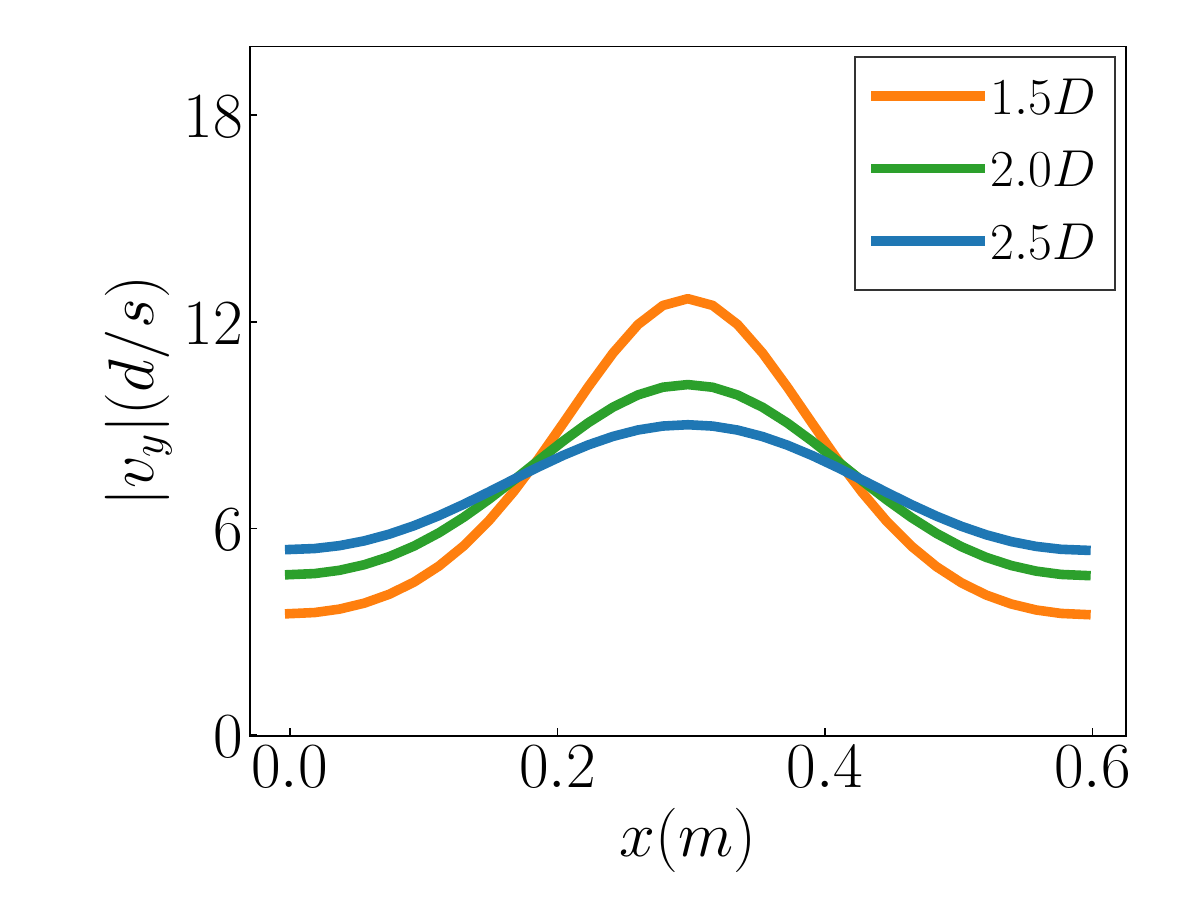}
 \includegraphics[height=5cm]
{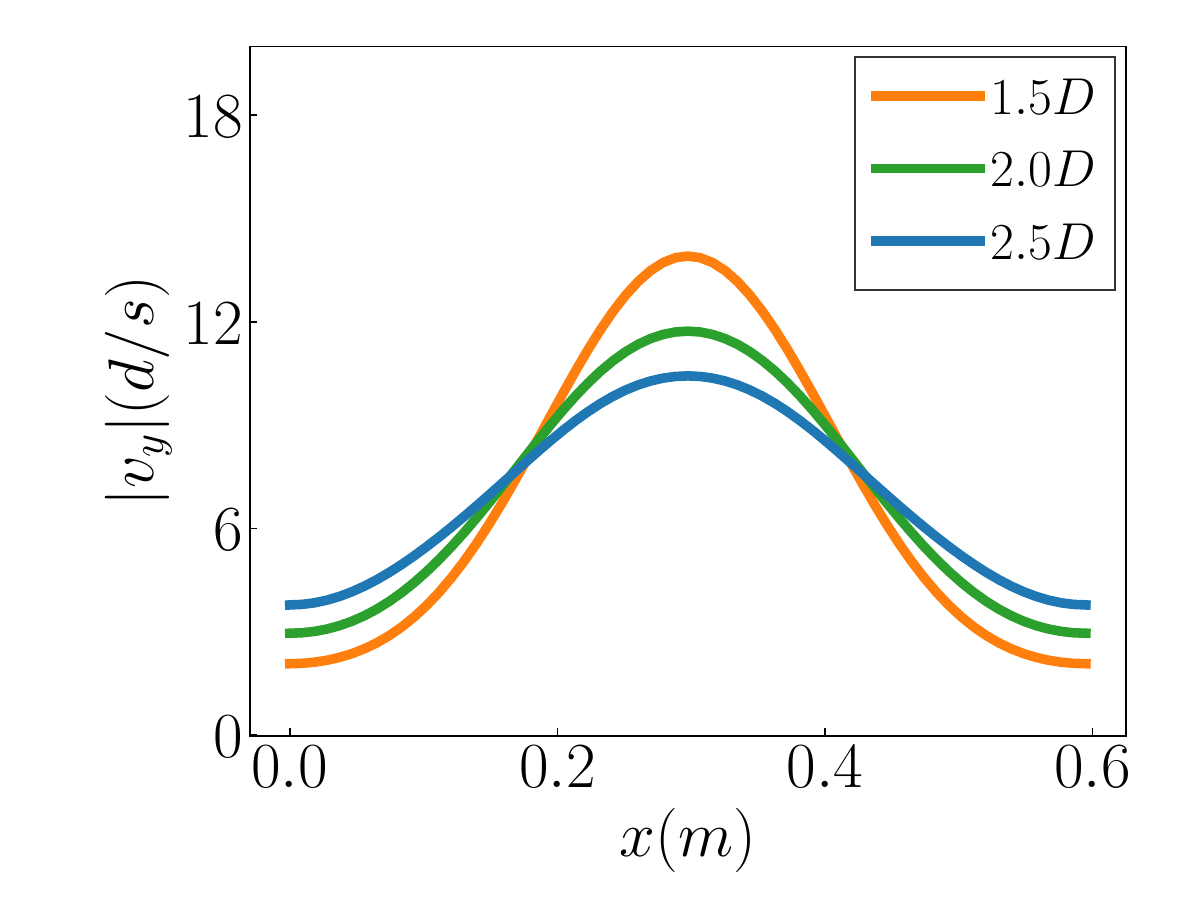}\\
 \hspace{-5.5cm} $(c)$ \hspace{6.cm} $(d)$\\
\includegraphics[height=5cm]
{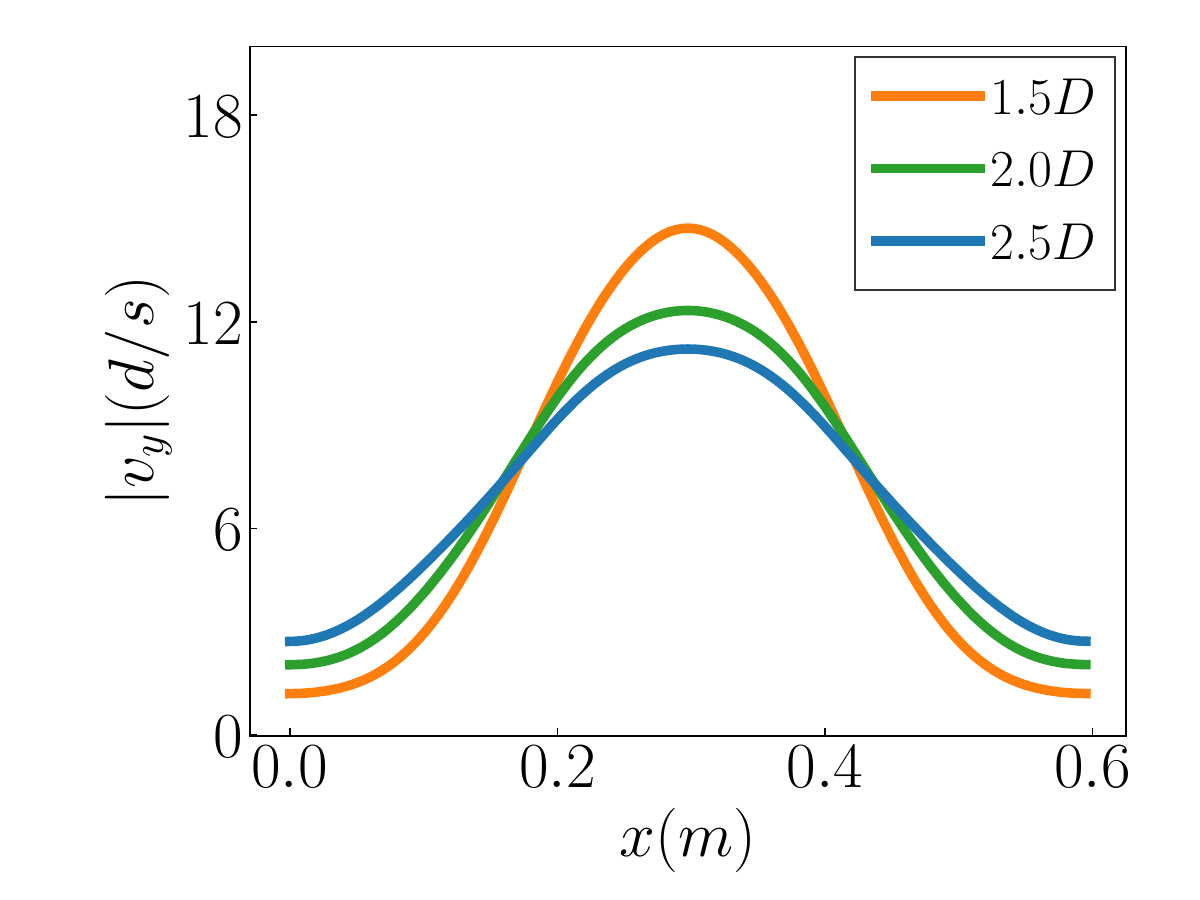}
\includegraphics[height=5cm]
{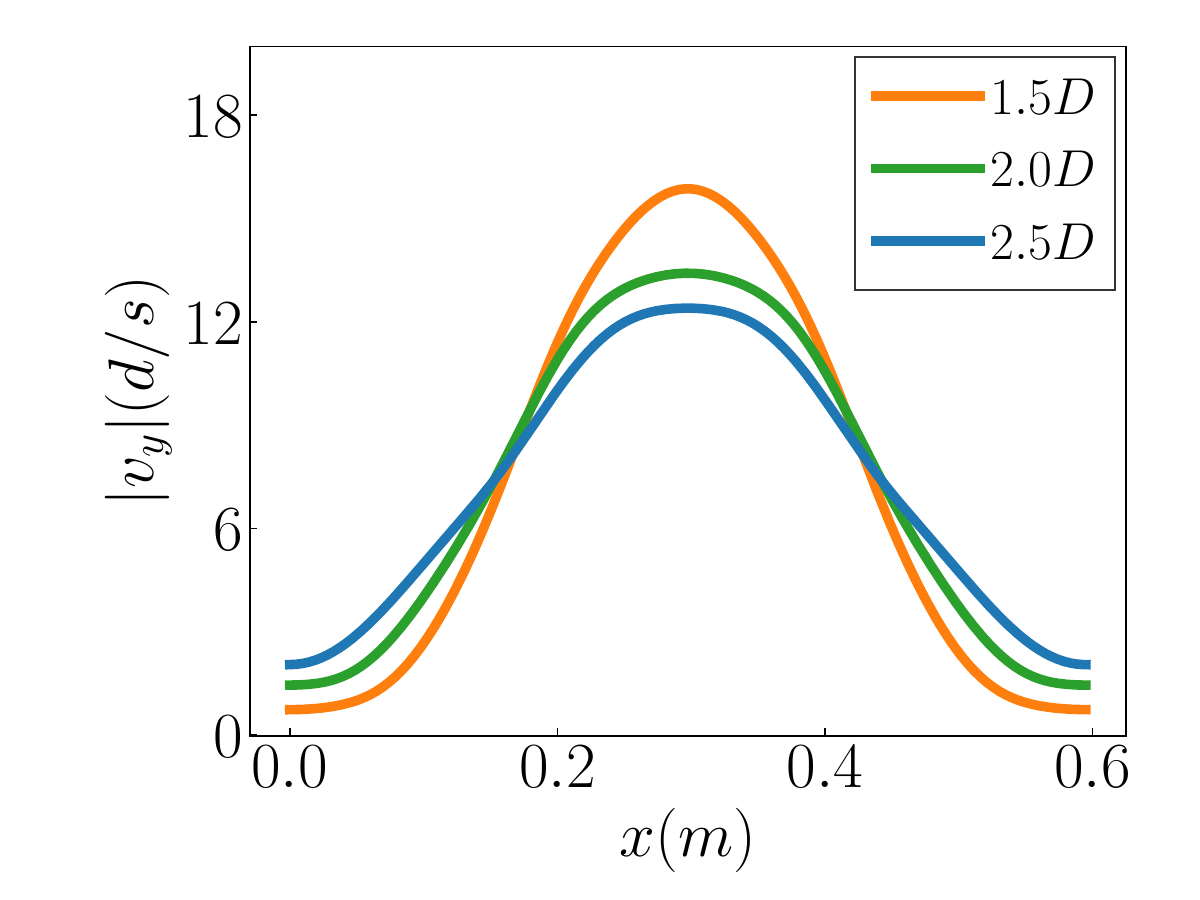}
\caption{\label{fig:convergence} Vertical velocity profiles $|v_y|$ versus $x$ at heights $1.5D$, $2.0D$ and $2.5D$ above the orifice, corresponding to the horizontal lines in figure \ref{fig:kamrin_comparison}(a), obtained from the dynamic NGF model using the rheological parameters listed in table~\ref{tableruns}, with different time steps and mesh sizes: 
(a) $\Delta x = L/2^5$, $\Delta t = 0.004$ s (shown at $t = 0.15$~s, beyond which the simulation becomes numerically unstable); 
(b) $\Delta x = L/2^6$, $\Delta t = 0.001$ s; 
(c) $\Delta x = L/2^7$, $\Delta t = 0.00025$ s; 
(d) $\Delta x = L/2^8$, $\Delta t = 0.0000625$ s. 
Panels (b)--(d) are shown at $t = 0.45$~s, where the flow is well-developed and numerically stable.}	
\end{centering}	
\end{figure}

In this section, we generally adopt the parameters used in \cite{Dunatunga2022}, except for minor adjustments based on the specifics of our numerical method. The orifice size is fixed at $D = 0.14~\text{m}$ and the grain diameter $d$ is varied. The total width of the silo $L$ is set to $0.595~\text{m}$. While the initial particle height in their work is equal to $L$, in our simulations it is set to $H_0 = 0.575~\text{m}$, slightly lower than the domain height. The material parameters used in this section are summarised in table~\ref{tableruns}. Following \cite{Dunatunga2022}, we set $\mu_s = 0.3819$ for glass beads.  The value of $\mu_2$ is set to a sufficiently large value ($50$) to approximate the linear $\mu(I)$ form, as discussed in \cite{Dunatunga2022}. For computational efficiency, we scale the solid density to $\rho_s = 1~\text{kg}/\text{m}^3$ instead of the physical value $2450~\text{kg}/\text{m}^3$. This scaling preserves all dimensionless quantities, since quantities such as pressure and flow rate scale proportionally with $\rho_s$, leaving all dimensionless groups unchanged. Consequently, the physical validity of the results remains intact, while numerical stability and convergence speed are improved. The switching time from the local to the non-local simulation is set to $t_{\text{switch}} = 0.1\,\text{s}$.

Following \cite{Dunatunga2022}, the boundary conditions are summarised as follows. For the fluidity field $g$, a homogeneous Dirichlet condition ($g = 0$) is imposed on the solid bottom boundaries, while a homogeneous Neumann condition ($\partial_n g = 0$) is applied on the free surface, the sidewalls and over the orifice. For the velocity field, a no-slip condition ($\mathbf{u} = \mathbf{0}$) is prescribed on the solid bottom boundaries; a perfect slip condition ($v_n = 0$ with free tangential velocity) is used on the sidewalls. For the pressure field, a zero gauge pressure ($p = 0$) is prescribed at the free surface and over the orifice.

\begin{figure}
\begin{centering}	
 \hspace{-10.5cm} $(a)$\\
 \hspace{0.4cm} \includegraphics[height=6.5cm]
{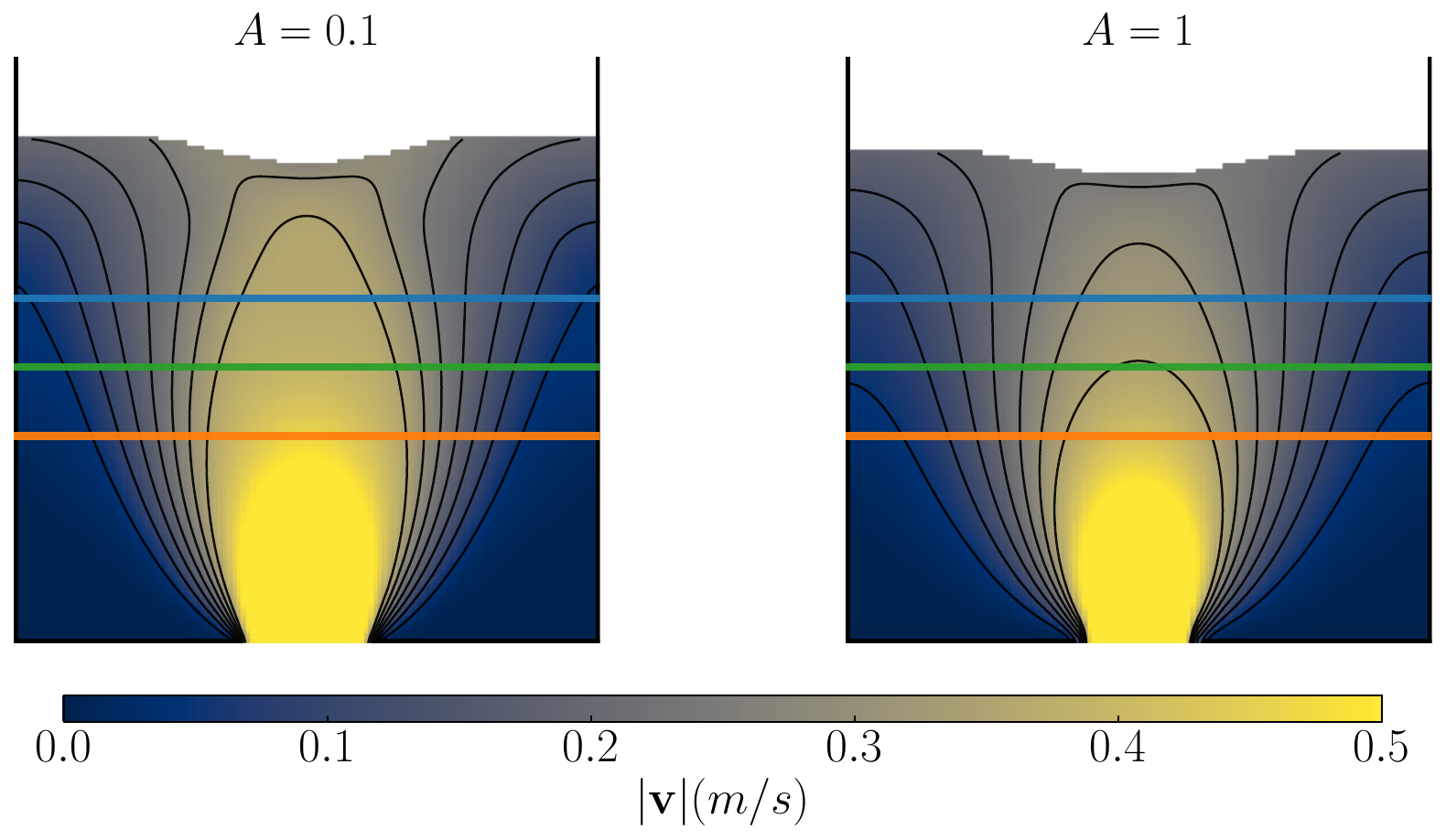}\\
 \hspace{-10.5cm} $(b)$\\
\includegraphics[height=5cm]
{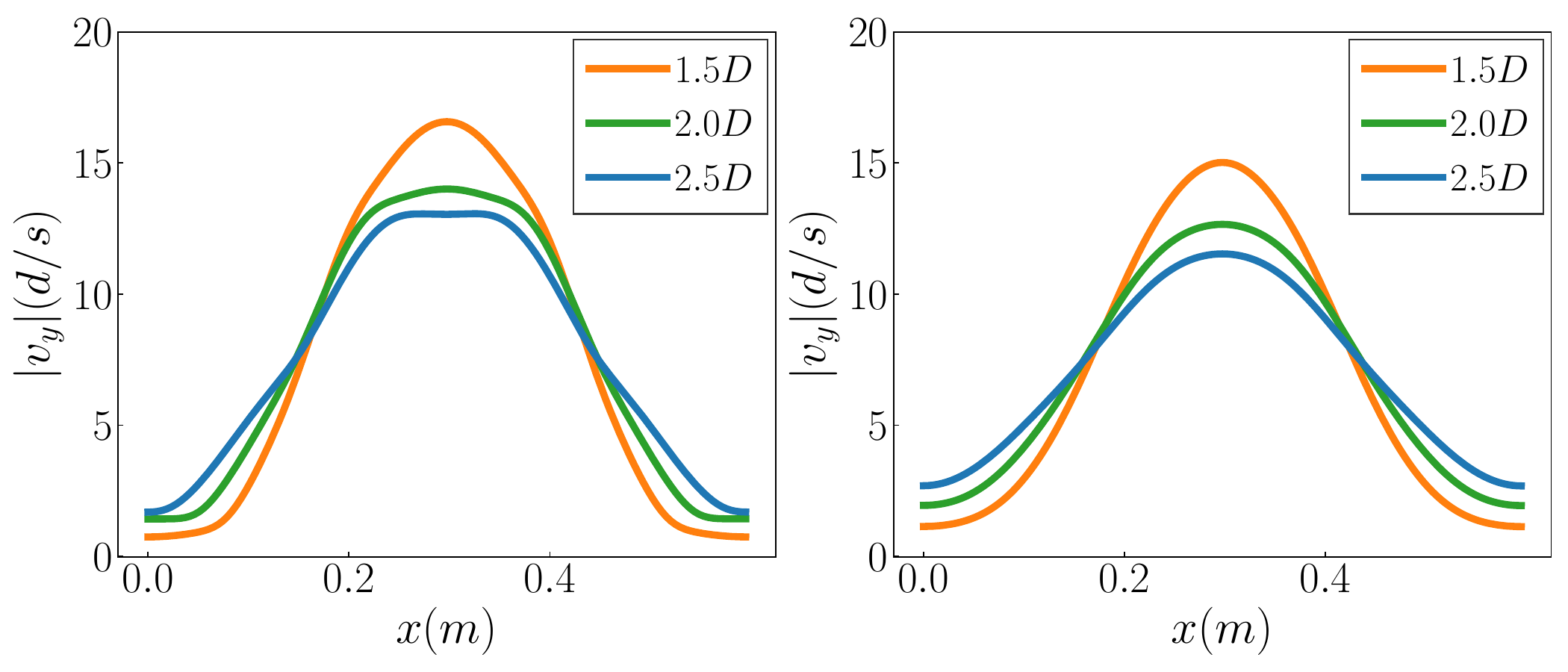}
\caption{\label{fig:kamrin_comparison} Comparison of the velocity fields and vertical velocity profiles obtained from the dynamic NGF model for $A = 0.1$ and $A = 1$, using the parameters listed in table~\ref{tableruns}. (a) Velocity fields for $A = 0.1$ (left) and $A = 1$ (right), with iso-speed lines at 0.05, 0.10, 0.15, 0.20, 0.25, 0.30 and 0.35 m s${}^{-1}$. The scale is the same as in figure 6(a) of \cite{Dunatunga2022}. (b) Vertical velocity profiles $|v_y|$ versus $x$ at heights $1.5D$, $2.0D$ and $2.5D$ above the orifice for $A = 0.1$ (left) and $A = 1$ (right). The snapshots for $A = 0.1$ and $A = 1$ were taken at $0.4$ s and $0.5$ s respectively with nearly the same remaining mass; both cases have reached steady state. The grain size is $d = 0.028$ m and the orifice size is $D = 0.14$ m.}	
\end{centering}	
\end{figure}

We first performed a convergence test to determine appropriate spatial and temporal resolutions. Figure~\ref{fig:convergence} presents the vertical velocity profiles $|v_y|$ versus $x$ at heights $1.5D$, $2.0D$ and $2.5D$ above the orifice, corresponding to the horizontal lines in figure~\ref{fig:kamrin_comparison}(a), obtained with different mesh sizes and time steps. The coarsest simulation ($\Delta x = L/2^5$, $\Delta t = 0.004$ s) becomes numerically unstable after $t = 0.15$ s, so its results are shown at that time. The other three simulations (with $\Delta x = L/2^6$, $L/2^7$ and $L/2^8$, and corresponding $\Delta t = 0.001$, $0.00025$ and $0.0000625$ s) are shown at $t = 0.45$ s, where the flow is well-developed and numerically stable. The results obtained with $\Delta x = L/2^5$ differ considerably from those with the finer grids, and a slight increase in the discharge rate is observed as the grid is refined from $\Delta x = L/2^6$ to $L/2^8$. A similar trend was noted by \cite{Dunatunga2022} in their convergence study, where they attributed the increase to improved resolution of the orifice geometry. In our case, however, we attribute this behaviour primarily to the splitting scheme used for the NGF evolution (equations~\eqref{gstar} and~\eqref{dynamicNGF2}); as $\Delta t$ is reduced, the explicit diffusive step in equation~\eqref{gstar} becomes less pronounced, and the solution more closely approaches the local $\mu(I)$ limit. This interpretation is supported by the observation that the convergence test performed with the local $\mu(I)$ model does not exhibit the same pronounced sensitivity to grid refinement. For the simulations in the present section, we adopt $\Delta x = L/2^7$ with $\Delta t = 0.00025$ s,  which provides a balance between accuracy and computational cost.

Figure~\ref{fig:kamrin_comparison} (a) shows the velocity fields for two silo simulations with the same parameters and initial conditions except for the value of $A$, using the same scale as figure~6(a) in \cite{Dunatunga2022}. The grain size is $d = 0.028$ m and the orifice size is $D = 0.14$ m, giving $D/d = 5$. Both simulations have nearly the same amount of mass remaining in the silo when imaged, with the snapshot for the $A = 0.1$ case taken at $0.4$ s and that for the $A = 1$ case taken at $0.5$ s. In both cases, the flow rate and velocity have already reached a steady state. Iso-speed lines are drawn at $0.05$, $0.10$, $0.15$, $0.20$, $0.25$, $0.30$ and $0.35$ m s$^{-1}$. In the $A = 1$ case, these contours are more widely spaced, indicating stronger non-local diffusion compared with $A = 0.1$. The vertical velocity profiles in figure~\ref{fig:kamrin_comparison}(b) are taken at horizontal slices at heights $1.5D$, $2.0D$ and $2.5D$ above the orifice, as indicated in figure~\ref{fig:kamrin_comparison}(a). For the $A = 0.1$ case, the flow is more plug-like with a thinner shear band near the sidewalls, consistent with the local $\mu(I)$ model solution as reported in \cite{Kamrin2010, STA12, Staron2014}, while the $A = 1$ case exhibits broader spreading. Our results agree well with those of \cite{Dunatunga2022}, despite the use of different numerical approaches.

 \begin{figure}
\begin{centering}	
 \hspace{-5.5cm} $(a)$ \hspace{6.cm} $(b)$\\
\includegraphics[height=5cm]
{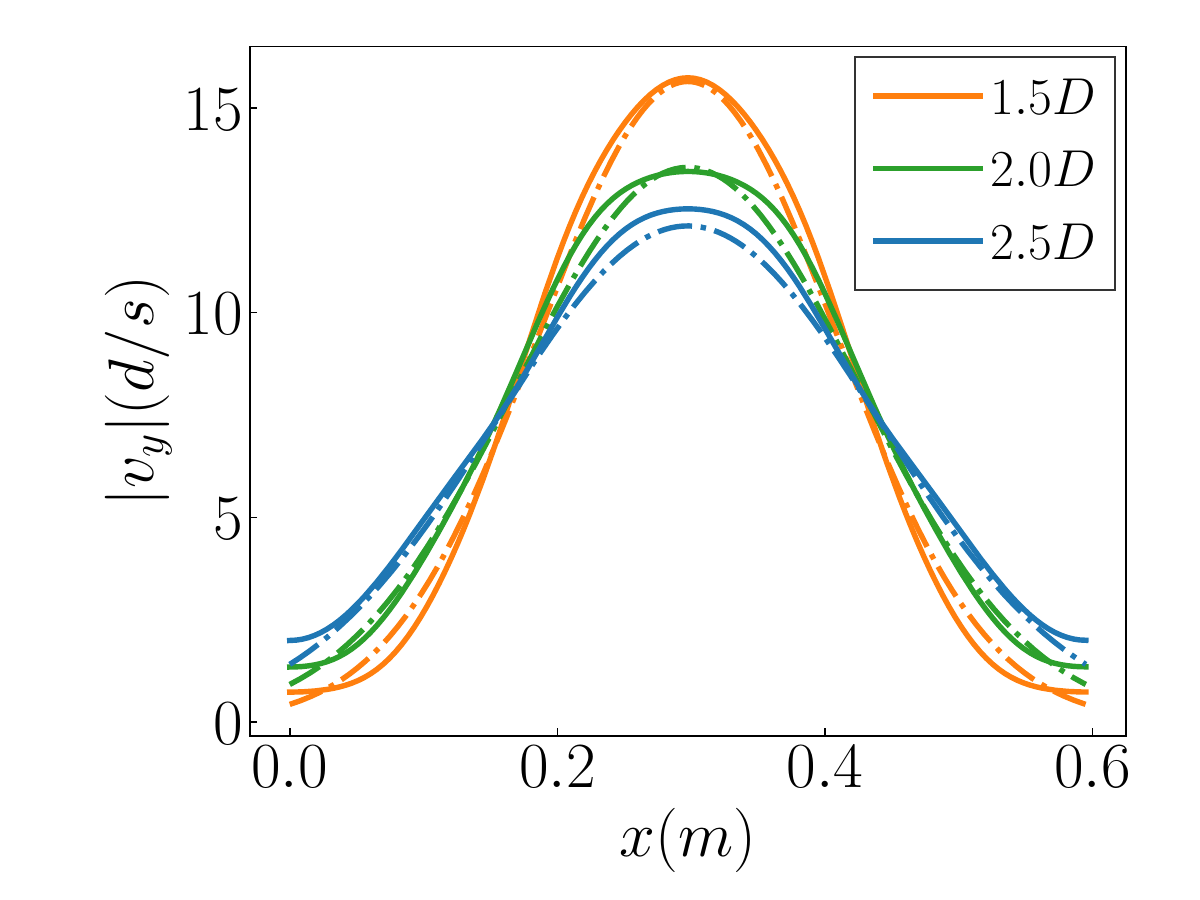}
\includegraphics[height=5cm]
{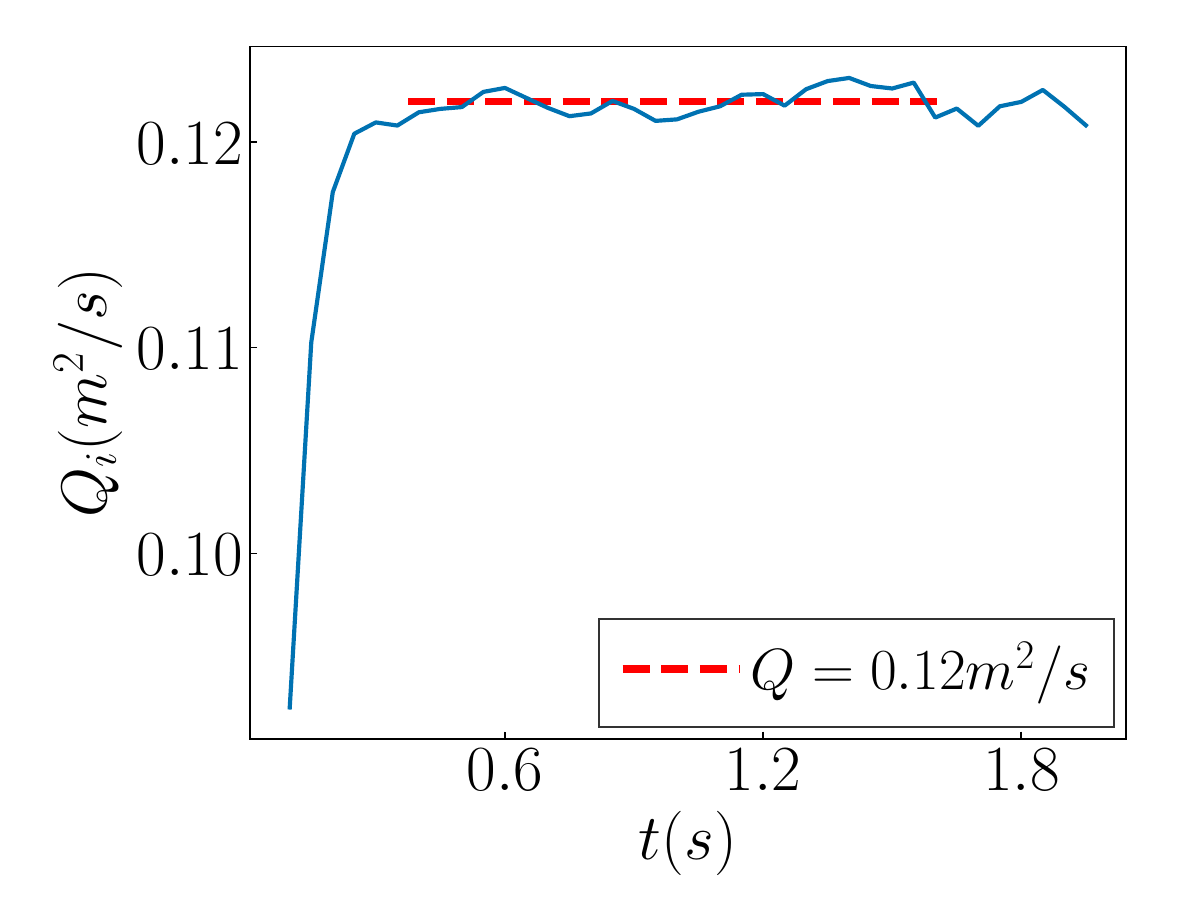}
\caption{\label{gaussianvelocity} Vertical velocity profiles and instantaneous flow rate obtained from the dynamic NGF model at $A = 0.5$, with all other parameters the same as in figure~\ref{fig:kamrin_comparison}. (a) Vertical velocity at heights $1.5D$, $2.0D$ and $2.5D$ ($t = 0.45~\mathrm{s}$). Simulation data are plotted with solid lines; dash-dot lines are Gaussian fits (equation~\eqref{gaussianvelocityeq}) with $B_{\text{k}} = 1.05d$. (b) Instantaneous flow rate $Q_i$ versus time $t$. The red dashed line represents the mean flow rate, which is $Q = 0.12~\mathrm{m}^2/\mathrm{s}$.}	
\end{centering}	
\end{figure}

Following \cite{Dunatunga2022}, we also compare the velocity profiles at $A = 0.5$ with a Gaussian-like fit. The vertical velocity profiles taken at horizontal slices at heights $1.5D$, $2.0D$ and $2.5D$ at time $t = 0.45$ s are plotted in figure~\ref{gaussianvelocity}(a). This Gaussian-like fit originates from the kinematic model proposed by Nedderman and T\"uz\"un, which is based on a constitutive law relating the horizontal and vertical velocity components (\cite{Tuzun1979, Nedderman1979, Choi2005, Irvine2023}). Combined with the incompressibility assumption, this yields a diffusion equation for the downward velocity, where the vertical coordinate $y$ plays the role of time. For a semi-infinite quasi-two-dimensional system with a point-like orifice at $y = 0$, the downward velocity profile near the orifice follows a Gaussian form:
\begin{equation}
v = -\displaystyle\frac{Q}{\sqrt{4\pi B_{\text{k}} y}} e^{-x^2/4 B_{\text{k}} y},
\label{gaussianvelocityeq}
\end{equation}
where $Q$ is the volumetric flow rate per unit thickness of the silo, $B_{\text{k}}$ controls the width of the Gaussian, and $y$ is the height above the orifice. The kinematic model has been tested extensively in silo discharge experiments, and the fitting parameter $B_{\text{k}}$ has been measured by various groups. Values ranging from $B_{\text{k}} \approx 1.3d$ to $4d$ have been reported (\cite{Nedderman1979, Tuzun1979, Medina1998, Samadani1999, Choi2005, Kamrin2007, Garcimartin2011, Zuriguel2019}). The success of the kinematic model lies in the fact that a single parameter $B_{\text{k}}$ suffices to reproduce the entire flow field (\cite{Choi2005}).

 \begin{figure}
\begin{center}
 \hspace{-13.cm} $(a)$\\
 \hspace{1.cm} \includegraphics[height=6.cm]{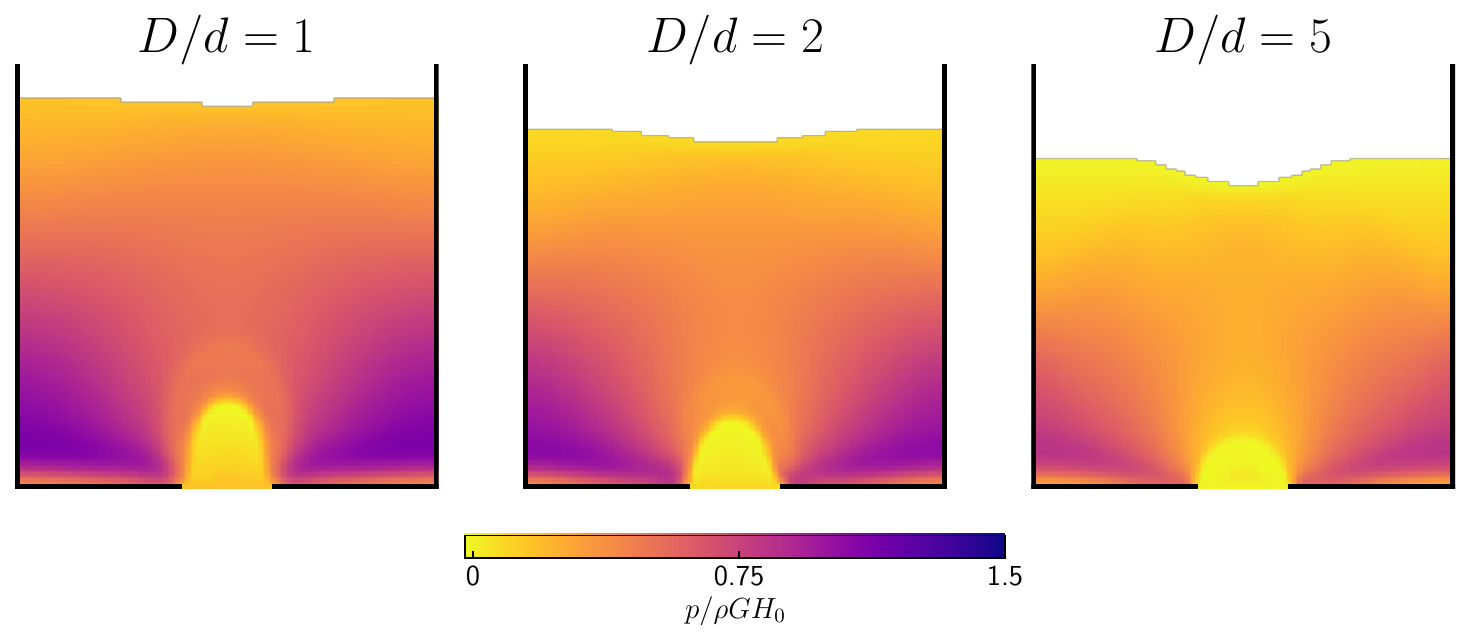}\\
 \hspace{-13.cm} $(b)$\\
\includegraphics[height=5cm]
{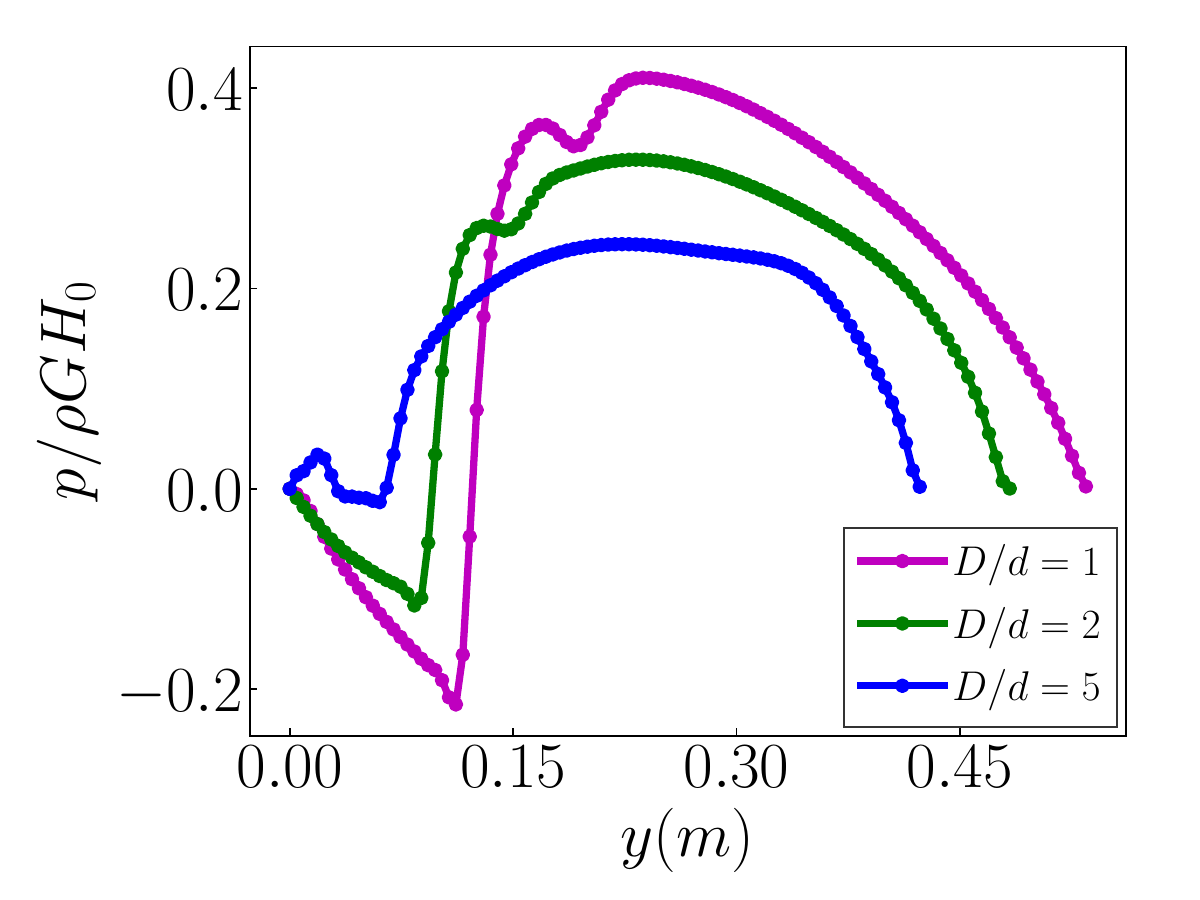}
\caption{\label{fig:pressure_comparison} Comparison of the pressure fields obtained from the dynamic NGF model for three particle sizes at $A = 1$, with a fixed orifice size $D = 0.14$ m, using the parameters listed in table~\ref{tableruns}. (a) Visualization of the pressure fields in silos with $D/d = 1$, $2$ and $5$ (left, centre and right, respectively) at $t = 0.7$ s. The scale is the same as in figure 5(a) of \cite{Dunatunga2022}. (b) Normalized pressure $p/\rho G H_0$ versus vertical position $y$ along the centreline at $t = 0.8$ s for the same three particle sizes (pink, green and blue, respectively).}	
\end{center}	
\end{figure}

Our continuum simulations successfully capture the full kinematic expression, including the $1/\sqrt{y}$ prefactor, indicating that our simulated velocity profiles are close to the experimental ones. In both cases, the lateral spreading and vertical decay of the velocity profiles are characterised by a single parameter $B_{\text{k}}$. Our value of $B_{\text{k}} = 1.05d$ is higher than the value reported in \cite{Dunatunga2022} ($0.6d$), and thus closer to the typical experimental values ($1.3d$ to $4d$), although still slightly lower. Three factors may explain the remaining discrepancy with the experiments. First, following \cite{Dunatunga2022}, we impose perfect-slip conditions on both sidewalls. These idealised conditions differ from those in experiments.~\cite{Fullard2019} showed that the wall response depends on surface roughness and that neither a simple slip nor a no-slip condition fully captures the interactions observed in their MRI experiments. Second, the front and back wall friction is not considered in our current simulations. As shown in \cite{Zhou2017}, incorporating such sidewall friction in the continuum simulation reduces the velocity, which would likely bring our results closer to experimental measurements. Third, the orifice size relative to the overall silo dimensions is not negligible in our simulations, whereas experiments typically employ a much smaller orifice relative to the silo width. Despite these differences, the non-local effect clearly makes the velocity distribution closer to a Gaussian profile compared with the low $A$ case shown in figure~\ref{fig:kamrin_comparison}(b). Figure~\ref{gaussianvelocity}(b) shows the instantaneous flow rate $Q_i$ versus time $t$. The red dashed line indicates the mean flow rate, averaged over the steady-state regime, with $Q = 0.12~\mathrm{m}^2/\mathrm{s}$.

Figure~\ref{fig:pressure_comparison}(a) shows the pressure distributions in the silo for three configurations with $A = 1$ at $t = 0.7$ s, using the same scale as figure~5 in \cite{Dunatunga2022}. The left, centre and right panels correspond to $D/d = 1$, $2$ and $5$, respectively. We observe a marked dip of pressure above the outlet and two high-pressure regions on either side of the outlet. These features are in qualitative agreement with the discrete and continuum simulations of \cite{Staron2014,Dunatunga2022}. As $D/d$ increases, this low-pressure region gradually diminishes, as seen in the corresponding centreline profiles (figure~\ref{fig:pressure_comparison}b), where negative values are visible (set to zero in the contour plots for visualisation). Unlike \cite{Dunatunga2022}, we do not truncate the pressure field near the orifice via a separated phase. Consequently, negative pressure emerges when $D/d$ is sufficiently small. Although negative pressure is physically impossible in granular materials, its emergence may be interpreted as an indicator of the potential for arch formation.

Negative pressure is unphysical in granular media. We employ a regularization technique: when $p \le 0$, we set $\eta_{eff} = 10^{-5}$. This typically occurs near the orifice for large $A$. Since the orifice is a region of high shear rate where $|\dot{\gamma}|$ is large, the effective viscosity is inherently very small, making this treatment physically plausible. In these regions, both $g$ and $\mu$ are also set to zero. These additional assignments do not affect the solution due to the extremely low viscosity, and therefore have no impact on the results presented in this study.

 \begin{figure}
  	\begin{center}
	 \includegraphics[height=6.5cm]{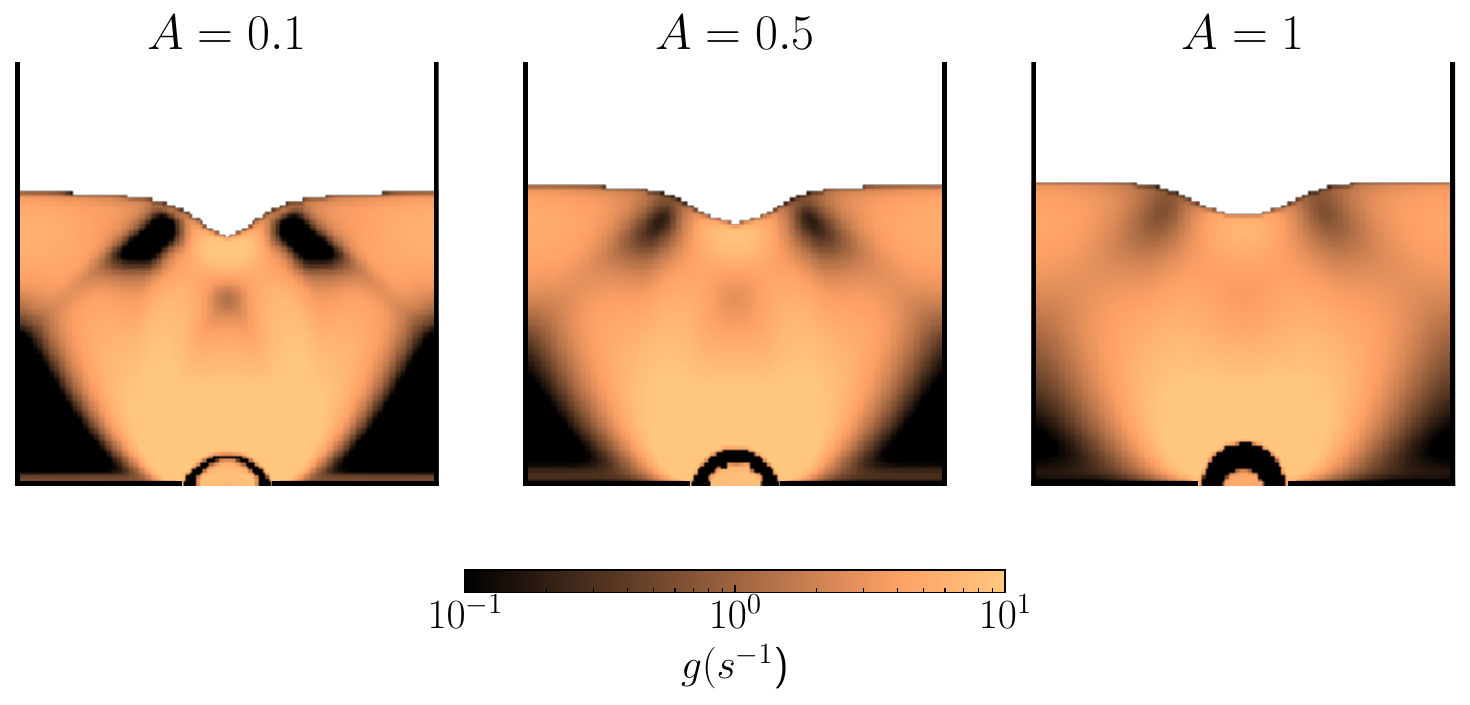}
	  \vspace{-5pt}
	\caption{\label{fig:gfield} Plots of $g$ (logarithmic scale) obtained from the dynamic NGF model for a silo with $D/d = 5$ using the same parameters as in figure \ref{fig:kamrin_comparison}, except for the value of $A$, which increases from $0.1$ to $0.5$ to $1$ from left to right. All snapshots are taken at $t = 0.9$ s. The scale is the same as in figure 8 of \cite{Dunatunga2022}.}
  	\end{center}
\end{figure}

We now examine the $g$ field, shown in figure~\ref{fig:gfield} using the same scale as figure~8 in \cite{Dunatunga2022}. The figure shows the $g$ field (on a logarithmic scale) for a silo with $D/d = 5$, with $A$ increasing from $0.1$ to $0.5$ to $1$ from left to right. The snapshots are taken at $t = 0.9$ s. As expected, lower values of $A$ exhibit less spreading of the $g$ field and a slightly larger flow rate (indicated by the lower free surface) compared with higher values of $A$, indicating that non-local diffusion is weaker. This observation is consistent with the velocity field results discussed earlier. As $A$ increases, the value of $g$ near the orifice becomes smaller, which is also consistent with the findings of \cite{Dunatunga2022}. The region just above the orifice where grains are in free fall is also clearly visible in these plots, indicated by a dark colour where $g$ is set to zero. In our simulations, we set $g = 0$ only when $p \le 0$, which results in a narrower region where $g = 0$ compared to \cite{Dunatunga2022}. Nevertheless, the overall spatial distribution and spreading of the $g$ field show good agreement, indicating that the underlying physical ideas are similar and that our regularization captures the essential non-local behaviour.

Our FVM implementation differs from the MPM framework of \cite{Dunatunga2022} in three respects. We do not introduce a separated phase, the diffusion step is solved implicitly, and the reaction step uses the local $\mu(I)$ value directly. The last choice avoids solving a quadratic equation for $g^{n+1}$. Despite these differences, our simulations reproduce their reported non-local features in velocity and pressure fields, with close quantitative agreement in the velocity profiles. However, because clogging in real granular systems is inherently probabilistic, we do not directly predict flow arrest. Instead, using continuum simulations, we establish a modified Beverloo relation for the discharge rate $Q$ as a function of $A$ and $D/d$, and extract the corresponding cutoff size. In parallel, we derive the clogging probability $J(D/d, A)$ based on the probabilistic framework of \cite{Janda2008}.

\section{Dynamic NGF results with modified parameters: flow rate, clogging probability, and field distributions}\label{DynamicNGFresults}

In the previous section, to facilitate a direct comparison with \cite{Dunatunga2022}, we adopted parameters and boundary conditions similar to theirs. In this section, we introduce a new set of parameters, which are summarised in table~\ref{tab:params}. Instead of using an artificially large $\mu_2 = 50$ to approximate the linear $\mu(I)$ form as in their work, we adopt a more physically realistic value $\mu_2 = 0.68$, consistent with typical glass beads. The friction coefficient $\mu_s = 0.4$ and the characteristic inertial number $I_0 = 0.4$ follow our previous work (\cite{Zhou2016, Zhou2017}). The boundary conditions are also chosen to be consistent with our previous work, in which the velocity is set to zero on the sidewalls. For the fluidity field $g$, we follow \cite{Kamrin2015SM, SalvadorVieira2017} and impose Dirichlet conditions $g = 0$ on the left and right walls and on the solid part of the bottom boundary, with Neumann conditions $\partial_n g = 0$ on the top free surface and over the orifice. The pressure is set to zero at both the orifice and the top free surface. As shown in \cite{Staron2014} and confirmed in our previous work (\cite{Zhou2016}), the choice of sidewall velocity boundary conditions does not affect the discharge flow rate. The initial fill height is set to $H_0/L = 0.9$. The switching time from the local to the non-local simulation is set to $t_{\text{switch}}/\sqrt{L/G} = 0.1$ for the present and subsequent sections.

\cite{Dunatunga2022} treated the non-local amplitude $A$ and grain diameter $d$ as the primary parameters controlling the critical clogging size. However, the orifice size $D$ generally has a stronger influence on silo discharge. We therefore fix $d$ in the main parametric study and examine how $A$ and $D$ affect the flow rate. The influence of the grain size $d$ is examined separately in Appendix~\ref{app:grain_size}. To balance computational cost and accuracy for the large number of parametric simulations performed here, we adopt a coarser grid with $\Delta x = L/2^6$ and $\Delta t/\sqrt{L/G} = 0.001$, whose accuracy has been validated by the convergence test in figure~\ref{fig:convergence} for the previous parameter set.

In the forthcoming analysis we use non dimensional variables hence quantities are normalized as follows: the silo orifice size, $\overline{D} = D/L$; the particle size, $\overline{d} = d/L$; the volume of material remaining in the silo, $\overline{V} = V/L^2$; the flow rate, $\overline{Q} = Q/\sqrt{GL^3}$; and the time, $\overline{t} = t/\sqrt{L/G}$.
Variables without dimension have a bar over them.

\begin{table}
  \begin{center}
\begin{tabular}{cc}
\textbf{Parameter} & \textbf{Value} \\
$\mu_s$ & 0.4 \\
$\mu_2$ & 0.68 \\
$I_0$ &  0.4\\
$t_0 /\sqrt{L/G}$ &  $0.001$\\
$\Delta x$ & $L/2^6$\\
$\Delta t /\sqrt{L/G}$ &  $0.001$

\end{tabular}
\caption{ \label{tab:params} Numerical parameters used for the dynamic NGF simulations with the modified parameter set.}

 \end{center}
\end{table}

\subsection{Flow rate dependence on $A$ and $D/d$, and the Beverloo cutoff size}
 \begin{figure}
\begin{centering}	
 \hspace{-5.5cm} $(a)$ \hspace{6.cm} $(b)$\\
 \includegraphics[height=5cm]
{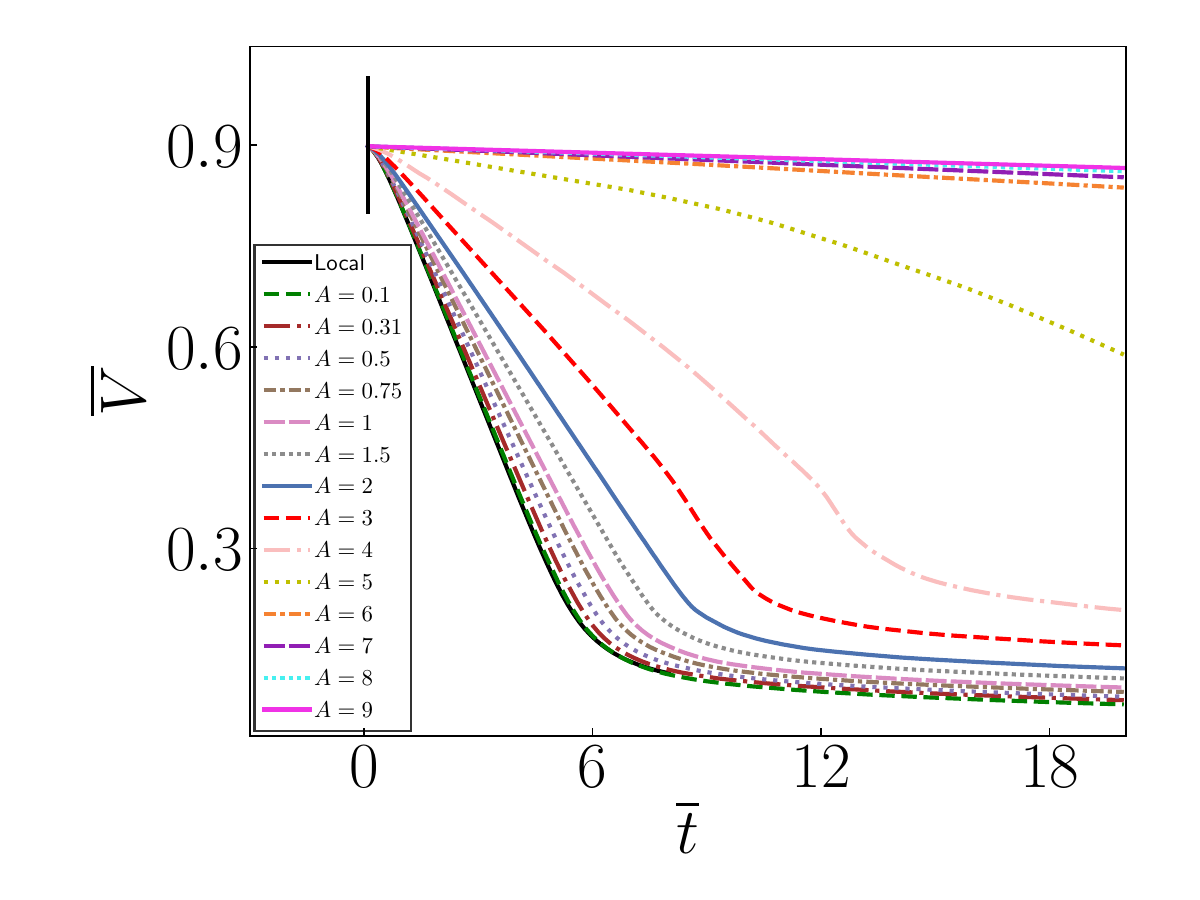}
 \includegraphics[height=5cm]
{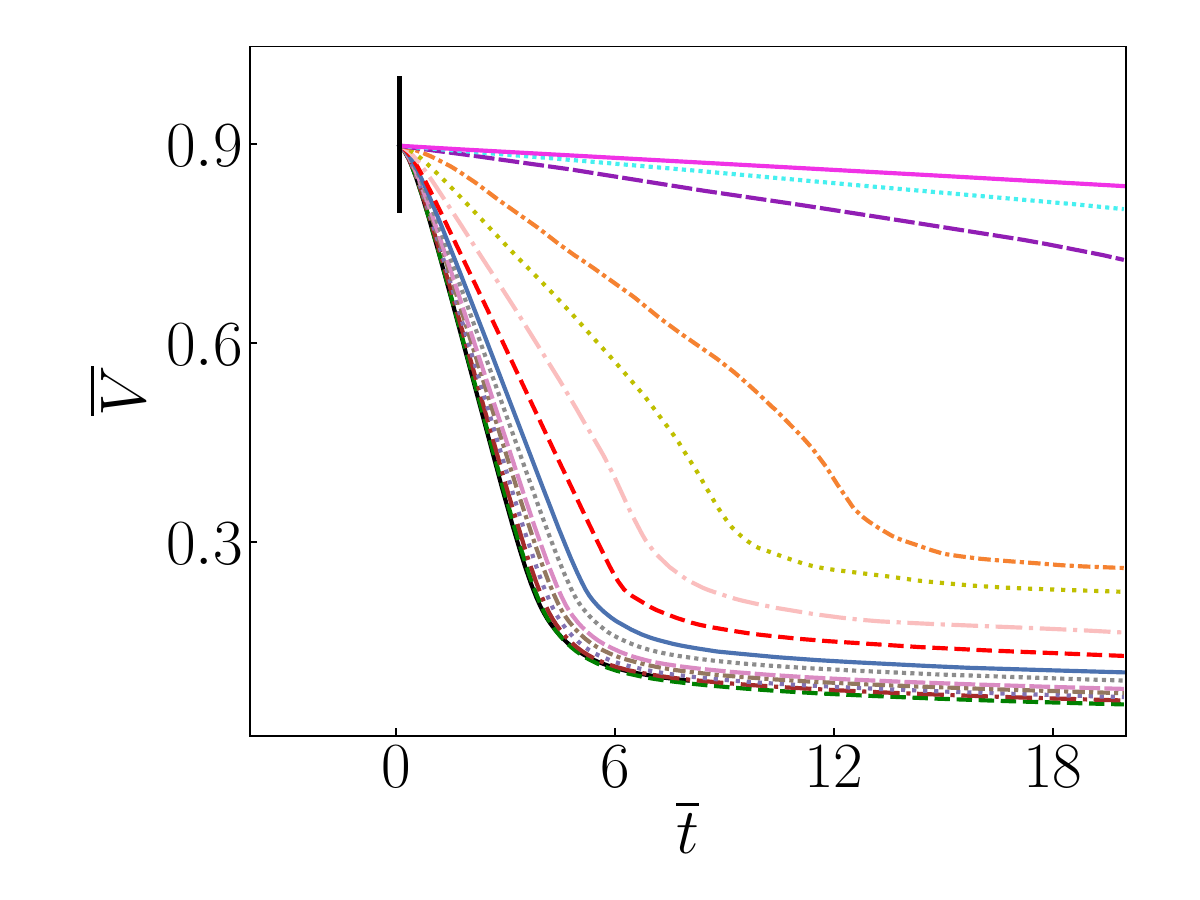}\\
 \hspace{-5.5cm} $(c)$ \hspace{6.cm} $(d)$\\
\includegraphics[height=5cm]
{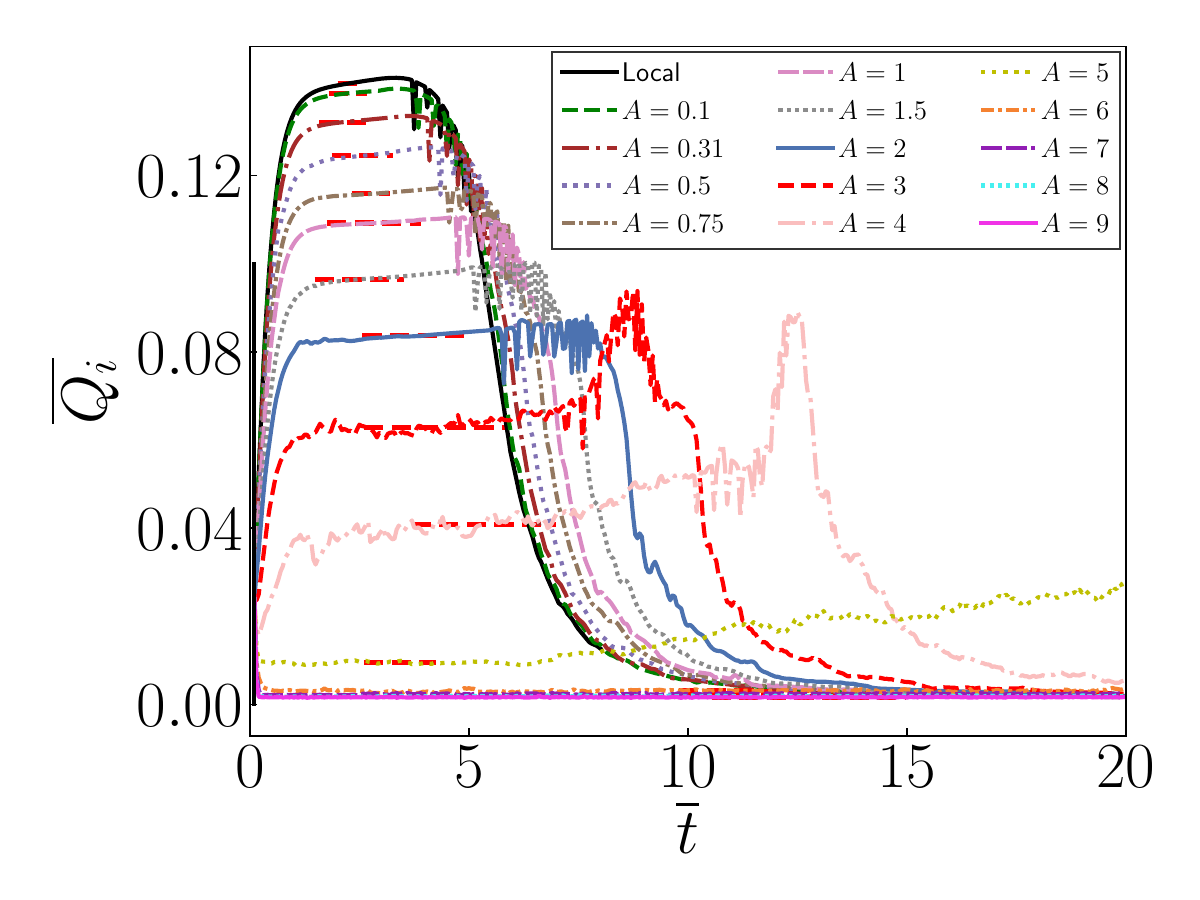}
 \includegraphics[height=5cm]
{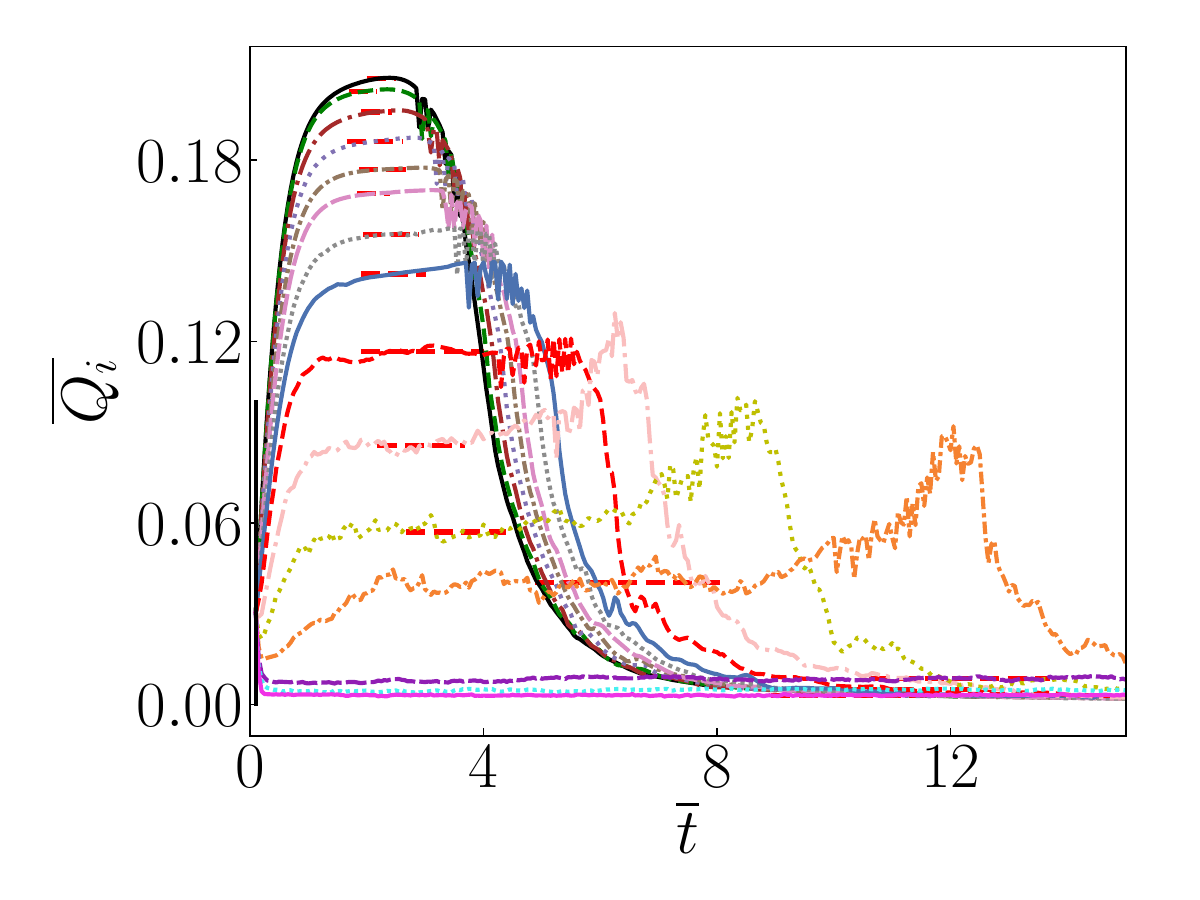}
\caption{\label{fig:flowrate1} The local $\mu(I)$ model and dynamic NGF model with various non-local amplitudes $A$ are compared in terms of transient discharge behaviour, using the modified parameter set (table~\ref{tab:params}). (a,b) Normalized volume remaining in the silo $\overline{V} = V/L^2$ as a function of dimensionless time $\overline{t} = t/\sqrt{L/G}$ for (a) $D/L = 0.25$ and (b) $D/L = 0.3125$. (c,d) Instantaneous dimensionless flow rate $\overline{Q}_i = Q_i/\sqrt{G L^3}$ versus $\overline{t}$ for (c) $D/L = 0.25$ and (d) $D/L = 0.3125$. The dashed lines represent the mean flow rate $Q/ \sqrt{GL^3}$. The vertical solid line in each panel indicates the switching time from the local to the non-local simulation, $t_{\text{switch}}/\sqrt{L/G} = 0.1$. 
}	
\end{centering}	
\end{figure}

Figure~\ref{fig:flowrate1} presents the temporal evolution of the normalized remaining volume $\overline{V}$ and the instantaneous dimensionless flow rate $\overline{Q}_i$ as functions of the dimensionless time $\overline{t}$ for both the local $\mu(I)$ model and the dynamic NGF model with various non-local amplitudes $A$. Panels (a) and (b) show $\overline{V}$ versus $\overline{t}$ for dimensionless orifice sizes $D/L = 0.25$ and $0.3125$, respectively. For the local $\mu(I)$ model and for the dynamic NGF model with small values of $A$, we observe a nearly linear evolution, indicating that the discharge rate is essentially constant and independent of the filling height. As $A$ increases, the magnitude of the slope decreases, indicating a significant reduction in the flow rate. For sufficiently large $A$, the evolution remains linear during most of the discharge, but deviates from linearity in the final stage when the silo is nearly empty. Panels (c) and (d) display the corresponding $\overline{Q}_i$ versus $\overline{t}$ for the same two orifice sizes, confirming that for large $A$ the flow rate increases in the final stage. In addition, pronounced oscillations appear near the end of the discharge, and become more severe as $A$ increases. Notably, the introduction of the dynamic NGF model does not improve stability compared to the local model; on the contrary, for very large $A$, the flow becomes even more unstable. This behaviour will be further discussed at the end of \S\ref{JDdA}. To compute the mean flow rate $Q$, we average only over the stable portion of the discharge, as indicated by the red dashed lines in panels (c) and (d).

\begin{figure}
\begin{centering}	

\includegraphics[height=5cm]
{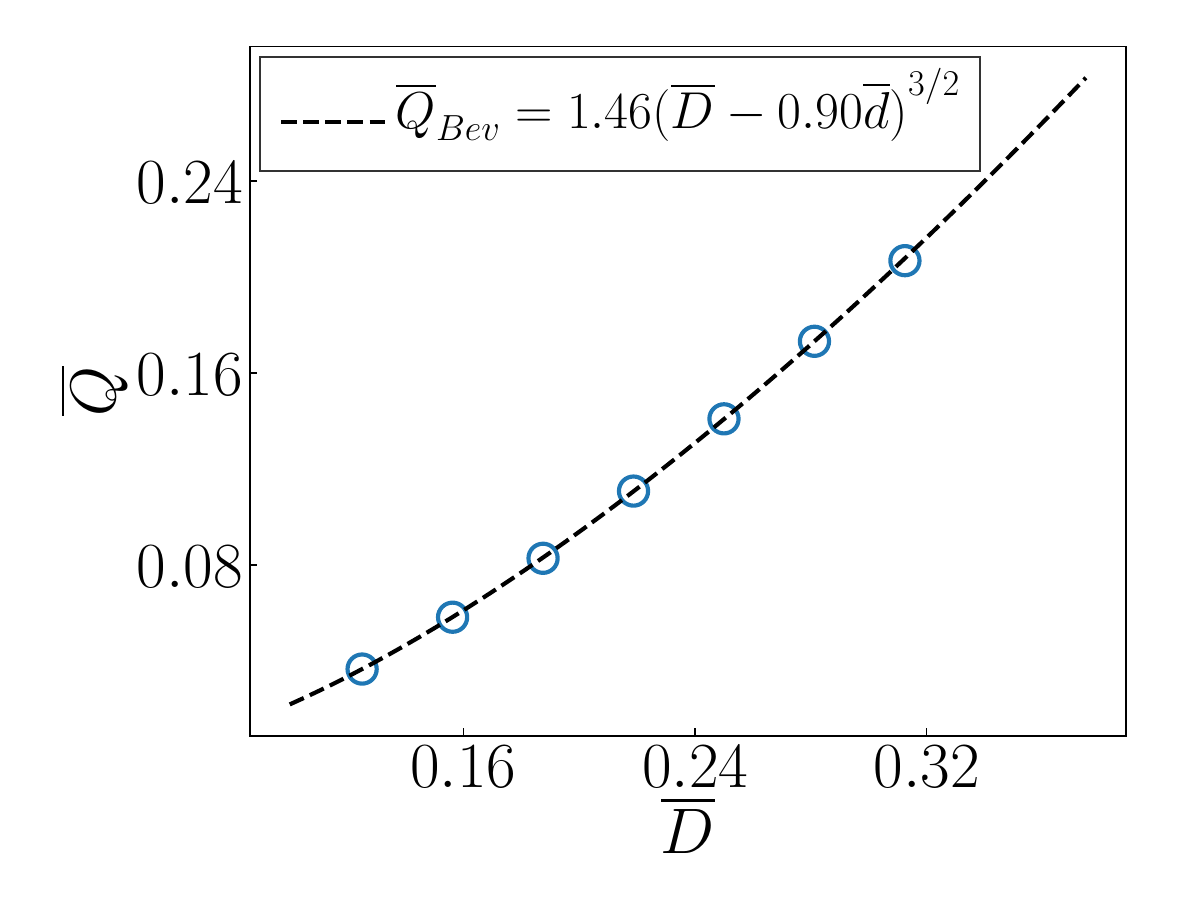}

\caption{\label{QvsDBev} Mean dimensionless flow rate $\overline{Q} = Q/\sqrt{G L^3}$ as a function of $\overline{D} = D/L$ for the local $\mu(I)$ model using the modified parameter set (table~\ref{tab:params}). The black dashed line corresponds to the Beverloo equation~\eqref{Qbev} with $C_{Bev} = 1.46$ and $k_{Bev} = 0.9$.}	
\end{centering}	
\end{figure}

As a reference, we first examine the Beverloo law using the local $\mu(I)$ model. The mean dimensionless flow rate $\overline{Q}$ is plotted as a function of $\overline{D}$ for the local $\mu(I)$ model in figure~\ref{QvsDBev}. The classical Beverloo correlation is given by
\begin{equation}
\label{Qbev}
\overline{Q}_{\text{Bev}} = C_{\text{Bev}} (\overline{D} - k_{\text{Bev}} \overline{d})^{3/2}.
\end{equation}
Our fitting parameters ($C_{\text{Bev}} = 1.46$, $k_{\text{Bev}} = 0.9$) are close to those reported by \cite{STA12} ($C_{\text{Bev}} = 1.49$, $k_{\text{Bev}} = 0.85$), who used a different finite-volume implementation and a lower static friction coefficient $\mu_s = 0.32$, compared with $\mu_s = 0.4$ in our simulations. The slightly smaller $C_{\text{Bev}}$ obtained here is consistent with the larger $\mu_s$ used in our study.

\begin{figure}
\begin{centering}	
 \hspace{-5.5cm} $(a)$ \hspace{6.cm} $(b)$\\
\includegraphics[height=5cm]
{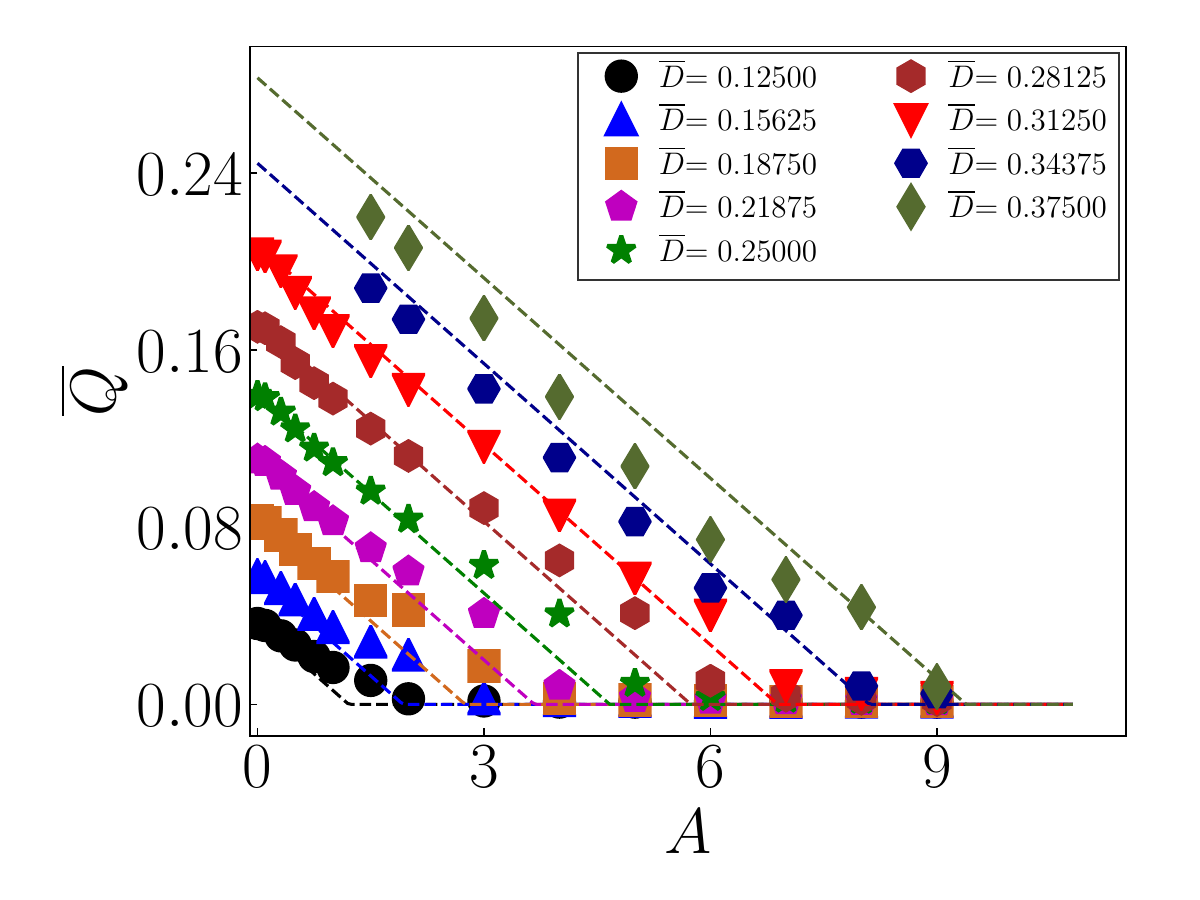}
\includegraphics[height=5cm]
{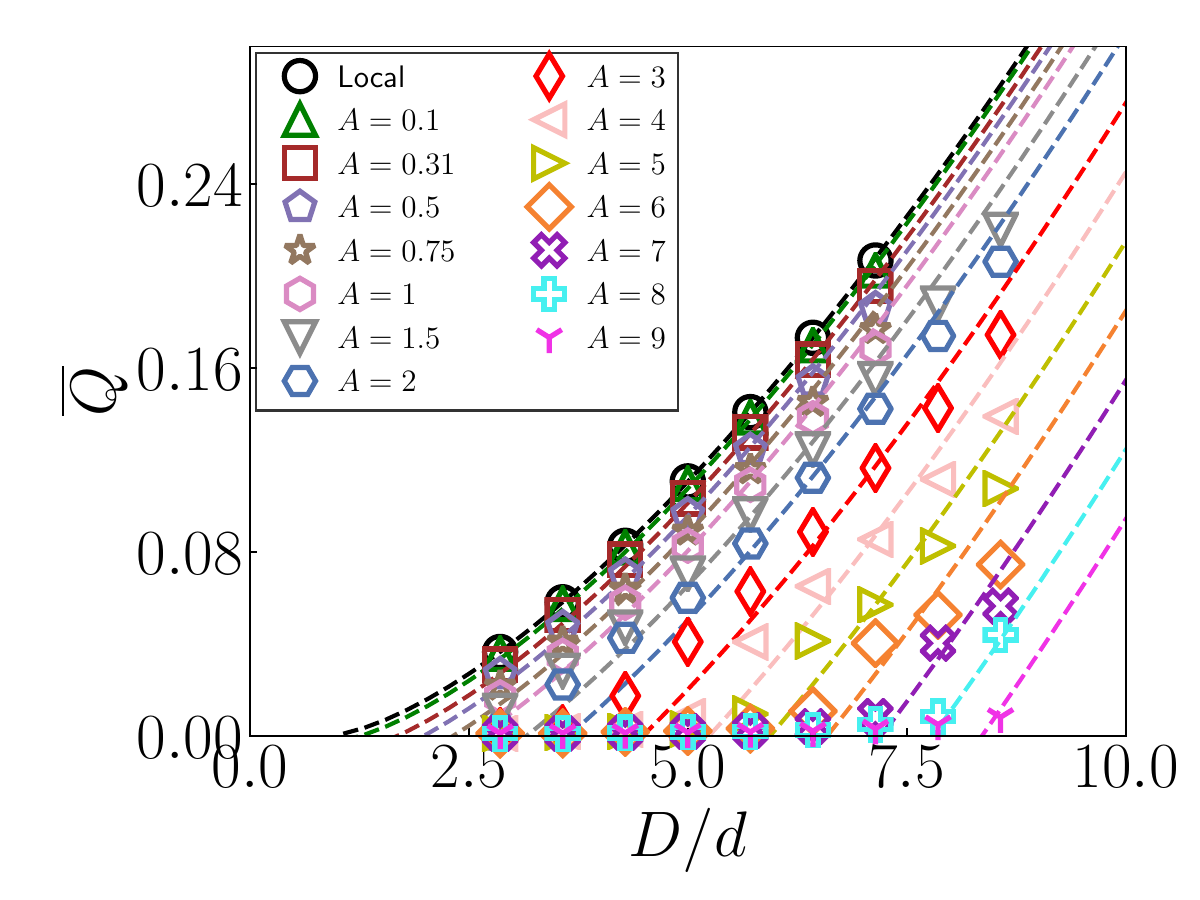}
\caption{\label{QvsDAdiffandvsA} Mean dimensionless flow rate $\overline{Q} = Q/\sqrt{G L^3}$ obtained from the dynamic NGF model using the modified parameter set (table~\ref{tab:params}): (a) versus $A$ for selected values of $\overline{D} = D/L$; (b) versus $D/d$ for selected values of $A$. The dashed lines in both panels correspond to equation~\eqref{Qalaw} truncated at $\overline{Q} = 0$ with $k_s = 3.35$.}	
\end{centering}	
\end{figure}

To investigate the relationship between the flow rate and the non-local amplitude $A$, we plot in figure~\ref{QvsDAdiffandvsA}(a) $\overline{Q}$ as a function of $A$ for selected values of $\overline{D}$. The results show that $\overline{Q}$ decreases approximately linearly with $A$. When $A = 0$, the flow rate reduces to the local $\mu(I)$ model prediction, which follows the Beverloo law (equation~\eqref{Qbev}). This suggests that for finite $A$, the flow rate can be expressed as the local Beverloo value minus a term proportional to $A$. By dimensional analysis and further examining the influence of the grain size $d$ (see Appendix~\ref{app:grain_size}), we propose that the flow rate under the dynamic NGF model can be written as
\begin{equation}
\label{Qalaw}
\overline{Q} = \overline{Q}_{\text{Bev}} - k_s \, \overline{d}^{3/2} A,
\end{equation}
where $k_s$ is an empirical constant. Figure~\ref{QvsDAdiffandvsA}(b) shows $\overline{Q}$ as a function of $D/d$ for selected values of $A$. The dashed lines in both panels correspond to equation~\eqref{Qalaw} truncated at $\overline{Q} = 0$ (i.e., $\max(\overline{Q}_{\text{Bev}} - k_s \overline{d}^{3/2} A, 0)$), with $k_s = 3.35$, which agrees well with the simulation data.

\begin{figure}
\begin{centering}	
 \hspace{-5.5cm} $(a)$ \hspace{6.cm} $(b)$\\
\includegraphics[height=5cm]
{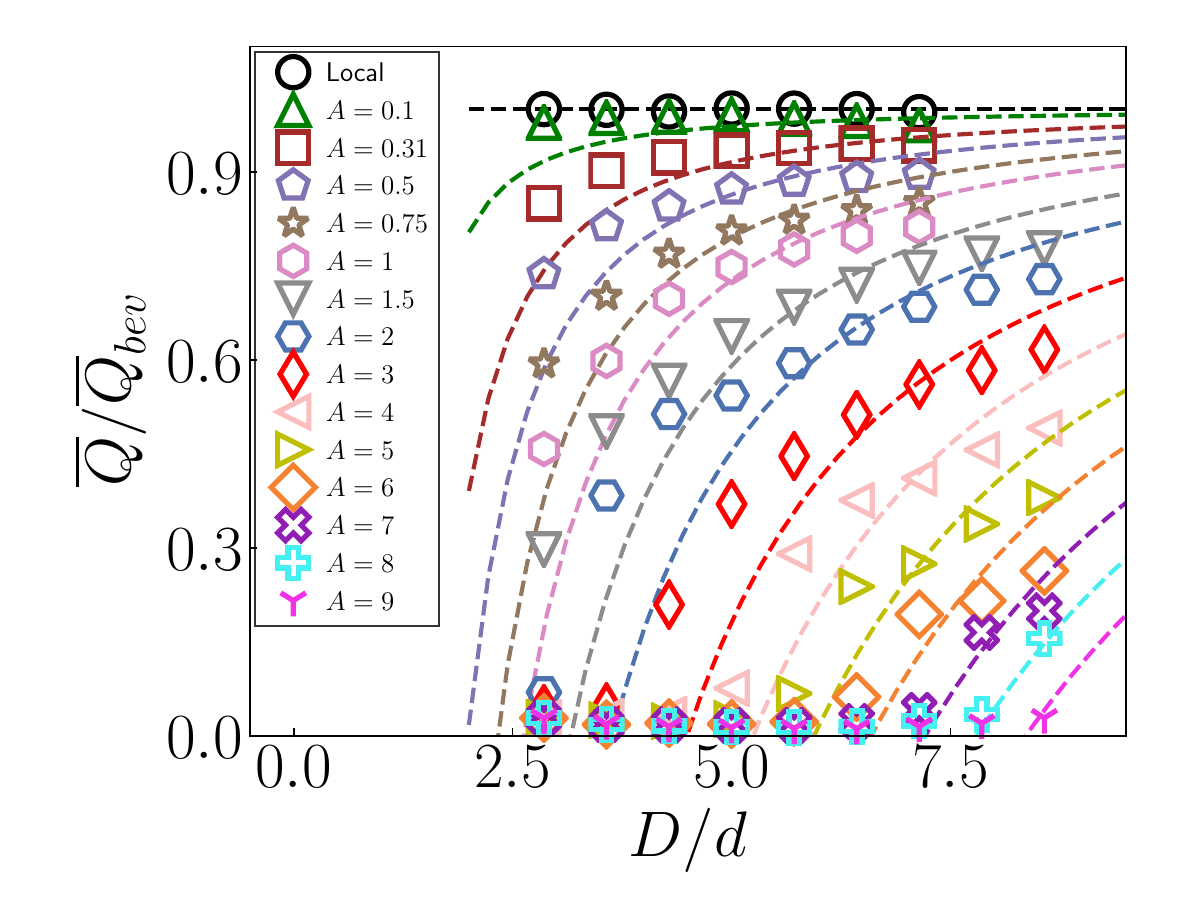}
\includegraphics[height=5cm]
{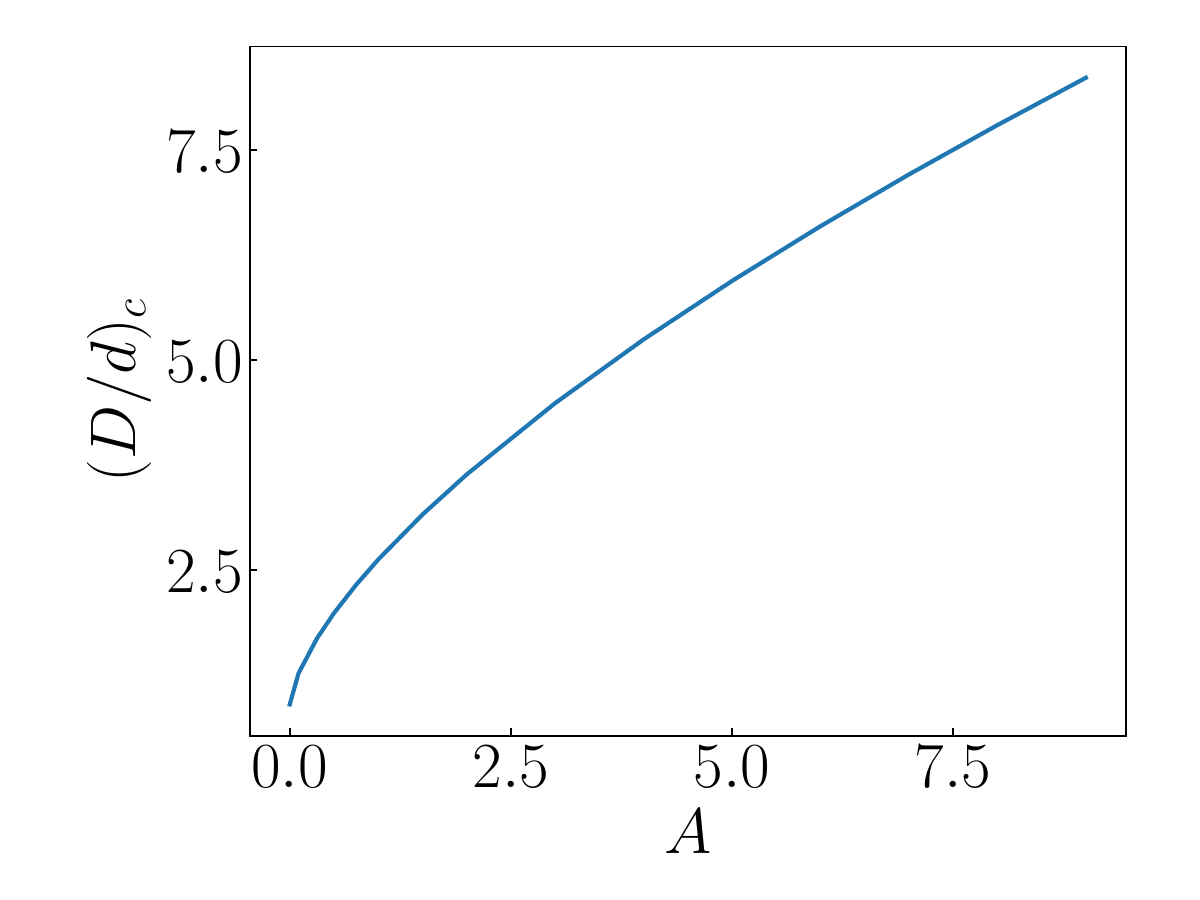}
\caption{\label{QoverQbev} Normalised flow rate and critical cutoff size for selected values of $A$, obtained from the dynamic NGF model using the modified parameter set (table~\ref{tab:params}). (a) Mean dimensionless flow rate $\overline{Q}$ scaled by $\overline{Q}_{\text{Bev}}$ as a function of $D/d$ for selected values of $A$. The dashed lines correspond to equation~\eqref{Qalaw} truncated at $\overline{Q} = 0$ with $k_s = 3.35$. (b) Theoretical critical cutoff size $(D/d)_c$ as a function of $A$ from equation~\eqref{Bevcutoff}.}	
\end{centering}	
\end{figure}

To better visualise the deviation from the local Beverloo law, figure~\ref{QoverQbev}(a) presents the normalized flow rate $\overline{Q}/\overline{Q}_{\text{Bev}}$ as a function of $D/d$ for selected values of $A$, with the same dashed lines as in figure~\ref{QvsDAdiffandvsA}. For small $A$, the normalized flow rate remains close to unity, indicating that non-local effects are weak. As $A$ increases, the curves deviate further from unity, reflecting the stronger influence of non-local diffusion on the flow rate.

Equation~\eqref{Qalaw} is valid for $\overline{Q} \ge 0$; a negative right-hand side implies flow arrest ($\overline{Q} = 0$). Setting $\overline{Q} = 0$ in equation~\eqref{Qalaw} leads to $\overline{Q}_{\text{Bev}} = k_s \overline{d}^{3/2} A$. Substituting the Beverloo relation (equation~\eqref{Qbev}) then gives the critical Beverloo cutoff orifice size 
\begin{equation}
\label{Bevcutoff}
(D/d)_c = (k_sA/C_{\text{Bev}})^{2/3} + k_{\text{Bev}},
\end{equation}
which corresponds to the critical opening size below which the ensemble-averaged flow rate vanishes and clogging is certain (\cite{Thomas2013, Dunatunga2022}). Figure~\ref{QoverQbev}(b) shows the theoretical critical cutoff size $(D/d)_c$ as a function of $A$ obtained from equation~\eqref{Bevcutoff}. When $A = 0.5$ (the value for glass beads \cite{Dunatunga2022}), we obtain $(D/d)_c = 1.996$. This value is consistent with experimental observations for quasi-two-dimensional rectangular silos. For instance, \cite{Benyamine2014} and \cite{Choi2005} reported Beverloo cutoff parameters of $k_{\text{Bev}} = 1.36$ and $k_{\text{Bev}} = 1$, respectively, for such geometries. Interestingly, it is widely observed that the critical $D/d$ for certain clogging in experiments agrees well with the Beverloo cutoff $k_{\text{Bev}}$. Experimental studies have shown that clogging is nearly certain (clogging probability $J \approx 1$) for $D/d$ between approximately 1 and 2 (\cite{To2001, Janda2008}), which is consistent with our result.

\subsection{A probabilistic approach toward clogging prediction}\label{JDdA}
 \begin{figure}
\begin{centering}	
 \hspace{-5.5cm} $(a)$ \hspace{6.cm} $(b)$\\
\includegraphics[height=5cm]
{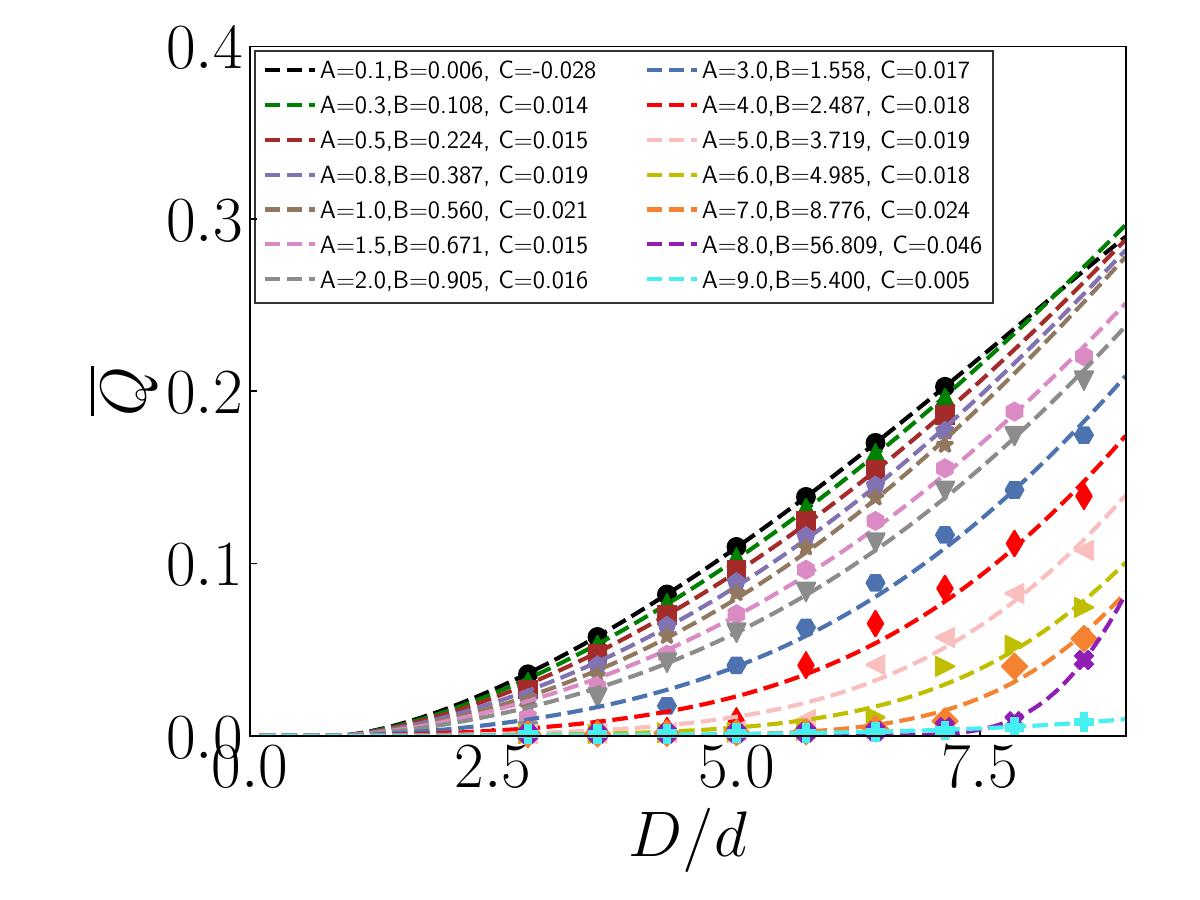}
\includegraphics[height=5cm]
{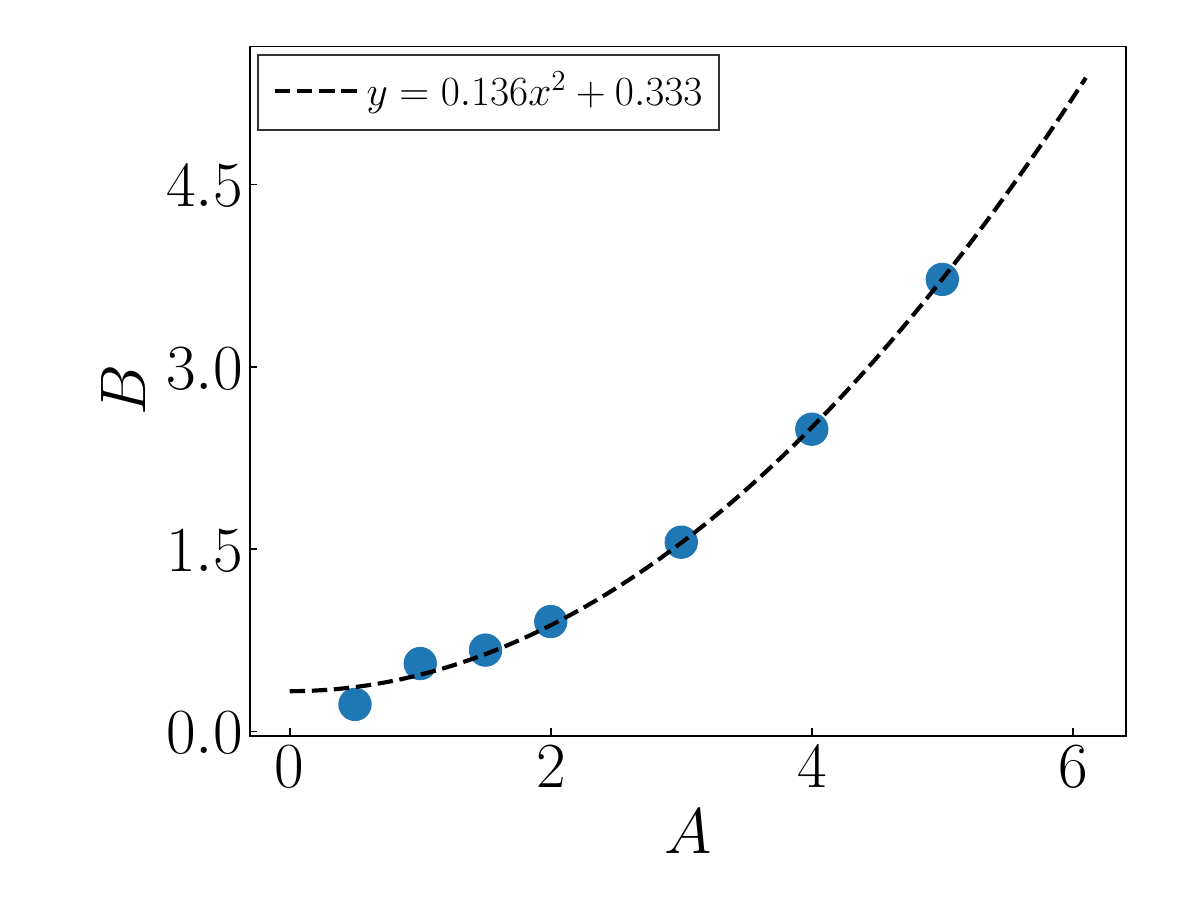}
\caption{\label{fig:Q_fit} Flow rate and fitting parameters for selected values of $A$, obtained from the dynamic NGF model using the modified parameter set (table~\ref{tab:params}). (a) Mean dimensionless flow rate $\overline{Q} =  Q/\sqrt{G L^3}$ as a function of $D/d$ for selected values of $A$. The dashed lines represent equation~\eqref{Qjanda} with fitting parameters $B$ and $C$ given in the legend. (b) Fitting parameter $B$ versus $A$. The black dashed line represents equation $B = 0.136A^2+0.333$.}	
\end{centering}	
\end{figure}

 \begin{figure}
\begin{centering}	
 \hspace{-5.5cm} $(a)$ \hspace{6.cm} $(b)$\\
\includegraphics[height=5cm]
{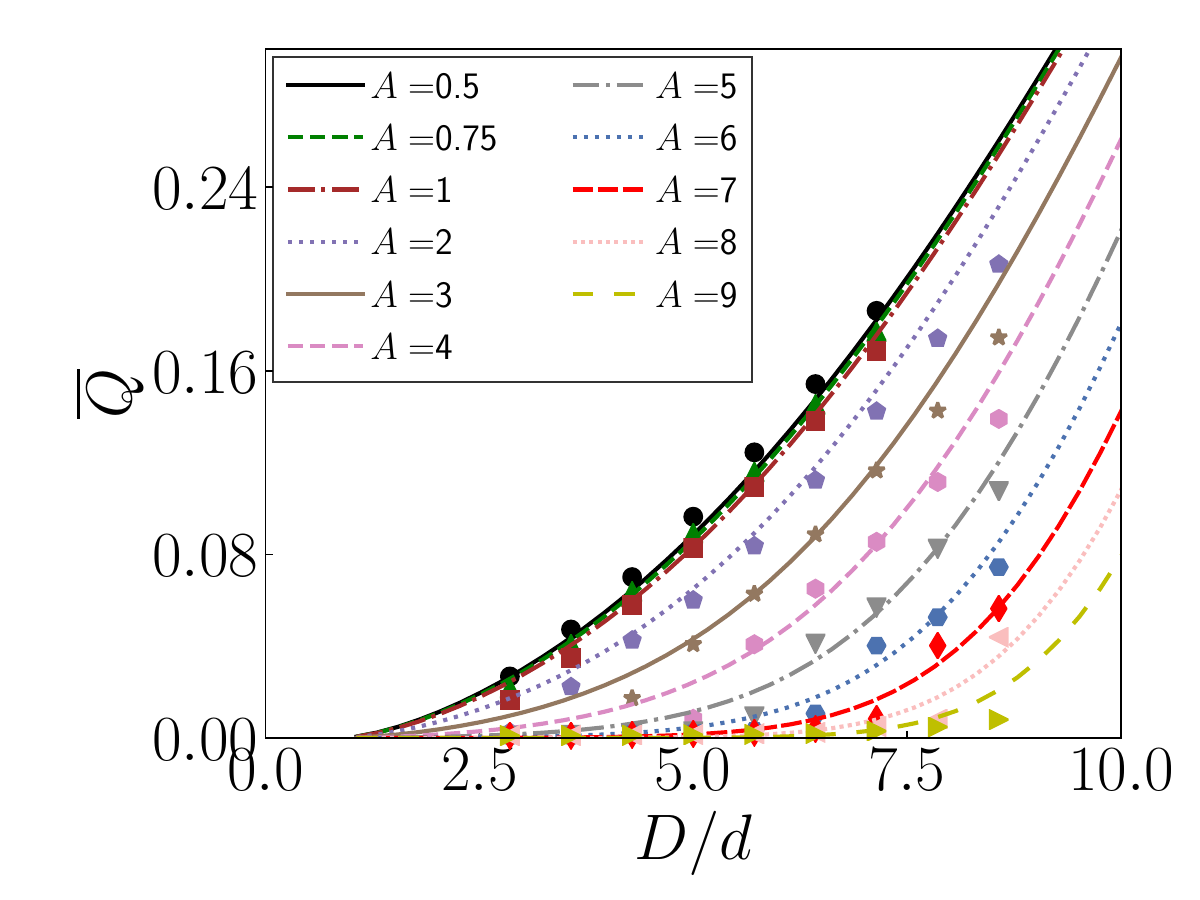}
\includegraphics[height=5cm]
{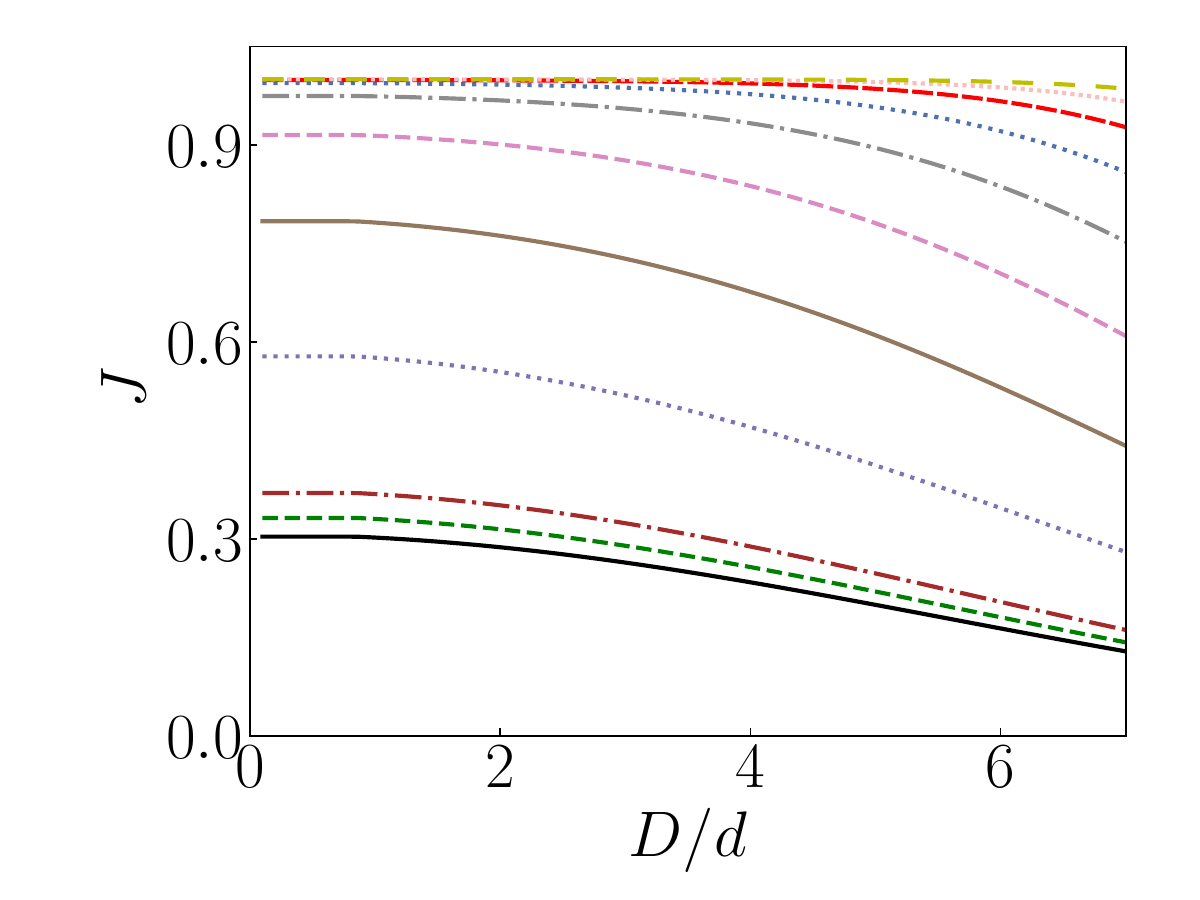}
\caption{\label{QJandatheory} Mean dimensionless flow rate and corresponding clogging probability for selected values of $A$, obtained from the dynamic NGF model using the modified parameter set (table~\ref{tab:params}). (a) Mean dimensionless flow rate $\overline{Q} = Q/\sqrt{G L^3}$ as a function of $D/d$. The lines represent equation~\eqref{Qjanda} with $B = 0.136A^2+0.333$ and $C=0.02$. (b) Theoretical clogging probability $J$ as a function of $D/d$, computed from equation~\eqref{JJanda} using the same $B$ and $C$ from (a).}	
\end{centering}	
\end{figure}

The cutoff size identified above marks the opening below which clogging is certain. For larger openings, however, clogging is no longer certain but probabilistic. To quantify this probability, we adopt the two-dimensional probabilistic arching model of \cite{Janda2008}. In this model, the probability of flow clogging before $N$ particles have been discharged follows the exponential form:

\begin{equation}
\label{Janda2008_eq}
J_{N}(D/d) = 1 - \exp \left[ -N A_J e^{-B_J (\eta_0 D/d)^2} \right],
\end{equation}
where $N$ is the number of grains, and $A_J$ and $B_J\eta_0^2$ are fitting parameters obtained from experimental data.

Based on this conceptual framework, we propose a direct relationship between the flow rate and the clogging probability. We assume that the flow rate obtained from our continuum simulations corresponds to the product of the flow rate in the absence of clogging (which, in the local model limit, follows the Beverloo relation) and the probability that no clogging occurs, i.e., $1 - J$:
\begin{equation}
\label{Q_from_J}
\overline{Q} = \overline{Q}_{\text{Bev}} (1 - J),
\end{equation}
where $\overline{Q}_{\text{Bev}}$ is the Beverloo flow rate for the local model (corresponding to zero clogging probability, $J = 0$). Substituting the exponential form of $J$ from equation~\eqref{Janda2008_eq} leads to
\begin{equation}
\label{Qjanda}
\overline{Q} = \overline{Q}_{\text{Bev}} \exp \left[ -B e^{-C (D/d)^2} \right],
\end{equation}
where $B$ and $C$ are fitting parameters. Figure~\ref{fig:Q_fit}(a) shows $\overline{Q}$ as a function of $D/d$ for selected values of the non-local amplitude $A$, where the dashed lines represent equation~\eqref{Qjanda} with individual fitting parameters $B$ and $C$ for each non-local amplitude $A$ (given in the legend). The dependence of the fitted parameter $B$ on $A$ is shown in figure~\ref{fig:Q_fit}(b), which follows a quadratic relation $B = 0.136A^2 + 0.333$, while $C$ is found to be nearly constant, so we set $C = 0.02$.

Using this unified fit, figure~\ref{QJandatheory}(a) presents $\overline{Q}$ as a function of $D/d$ for selected values of $A$, where the lines correspond to equation~\eqref{Qjanda} with the unified parameters $B = 0.136A^2 + 0.333$ and $C = 0.02$. The agreement between the simulation data and the unified fit remains excellent. Figure~\ref{QJandatheory}(b) shows the corresponding theoretical clogging probability $J$ computed from

\begin{equation}
\label{JJanda}
J = 1 - \exp \left[ -B e^{-C (D/d)^2} \right].
\end{equation}

  \begin{figure}
  	\begin{center}
	 \hspace{-12.5cm} $(a)$\\
 \includegraphics[height=9cm]{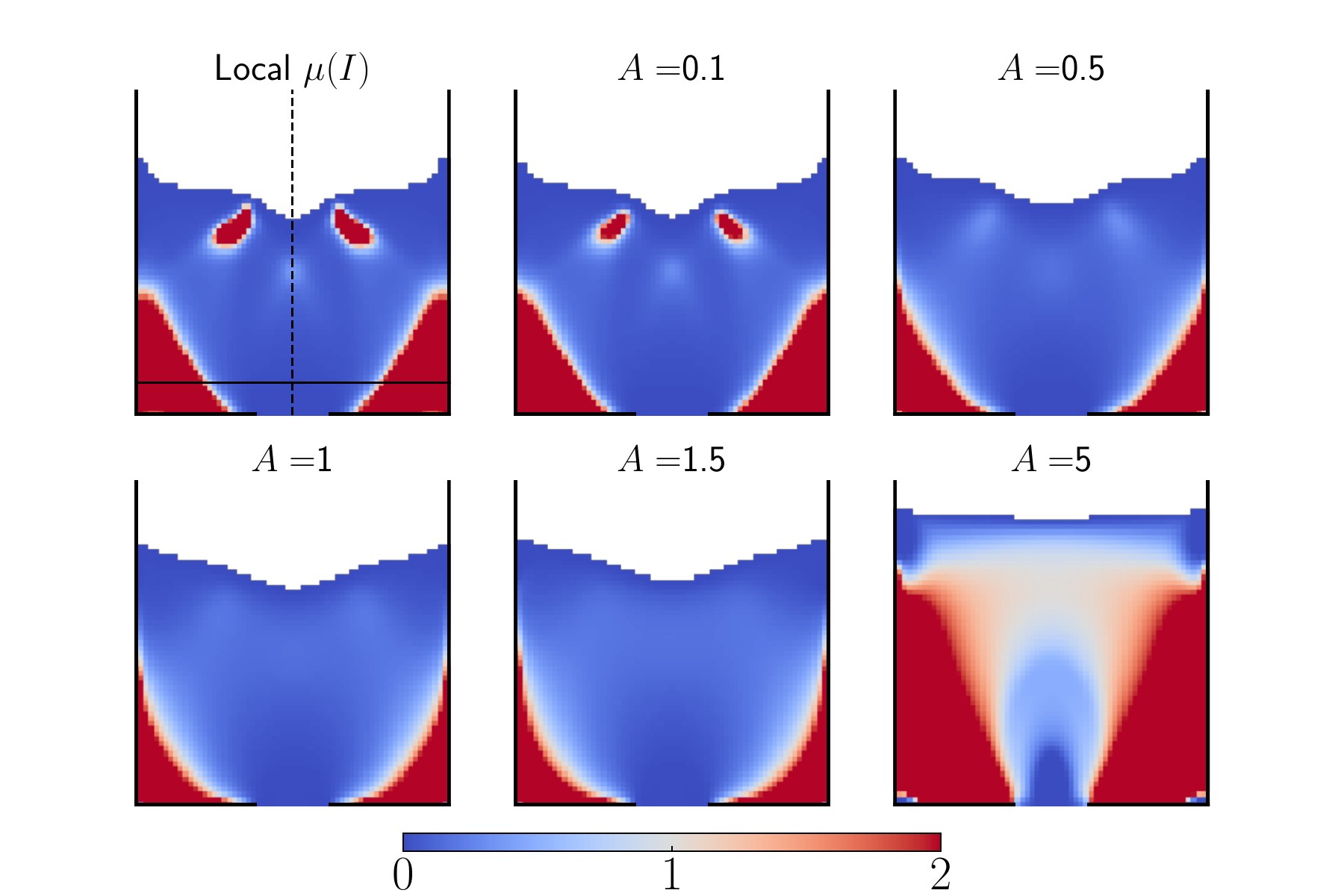}\\
	 \hspace{-3.5cm} $(b)$ \hspace{4.cm} $(c)$ \hspace{4.cm} $(d)$\\
   \includegraphics[height=3.2cm]
{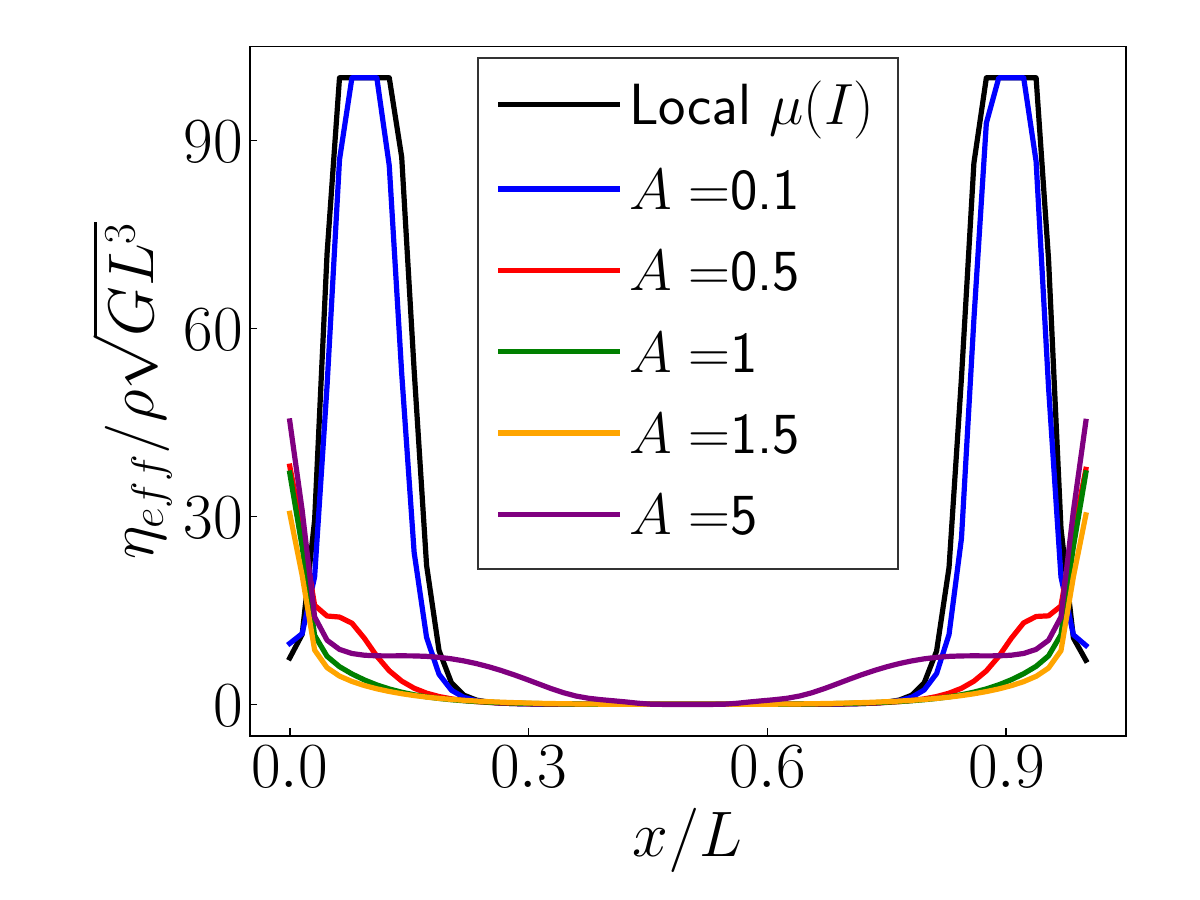}
   \includegraphics[height=3.2cm]
{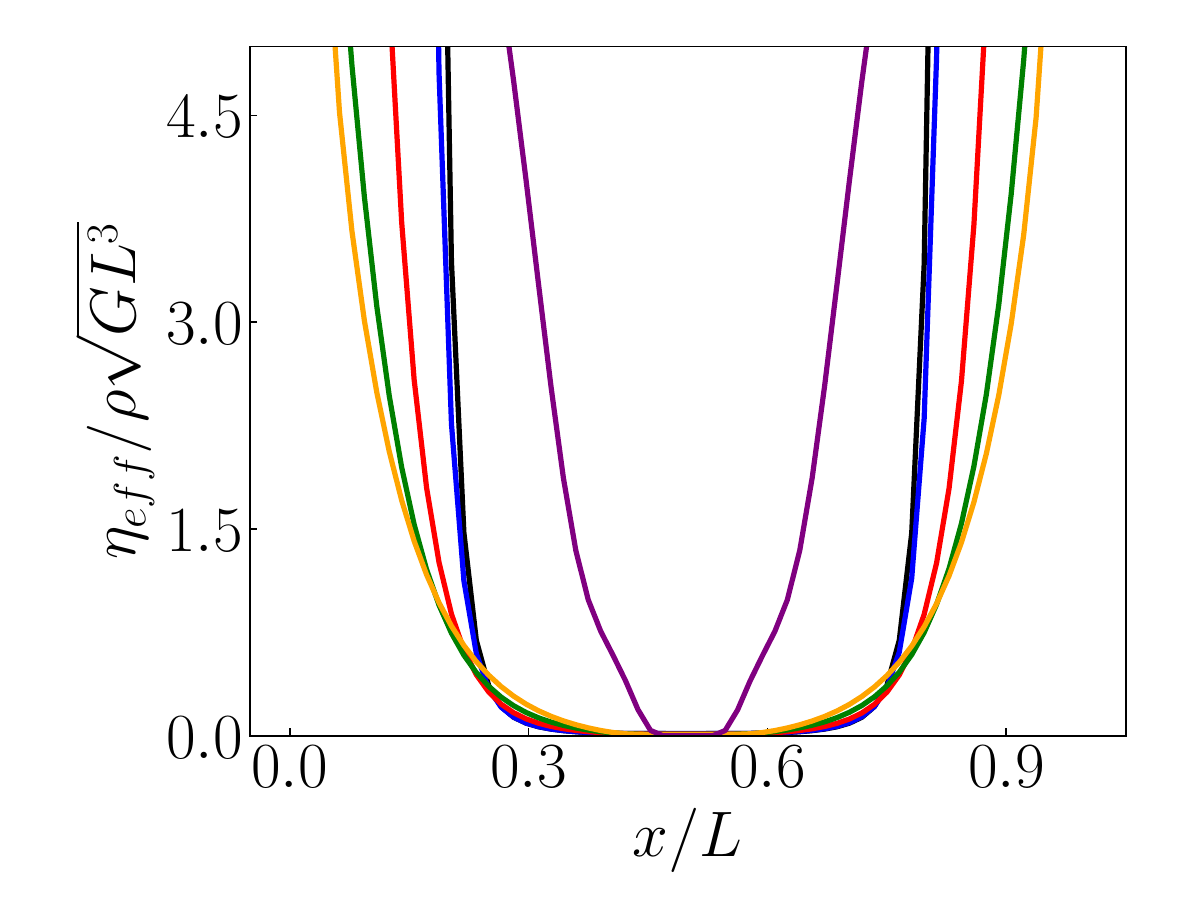}
\includegraphics[height=3.2cm]
{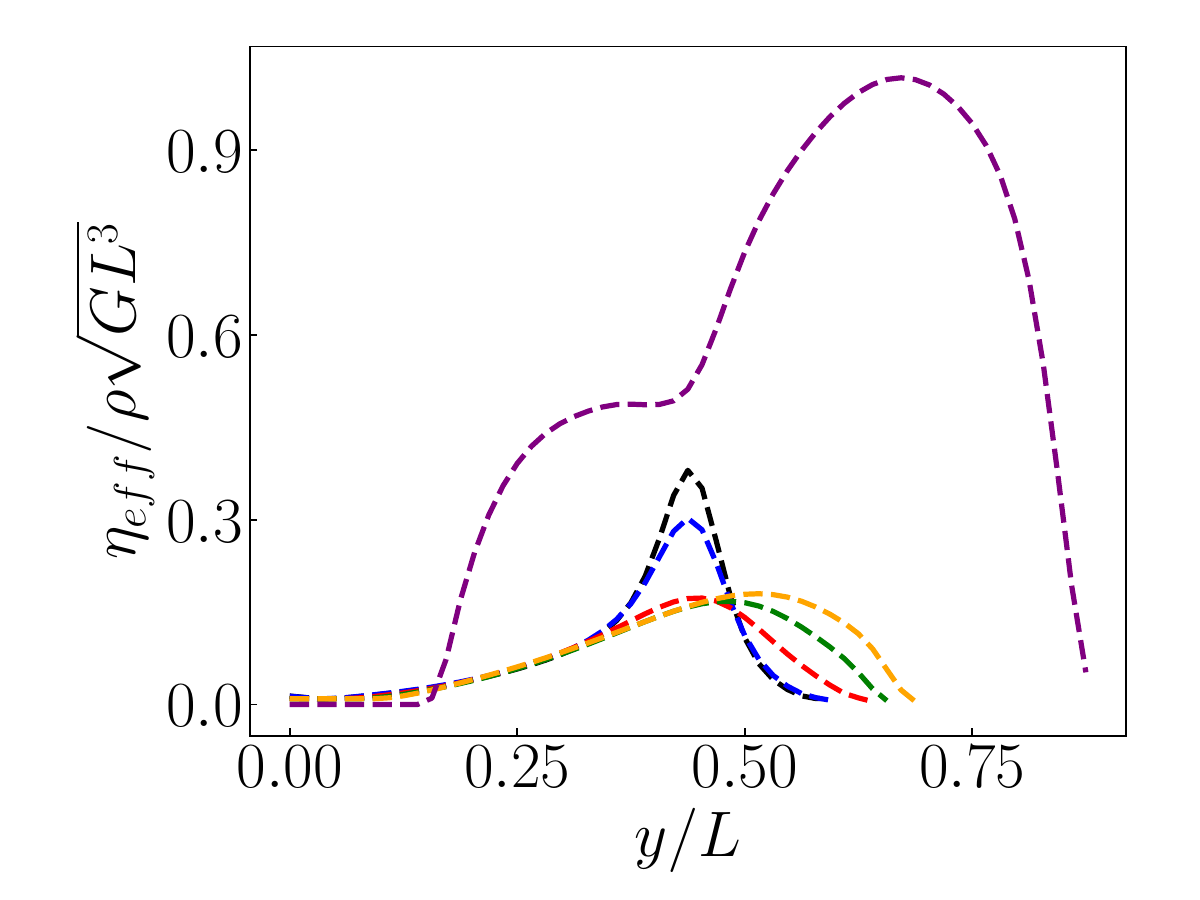}
	\caption{\label{fig:viscosity_fields}  Normalized viscosity $\eta_{\text{eff}} / (\rho \sqrt{G L^3})$ fields and profiles at $\overline{t} = 2$ for $\overline{D} = 0.25$, comparing the local $\mu(I)$ model and the dynamic NGF model with various non-local amplitudes $A$, using the modified parameter set (table~\ref{tab:params}). 
        (a) Field visualization. The solid horizontal line indicates $y/L = 0.1$, and the dashed vertical line indicates $x/L = 0.5$. 
        (b) Profiles of normalized viscosity along the horizontal line at $y/L = 0.1$. 
        (c) Same as (b) but with a reduced vertical axis range $(0, 5)$. 
        (d) Corresponding profiles along the vertical centerline at $x/L = 0.5$.}
  	\end{center}
\end{figure}

The trend of $J$ as a function of $D/d$ is qualitatively similar to the experimental observations of \cite{Janda2008}, with larger $A$ leading to higher clogging probabilities for a given $D/d$, consistent with enhanced non-local effects promoting clogging. 
However, a quantitative comparison with experiments is not attempted here, as the experimental clogging probability depends explicitly on the number of discharged particles $N$, whereas our continuum simulations do not track individual particles. Moreover, the present simulations do not account for variations in the volume fraction $\phi$, which are known to play an important role in the clogging process.

To assess the numerical robustness of the present results, we conducted additional convergence tests by varying $\Delta x$ and $\Delta t$ with the modified parameters, following the same procedure as in figure~\ref{fig:convergence} for the previous parameter set. Refining these parameters also leads to a slight increase in the discharge rate, as the diffusive term becomes less pronounced and the solution approaches the local $\mu(I)$ limit. This behaviour is consistent with the operator-splitting scheme (equation~\eqref{gstar}). This also explains the mild oscillations observed in figures~\ref{fig:flowrate1} and~\ref{fig:flowrate}, the latter of which will be shown in \S\ref{comparenonlocal}. These oscillations are more pronounced at larger $A$ because the diffusion term scales as $A^2$ and the numerical discretisation becomes increasingly sensitive to $\Delta t$. Nevertheless, the relative trends of the discharge ratio $\overline{Q}/\overline{Q}_{\text{Bev}}$ with respect to $A$ and $D/d$ remain largely unchanged, confirming that the conclusions drawn from the unified fit are not affected by numerical discretisation.

\subsection{Field distributions of $\eta_{\text{eff}}$, $\mu$, $|\dot{\gamma}|$ and $I$}
\label{FielddistributionsDNGF}
Having established the quantitative relationships between $\overline{Q}$, $A$ and $D/d$, we now examine the spatial distributions of several key physical quantities. This analysis provides further insight into the underlying mechanisms. The velocity and pressure fields have already been presented and discussed in the previous section in comparison with \cite{Dunatunga2022}; these are provided in Appendix~\ref{app:fields} for completeness. Here, we focus on quantities that have not been shown before, namely the effective viscosity $\eta_{\text{eff}}$, the friction coefficient $\mu$, the shear rate $|\dot{\gamma}|$ and the inertial number $I$. We select the case $\overline{D} = 0.25$ at $\overline{t} = 2$, which corresponds to the steady-state regime (see figure~\ref{fig:flowrate1}(c)).

 \begin{figure}
  	\begin{center}
	 \hspace{-12.5cm} $(a)$\\
 \includegraphics[height=9cm]{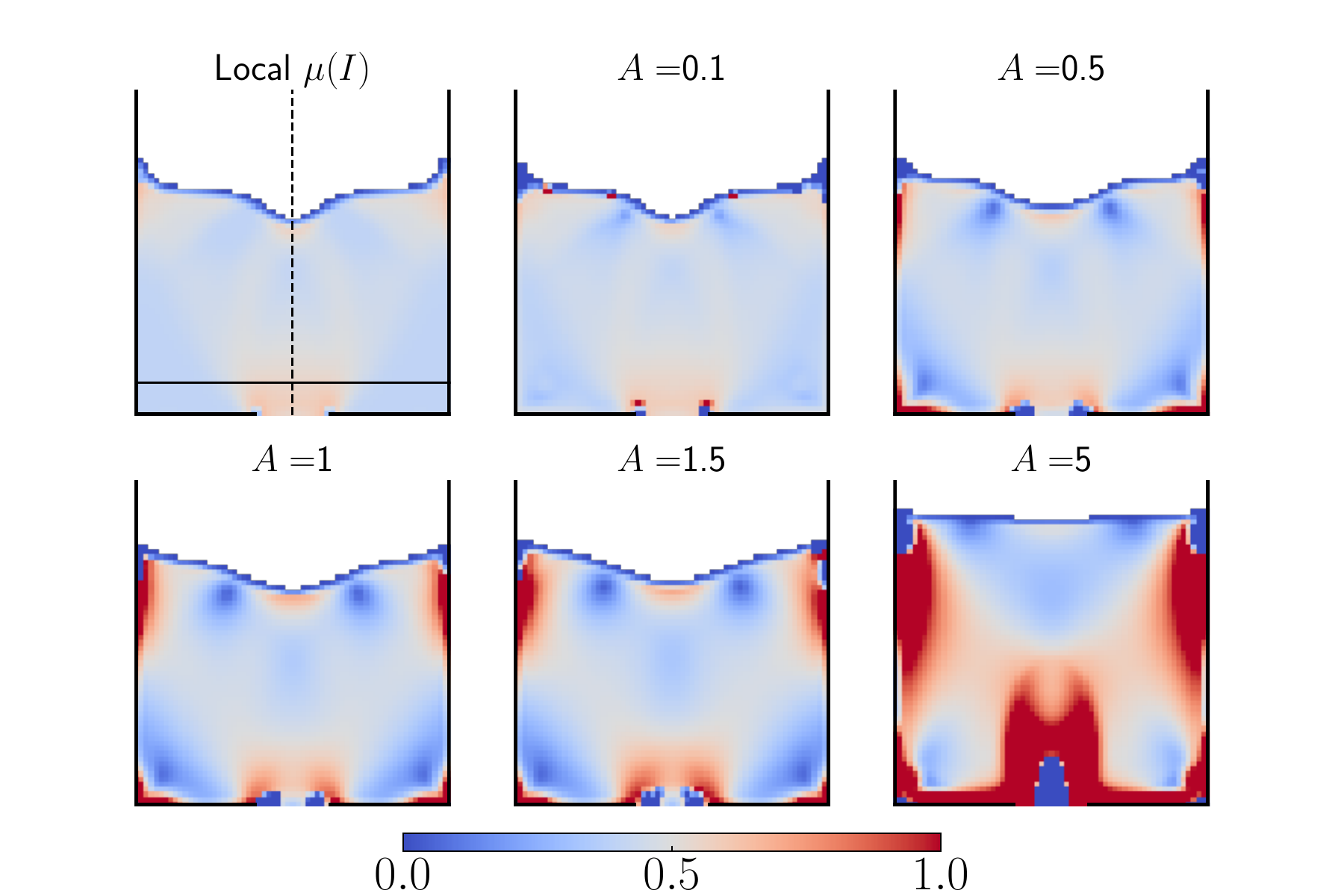}\\
	 \hspace{-3.5cm} $(b)$ \hspace{4.cm} $(c)$ \hspace{4.cm} $(d)$\\
   \includegraphics[height=3.2cm]
{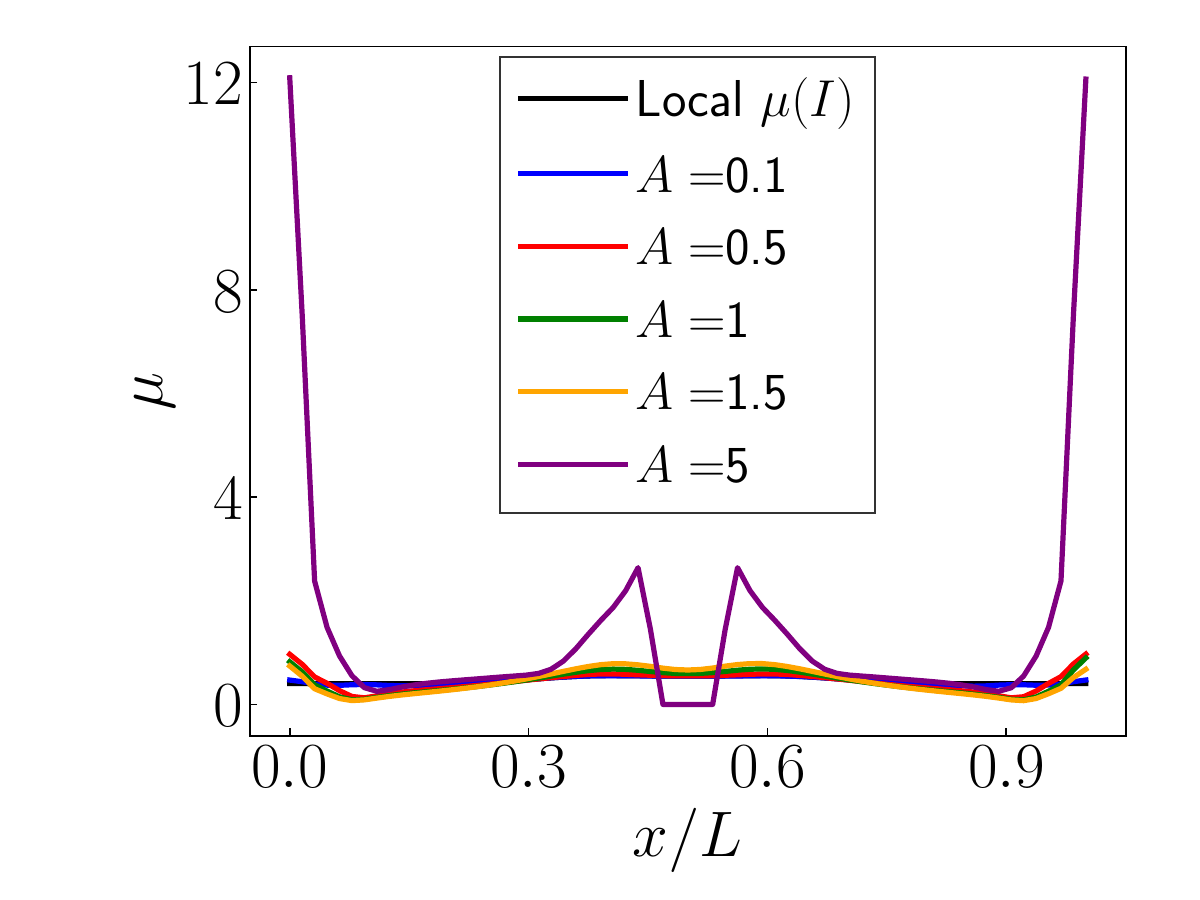}
   \includegraphics[height=3.2cm]
{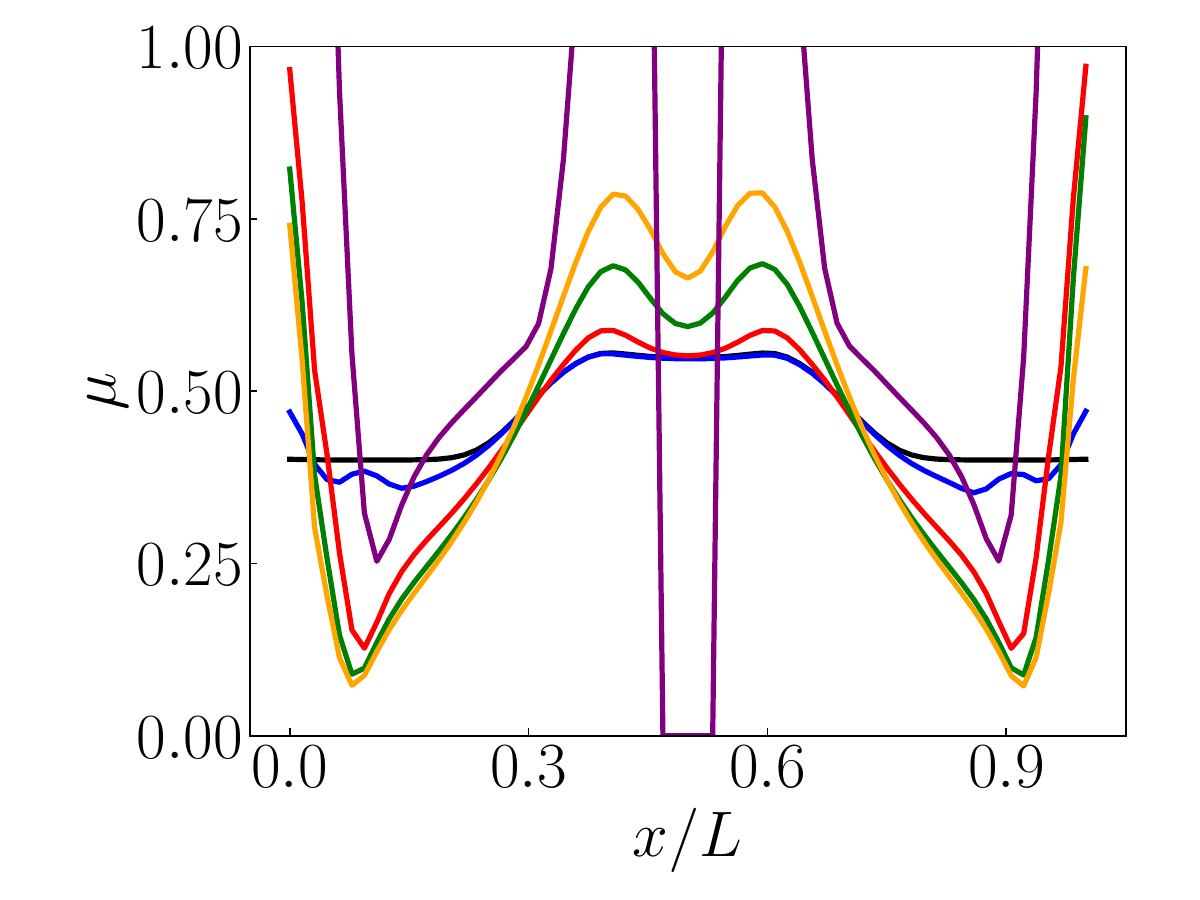}
\includegraphics[height=3.2cm]
{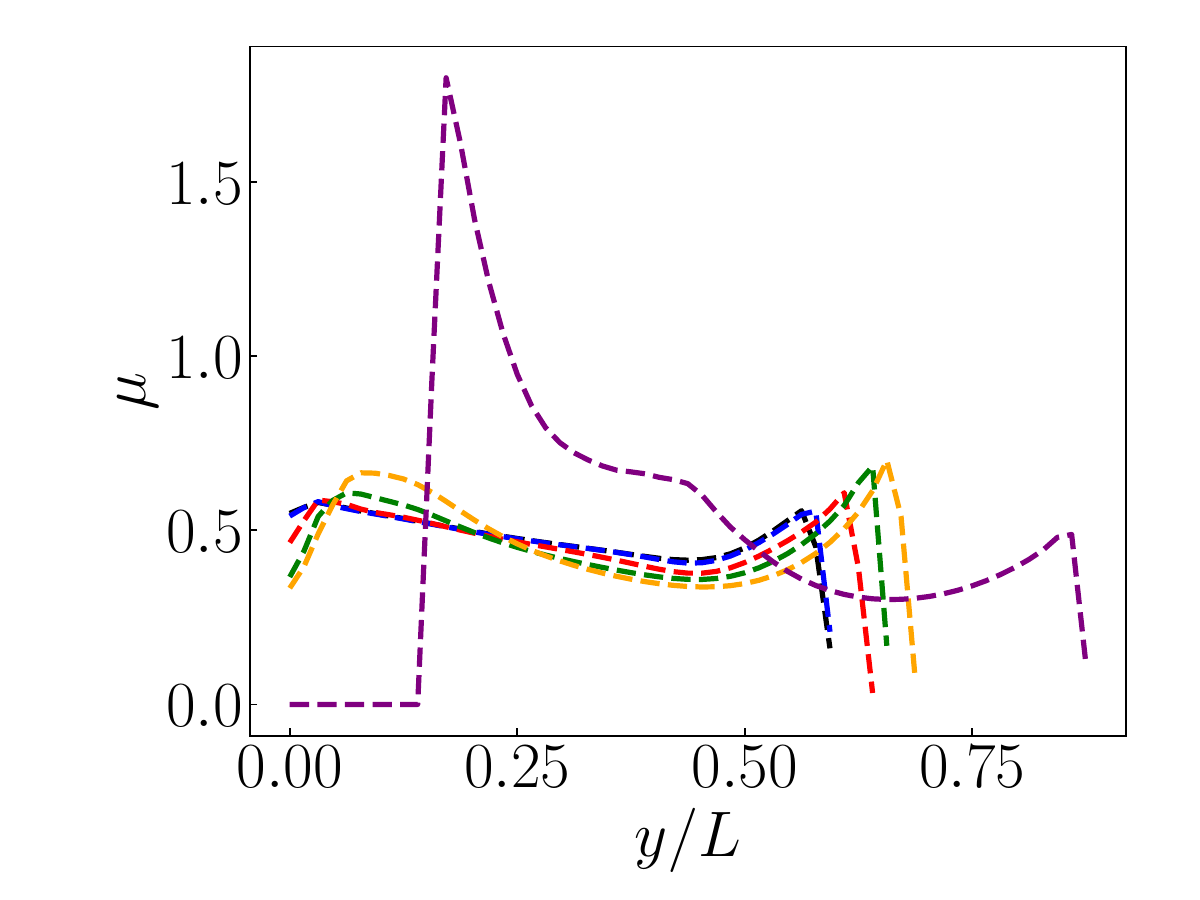}
	\caption{\label{fig:mu_fields}  Friction coefficient $\mu$ fields and profiles at $\overline{t}=2$ for $\overline{D}=0.25$, comparing the local $\mu(I)$ model and the dynamic NGF model with various non-local amplitudes $A$, using the modified parameter set (table~\ref{tab:params}). 
        (a) Field visualization. The solid horizontal line indicates $y/L = 0.1$, and the dashed vertical line indicates $x/L = 0.5$. 
        (b) Profiles of friction coefficient $\mu$ along the horizontal line at $y/L = 0.1$. 
        (c) Same as (b) but with a reduced vertical axis range $(0, 1)$. 
        (d) Corresponding profiles along the vertical centerline at $x/L = 0.5$.}
  	\end{center}
\end{figure}

Figure~\ref{fig:viscosity_fields}(a) shows the normalized effective viscosity $\eta_{\text{eff}} / (\rho \sqrt{G L^3})$ field for $\overline{D} = 0.25$ at $\overline{t} = 2$, comparing the local $\mu(I)$ model (equivalent to the non-local model at $A = 0$) with the dynamic NGF model for $A = 0.1$, $0.5$, $1$, $1.5$ and $5.0$. The colour scale is fixed between $0$ and $2$. As $A$ increases, the region where $\eta_{\text{eff}} / (\rho \sqrt{G L^3}) > 2$ first decreases slightly and then expands significantly for larger $A$. However, because the colour scale is truncated, the quantitative values above $2$ are not visible in the contour plots. Therefore, we examine the horizontal and vertical profiles in panels (b)-(d). Figure~\ref{fig:viscosity_fields}(b) presents the horizontal profiles of $\eta_{\text{eff}} / (\rho \sqrt{G L^3})$ along the line $y/L = 0.1$, which is indicated by the black solid line in figure~\ref{fig:viscosity_fields}(a). 
For the local model, the viscosity is nearly zero in the central region ($0.25 < x/L < 0.75$) but rises sharply towards the sidewalls. This produces narrow bands of very high viscosity near the walls, around $0.05 < x/L < 0.15$ and in the corresponding symmetric region on the right. This localized high-viscosity region corresponds to the stagnant corners where the material is almost rigid. Upon introducing non-local effects, the profiles become significantly smoother, with a more gradual transition between the low-viscosity central zone and the high-viscosity sidewall regions. This behaviour appears more physical. In a real granular flow, the transition from the rapidly flowing central region to the nearly stagnant sidewall regions should be smooth rather than abrupt. The influence of particle motion can propagate into adjacent regions through grain-grain interactions. The non-local model captures this cooperative effect by diffusing the fluidity, thereby eliminating the unphysical sharp viscosity gradient predicted by the local model. Figure~\ref{fig:viscosity_fields}(c) shows the same horizontal profiles but with a reduced vertical axis range $(0, 5)$, revealing that at positions near $x/L \approx 0.3$ (and the symmetric location on the right), larger $A$ leads to higher viscosity values. Figure~\ref{fig:viscosity_fields}(d) shows the vertical profiles along the centreline ($x/L = 0.5$), indicated by the black dashed line in figure~\ref{fig:viscosity_fields}(a). In the region near the top ($y/L > 0.5$), the viscosity increases with $A$. The introduction of non-local effects makes the spatial variation of $\eta_{\text{eff}}$ along the $y$ direction more gradual as $A$ increases. However, for very large $A$ ($A = 5.0$), the variation becomes steeper again, indicating that excessive non-local diffusion may lead to a different flow regime.

 \begin{figure}
  	\begin{center}
	 \hspace{-12.5cm} $(a)$\\
 \includegraphics[height=9cm]{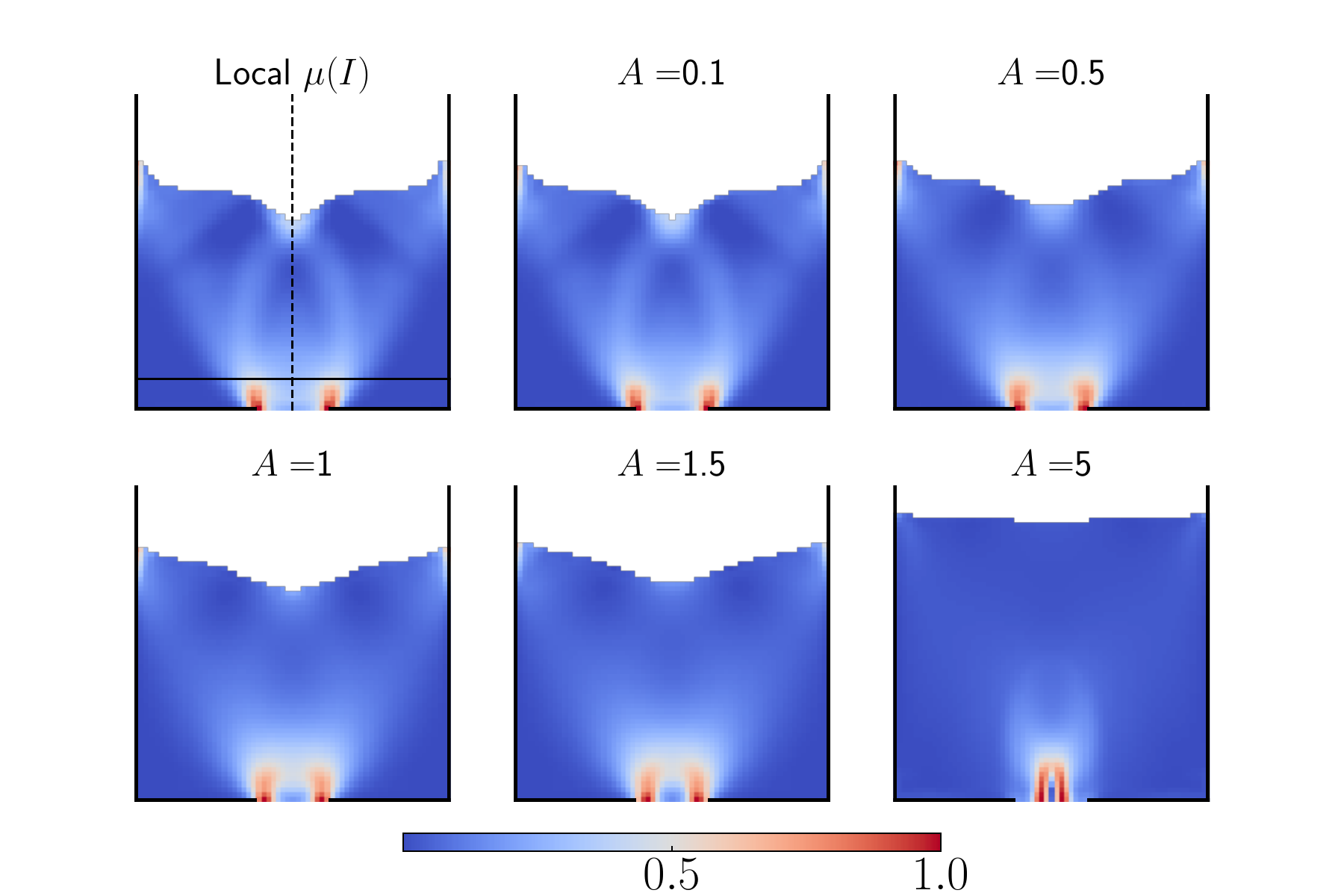}\\
	 \hspace{-5.5cm} $(b)$ \hspace{6.cm} $(c)$\\
   \includegraphics[height=4.5cm]
{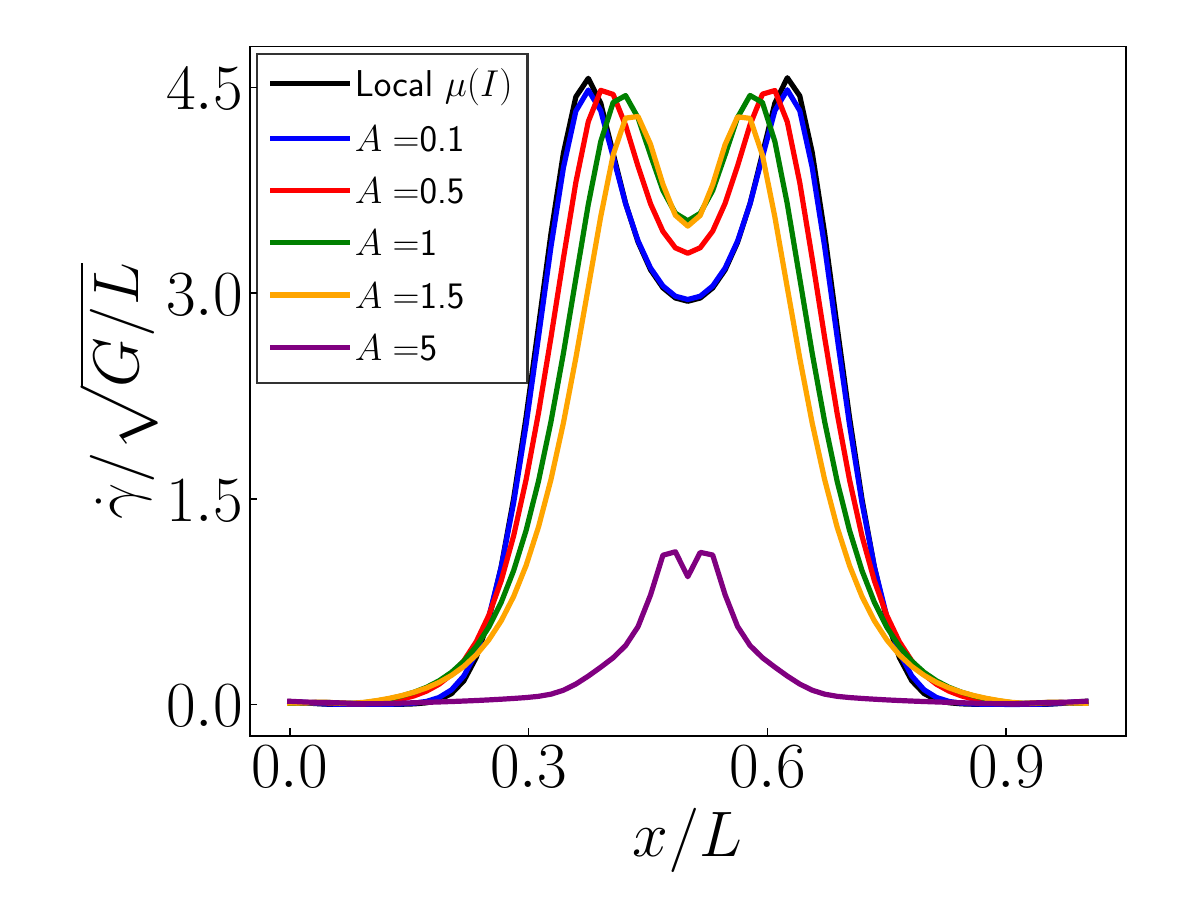}
\includegraphics[height=4.5cm]
{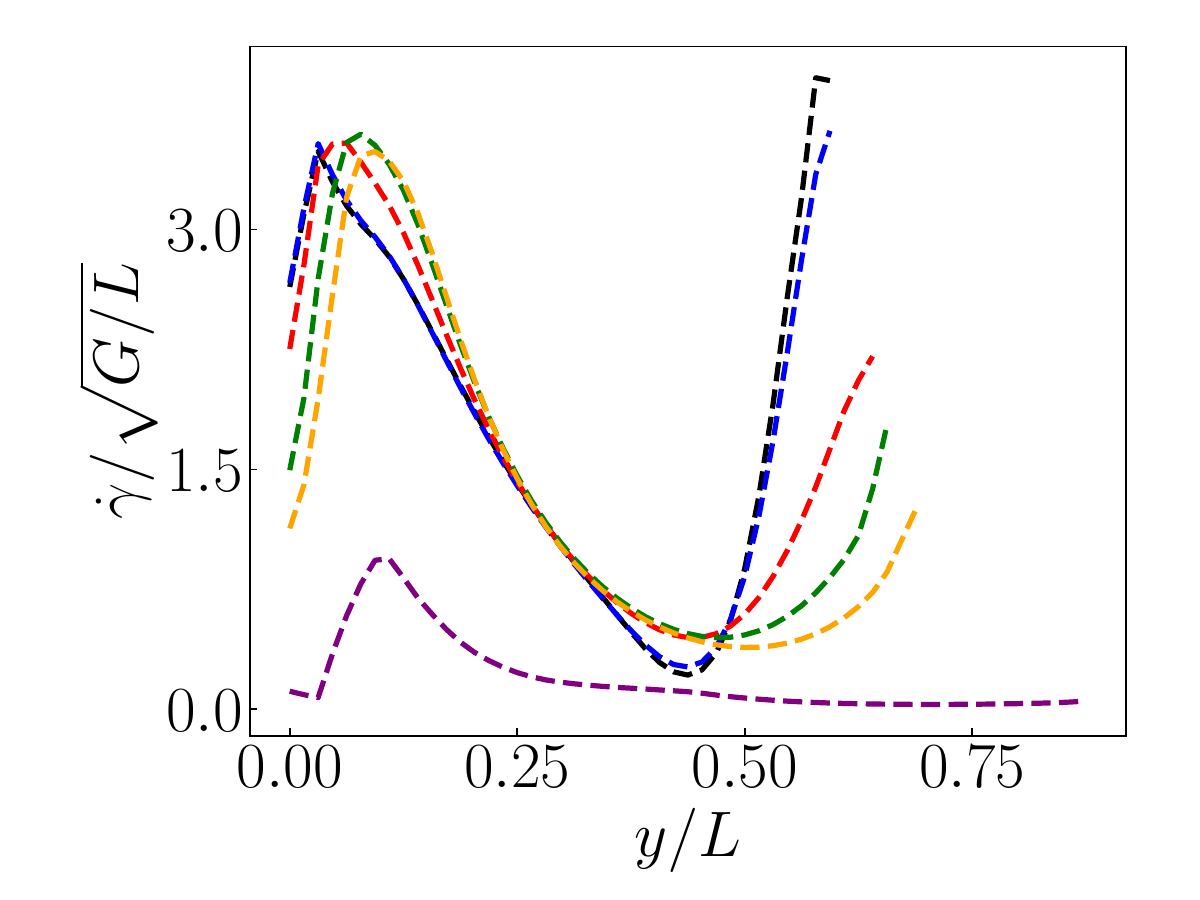}
	\caption{\label{fig:gammatot_fields} Normalized shear rate fields and profiles at $\overline{t}=2$ for $\overline{D}=0.25$, comparing the local $\mu(I)$ model and the dynamic NGF model with various non-local amplitudes $A$, using the modified parameter set (table~\ref{tab:params}). (a) Normalized shear rate $|\dot{\gamma} |/|\dot{\gamma} |_{\max}$ field visualization. The solid horizontal line represents $y/L = 0.1$, and the dashed vertical line represents $x/L = 0.5$. (b) Profiles of the dimensionless shear rate $|\dot{\gamma} |//\sqrt{G/L}$ along the horizontal line $y/L=0.1$. (c) Corresponding profiles along the vertical centreline $x/L=0.5$.}
  	\end{center}
\end{figure}

Figure~\ref{fig:mu_fields} presents the friction coefficient $\mu$ fields and profiles under the same conditions. From the contour plot in figure~\ref{fig:mu_fields}(a), the local model ($A=0$) exhibits $\mu$ values in the range $0.4$ to $0.68$ in flowing regions with non-zero pressure, consistent with the prescribed bounds $\mu_s = 0.4$ and $\mu_2 = 0.68$. As $A$ increases, the range of $\mu$ expands significantly, indicating that non-local effects allow the friction coefficient to deviate from the local bounds. Figure~\ref{fig:mu_fields}(b) shows the horizontal profiles of $\mu$ along $y/L = 0.1$. For $A = 5.0$, $\mu$ can reach values as high as $12$, far exceeding the local upper bound. To better visualise the behaviour for smaller $A$, Figure~\ref{fig:mu_fields}(c) presents the same horizontal profiles with a reduced vertical axis range $(0, 1)$. It can be observed that as $A$ increases, $\mu$ becomes larger both in the central region and near the sidewalls, while a region of lower $\mu$ develops in between ($0.1 < x/L < 0.3$). Figure~\ref{fig:mu_fields}(d) shows the vertical profiles of $\mu$ along the centreline ($x/L = 0.5$). For the local model, $\mu$ varies around $0.5$ along the centreline. As $A$ increases, the variation of $\mu$ becomes slightly more pronounced compared with the local model, with a slightly wider range between the maximum and minimum values. For $A = 5.0$, a region of negative pressure appears near the orifice, where we set $\mu = 0$ for numerical consistency.

Figure~\ref{fig:gammatot_fields} presents the shear rate fields and profiles under the same conditions. From the contour plot in figure~\ref{fig:gammatot_fields}(a), which shows the normalized shear rate $|\dot{\gamma}|/|\dot{\gamma}|_{\max}$, the high-shear-rate zone is initially concentrated at the edges of the orifice for the local $\mu(I)$ model. As $A$ increases, these regions gradually shift towards the centre of the orifice, and the shear-rate distribution becomes more uniform across the silo width. Figure~\ref{fig:gammatot_fields}(b) shows the horizontal profiles of the dimensional shear rate $|\dot{\gamma}|/\sqrt{G/L}$ along $y/L = 0.1$. For the local model, the maximum shear rate is sharply peaked near the sidewalls, indicating a thinner shear band. As $A$ increases, the peak magnitude decreases, and the distribution widens, indicating that the shear band broadens. This is consistent with the velocity profiles shown in figure~\ref{fig:kamrin_comparison}(b). For $A=5.0$, the shear rate becomes significantly lower and more uniform across the width. Figure~\ref{fig:gammatot_fields}(c) presents the vertical profiles of the dimensional shear rate $|\dot{\gamma}|/\sqrt{G/L}$ along the centreline ($x/L = 0.5$). Near the orifice, the location of the maximum shear rate shifts slightly upward with increasing $A$, while the shear rate near the top centre decreases, primarily due to the overall reduction in velocity as $A$ increases.

 \begin{figure}
  	\begin{center}
 \includegraphics[height=9cm]{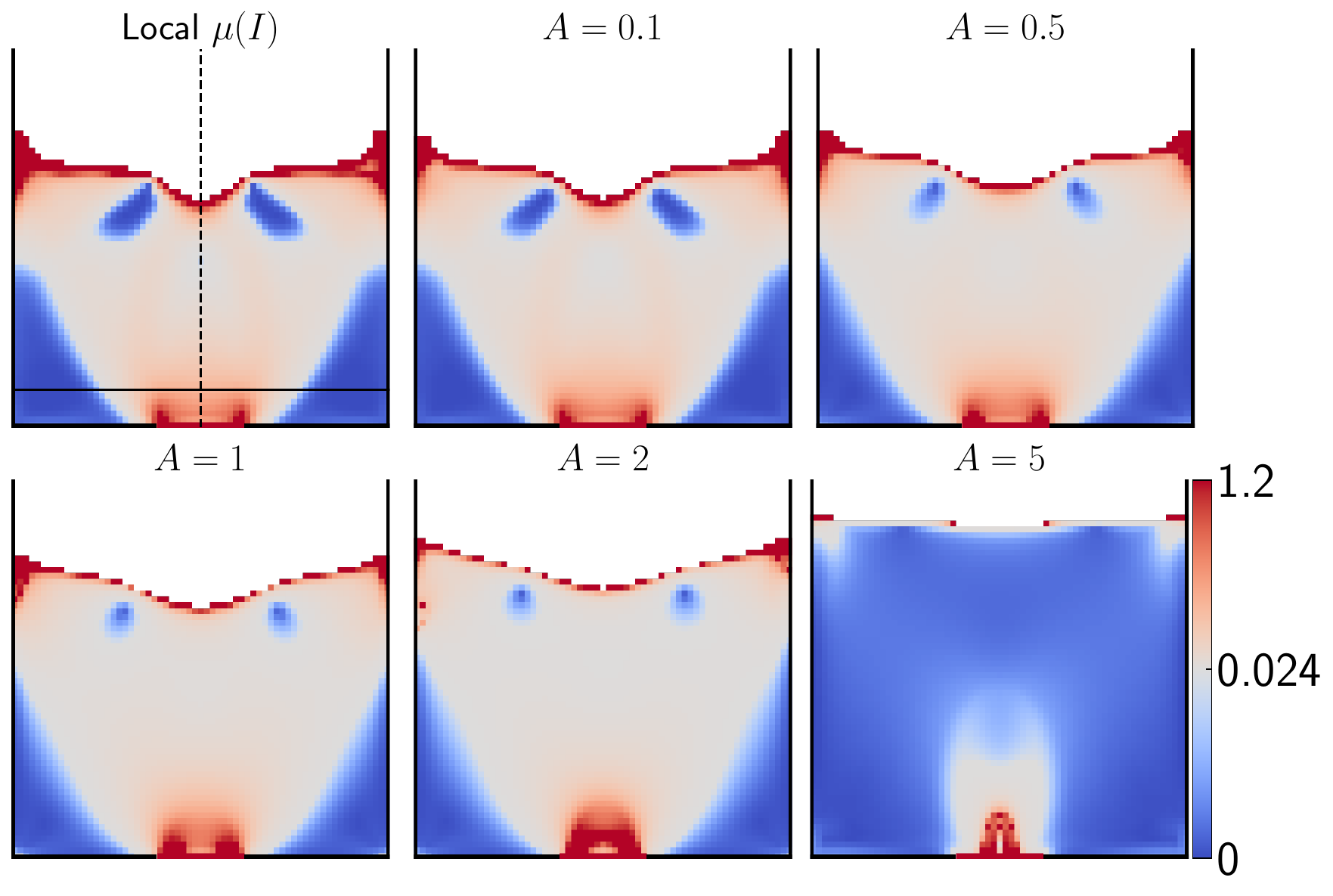}
 	\caption{\label{fig:I_fields} Inertial number $I$ field visualization at $\overline{t}=2$ for $\overline{D}=0.25$, comparing the local $\mu(I)$ model and the dynamic NGF model with various non-local amplitudes $A$, using the modified parameter set (table~\ref{tab:params}).}
  	\end{center}
\end{figure}

Figure~\ref{fig:I_fields} presents the inertial number $I$ field under the same conditions as in figure~\ref{fig:gammatot_fields}. Since $I$ is proportional to the shear rate $|\dot{\gamma}|$, the distribution of $I$ directly reflects the shear localisation behaviour discussed above. Here we focus on the $I$ field primarily to examine the well-posedness of the governing equations in different regions of the flow domain. Based on the stability analysis of \cite{Barker2015}, for our parameters ($\mu_s = 0.4$, $\mu_2 = 0.68$, $I_0 = 0.4$), the local $\mu(I)$ rheology is well-posed approximately in the range $0.024 \lesssim I \lesssim 1.2$. To clearly distinguish the well-posed region from the ill-posed regions, the color scale in figure~\ref{fig:I_fields} is adjusted such that the range $0.024 \lesssim I \lesssim 1.2$ is highlighted, with values below $0.024$ and above $1.2$ shown in distinct colors. 
For the local model, the high-$I$ regions are concentrated near the edges of the orifice. By contrast, the low-$I$ region occupies a large portion of the domain, particularly near the sidewalls. As $A$ increases, the high-$I$ regions gradually shift from the orifice edges towards its centre. Importantly, the ill-posed low-$I$ region remains confined to the vicinity of the sidewalls. The ill-posed high-$I$ region is restricted to a small area near the orifice in all cases. For $A \leq 2$, a significant portion of the flow domain falls within the well-posed range $0.024 \lesssim I \lesssim 1.2$. For $A = 5$, the low-$I$ region expands considerably. In these regions, the effective viscosity becomes very large, but our regularization technique limits it to a maximum value.

The fact that the ill-posed regions are spatially limited, together with the viscosity regularization strategy, explains the numerical robustness observed in these simulations. 
Although minor oscillations persist in some cases, the simulations remain relatively stable over the range of parameters considered and do not exhibit severe numerical divergence. Notably, the non-local formulations do not significantly improve numerical stability relative to the local model, even though the additional diffusive terms are expected to have a regularising effect. This result suggests that the viscosity regularisation already provides sufficient control over the ill-posedness under the present numerical conditions. The non-local effects therefore primarily modify the flow structure, such as the shear-band width and velocity profiles, rather than the overall numerical stability.

\section{Comparison of the different non-local models  and discussion} \label{comparenonlocal}

Having established the behaviour of the dynamic NGF model under various parameters, we now turn to a comparison of different non-local constitutive models.  The local model and the seven non-local models presented in \S\ref{sec:numericalscheme} have been implemented in the {\usefont{T1}{pzc}{m}{n}Basilisk}  solver. The numerical and rheological parameters used in this section are the same as those listed in table~\ref{tab:params} (see \S\ref{DynamicNGFresults}). For consistency, the non-local amplitude is denoted by $A$ for all non-local models, and identical $A$ values are used when comparing their predictions. This enables us to assess the effect of each model on the silo discharge rate and other flow quantities.

Regarding the boundary conditions for the comparative study, the velocity field $\mathbf{u}$ and the pressure field $p$ follow the same configurations as described in \S\ref{DynamicNGFresults}. For the fluidity field $g$, we follow \cite{Kamrin2015SM, SalvadorVieira2017} and impose homogeneous Dirichlet conditions $g = 0$ on the solid bottom boundaries and sidewalls. For the inertial number $I$, we adopt the same type of boundary condition $I = 0$, following \cite{Lin2020, Lin2021}. Homogeneous Neumann conditions ($\partial_n g = 0$, $\partial_n I = 0$) are imposed on the top free surface and over the orifice. For the granular temperature field $\Theta$ in the $\mu(I, \Theta)$ model, a homogeneous Neumann boundary condition ($\partial_n \Theta = 0$) is imposed on all solid boundaries, following \cite{Yuan2026}. This choice avoids the artificial dissipation that would result from enforcing $\Theta = 0$ at the wall. Unlike externally vibrated boundaries, where Dirichlet conditions may be prescribed based on the vibration amplitude (\cite{Irmer2025}), the present configuration involves no external energy input.

\begin{figure}
\begin{centering}	
 \hspace{-5.5cm} $(a)$ \hspace{6.cm} $(b)$\\
\includegraphics[height=5cm]
{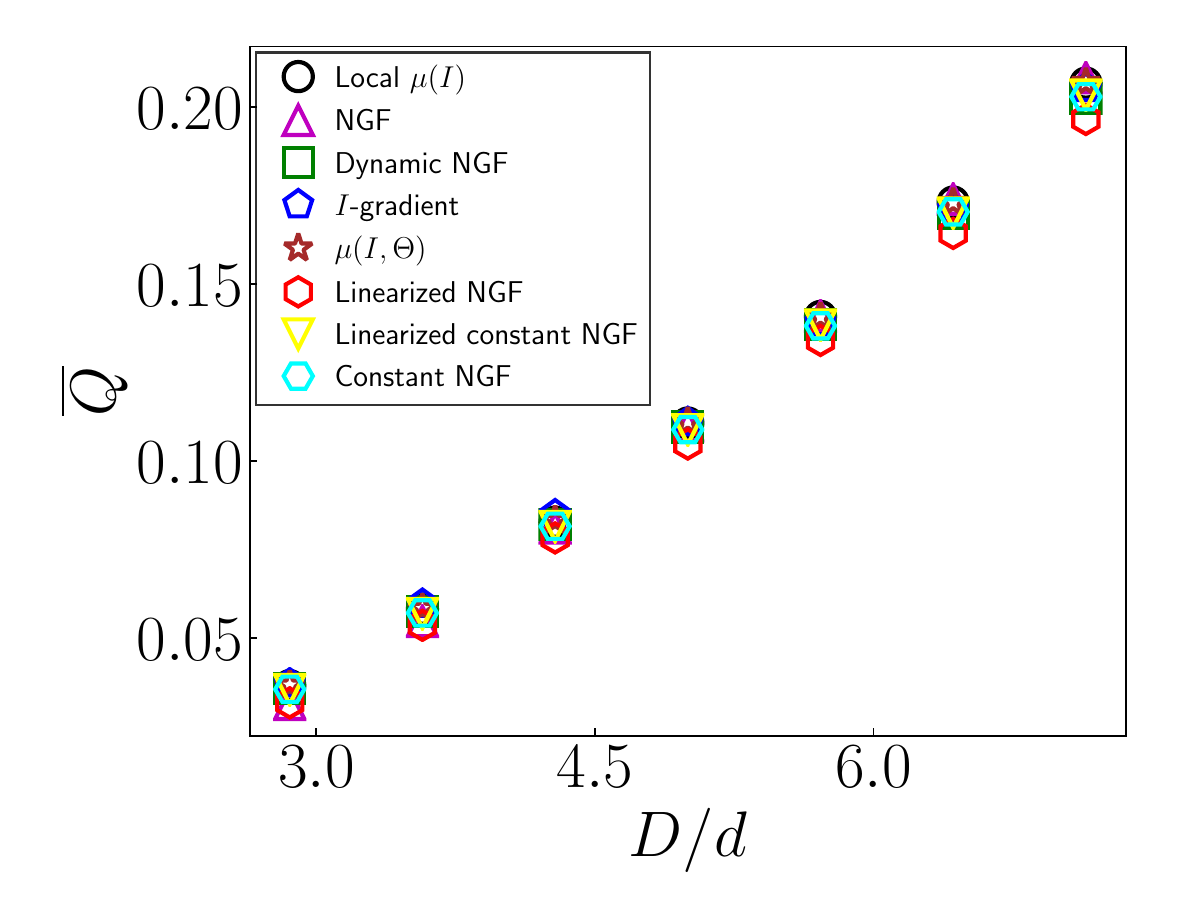}
\includegraphics[height=5cm]
{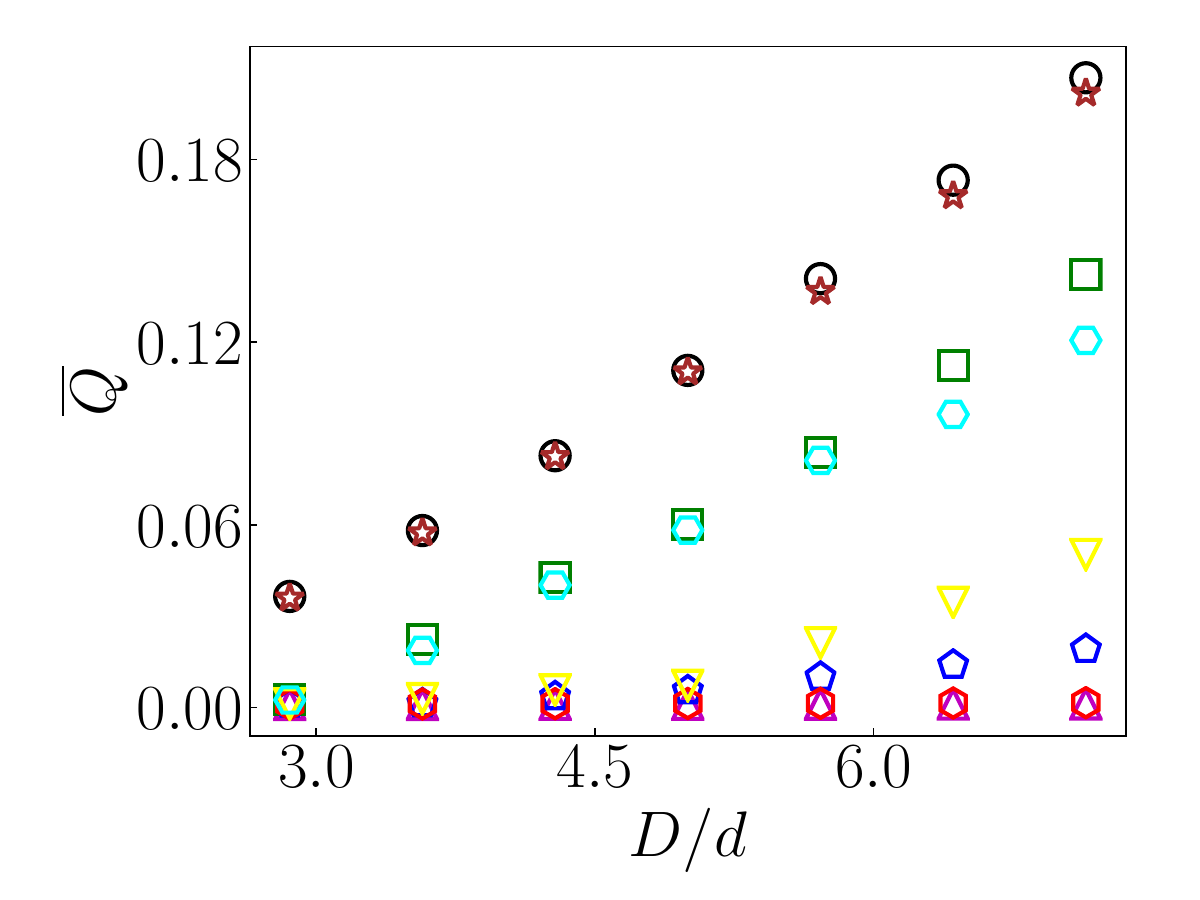}
\caption{\label{flowratecompareallAdiff} Mean dimensionless flow rate $\overline{Q}$ versus $D/d$ for different constitutive models at (a) $A = 0.1$ and (b) $A = 2$.}	
\end{centering}	
\end{figure}

To validate our numerical implementation, we first compare the mean dimensionless flow rate $\overline{Q}$ as a function of $D/d$ obtained from the different methods described in \S\ref{sec:numericalscheme} for two values of $A$. At $A = 0.1$ (figure~\ref{flowratecompareallAdiff}(a)), all methods produce nearly identical results, confirming that the implementation correctly reproduces the expected behaviour when non-local effects are negligible. 
At $A = 2$ (figure~\ref{flowratecompareallAdiff}(b)), the $\mu(I,\theta)$ model (\S\ref{sec:muItheta}) predicts flow rates slightly below those of the local model (\S\ref{sec:local}). The dynamic NGF (\S\ref{sec:algorithmDymNGF}) and constant NGF (\S\ref{sec:constantNGF}) models produce similar results because of their closely related mathematical formulations. However, their predicted flow rates are significantly lower than those obtained from the local and $\mu(I,\theta)$ models. The linearised constant NGF model (\S\ref{sec:linearisedConstantNGF}) produces slightly smaller flow rates than the constant NGF model for large $A$. This is because the linearisation approximation, which assumes $\xi^2 \nabla^2 g / g_{\text{loc}} \ll 1$, becomes inaccurate as $A$ increases; it overestimates the effective viscosity $\eta_{\text{eff}}$ relative to the Laplacian form, thereby reducing the discharge rate. The $I$-gradient results (\S\ref{sec:Igradient}) lie close to, but marginally below, those of the linearised constant NGF model, consistent with the similarity of their mathematical formulations. When non-local effects become significant, both the NGF and linearised NGF models predict an almost immobile flow. 
Moreover, solving the governing equation for $g$ in the NGF model (equation~\eqref{gnonlocalequation} in \S\ref{sec:NGF}) can yield negative values of $g$. These non-physical values indicate a limitation of the NGF and linearised NGF models at large $A$. Overall, the observed trends are consistent with the mathematical formulations of the respective models. As $A$ increases, all models predict a reduction in the discharge rate, although the magnitude of this reduction varies among the different formulations.

\begin{figure}
\begin{centering}	
 \hspace{-5.5cm} $(a)$ \hspace{6.cm} $(b)$\\
 \includegraphics[height=5cm]
{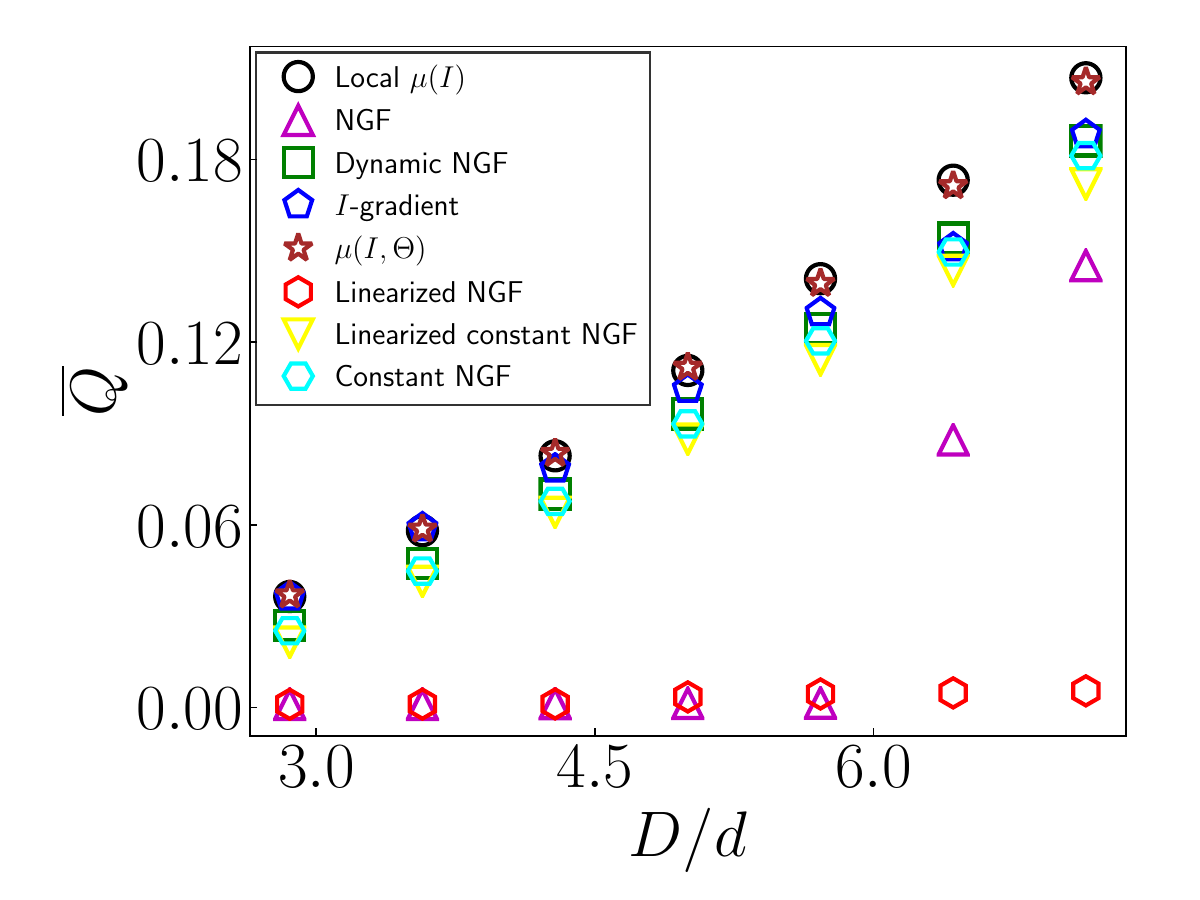}
\includegraphics[height=5cm]
{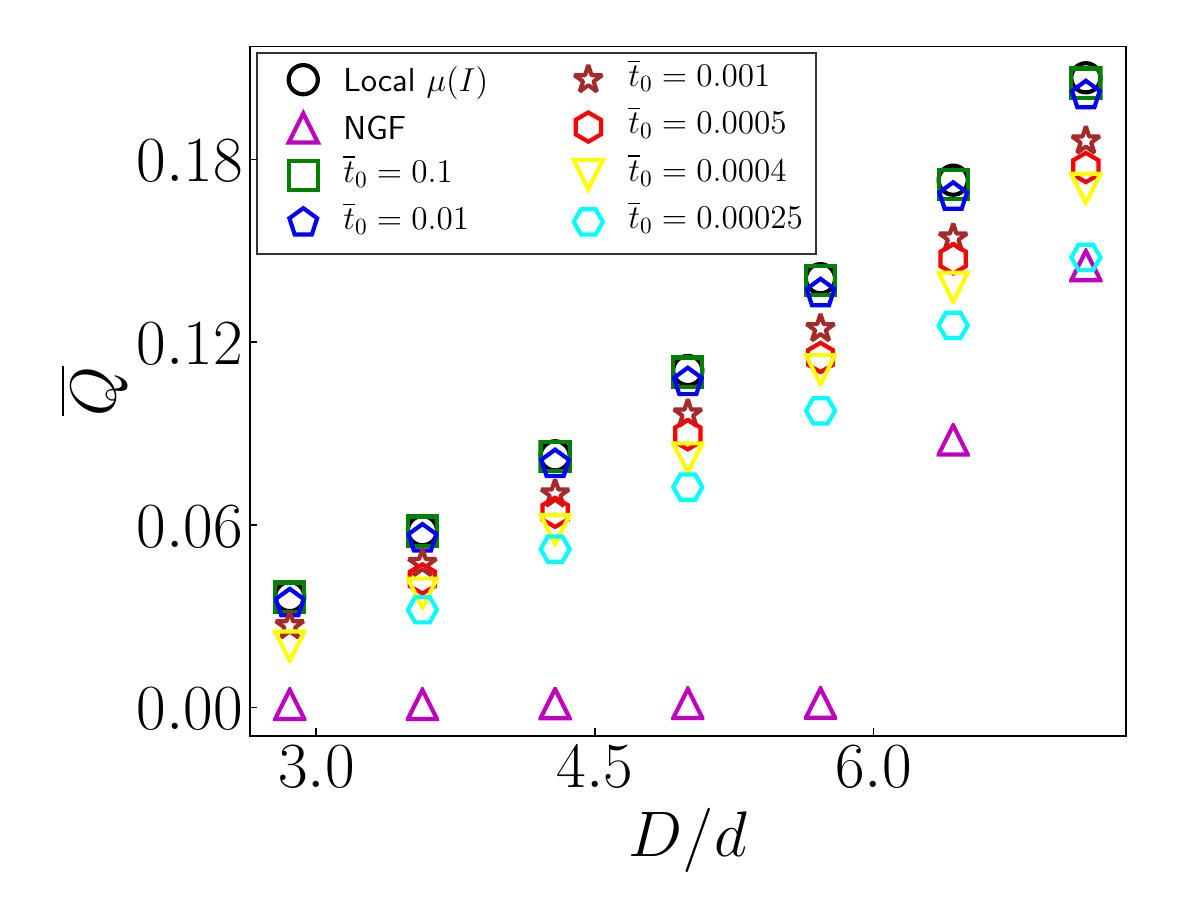}

\caption{\label{T0diff}  Mean dimensionless flow rate $\overline{Q}$ versus $D/d$ for different constitutive models and time scales $t_0$. (a) Mean dimensionless flow rate $\overline{Q}$ versus $D/d$ at  $A = 0.5$. (b) Mean dimensionless flow rate $\overline{Q}$ as a function of $D/d$ for the local model, the NGF model and the dynamic NGF model with different values of $\overline{t}_0 = t_{0}/\sqrt{L/G}$. All parameters are the same as in (a). Additionally, dynamic NGF results with different $\overline{t}_0$ values (ranging from $0.00025$ to $0.1$) are shown. The case $\overline{t}_0 = 0.001$ corresponds to the dynamic NGF curve already shown in (a).}	
\end{centering}	
\end{figure}

We now examine the case $A = 0.5$ in figure~\ref{T0diff}. Figure~\ref{T0diff}(a) shows that the $\mu(I,\Theta)$ model again produces results close to those of the local model. By contrast, the NGF and linearised NGF models predict significantly lower flow rates. This trend is similar to that observed at $A = 2$ in figure~\ref{flowratecompareallAdiff}. This behaviour can be attributed to the form of the cooperativity length $\xi(\mu)$ in the NGF-type models (see equation~\eqref{eq:xi} in \S\ref{sec:NGF}). The term $(\mu - \mu_s)$ in the denominator causes $\xi$ to diverge as $\mu$ approaches the yield friction $\mu_s$. In regions where the flow becomes nearly static, $\mu \to \mu_s$ and $\xi \to \infty$, making the diffusion term $-\xi^2 \nabla^2 g$ dominant in the governing equation $-\xi^2 \nabla^2 g + g = g_{\text{loc}}$. For sufficiently large $\xi$, the numerical solution tends towards $g = 0$, which through $\eta_{\text{eff}} = p/g$ corresponds to an infinite effective viscosity and thus a nearly rigid, immobile state. Indeed, our numerical solutions yield $g = 0$ throughout the domain when $A$ exceeds a geometry-dependent threshold: for $D/L = 0.25$, this occurs at $A \geq 0.5$; for $D/d = 0.3125$, at $A \geq 0.75$. This explains why the NGF and linearised NGF models produce almost zero flow rates in these regimes.

In the dynamic NGF model, the relaxation time $\bar{t}_0$ governs the rate at which the fluidity field $g$ evolves.  \cite{Dunatunga2022} note that experimental values of $t_0$ are unknown, and they choose a small fixed value for convenience. The evolution of $g$ is updated via a splitting scheme, as in their implementation. The diffusion step depends on the ratio $\Delta t / t_0$ (see equation (5.3) in  \cite{Dunatunga2022}; also equation~\eqref{gstar} in the present work). Importantly, this diffusion step does not contain the term $(\mu - \mu_s)$ in the denominator, which is present in the cooperativity length $\xi$ of the NGF model. Consequently, the dynamic formulation avoids the singularity that arises when $\mu \to \mu_s$ in the NGF, which often leads to non-physical solutions. Figure~\ref{T0diff}(b) examines the influence of $\bar{t}_0$ on the predicted flow rate, showing $\overline{Q}$ as a function of $D/d$ for the local model, the NGF model, and the dynamic NGF model with $\bar{t}_0$ ranging from $0.00025$ to $0.1$ at $A = 0.5$. For the same values of $A$ and $D/d$, both a high-flow-rate solution resembling the local model and a low-flow-rate solution approaching the NGF result appear to be attainable, depending on the choice of $\bar{t}_0$. The absence of the $(\mu - \mu_s)$ singularity in the dynamic model allows the solution to remain well-behaved across the entire parameter range, while the NGF limit is only approached as $\bar{t}_0$ becomes very small. Thus, $\bar{t}_0$ plays a role similar to that of an inertial parameter and controls the selection between coexisting steady states in non-local granular flow. Although its physical value is unknown, it has a significant influence on the predicted flow rate.

However, a numerical caveat must be noted regarding this splitting scheme. The diffusion step in equation~(\ref{gstar}) is implicit, but the source term update in equation~(\ref{dynamicNGF2}) is semi-implicit, which imposes a stability constraint on $\Delta t / t_0$. For the smallest $\bar{t}_0$ values in figure~\ref{T0diff}(b) (e.g., $\bar{t}_0 = 0.001$ and $0.00025$), this condition may be violated. Our main conclusion nevertheless concerns the steady-state discharge rate rather than the transient dynamics. The observation that the solution approaches the NGF limit as $\bar{t}_0$ decreases therefore remains valid, although the transient evolution may not be accurately resolved.
 
 \begin{figure}
\begin{centering}	
 \hspace{-5.5cm} $(a)$ \hspace{6.cm} $(b)$\\
 \includegraphics[height=5cm]
{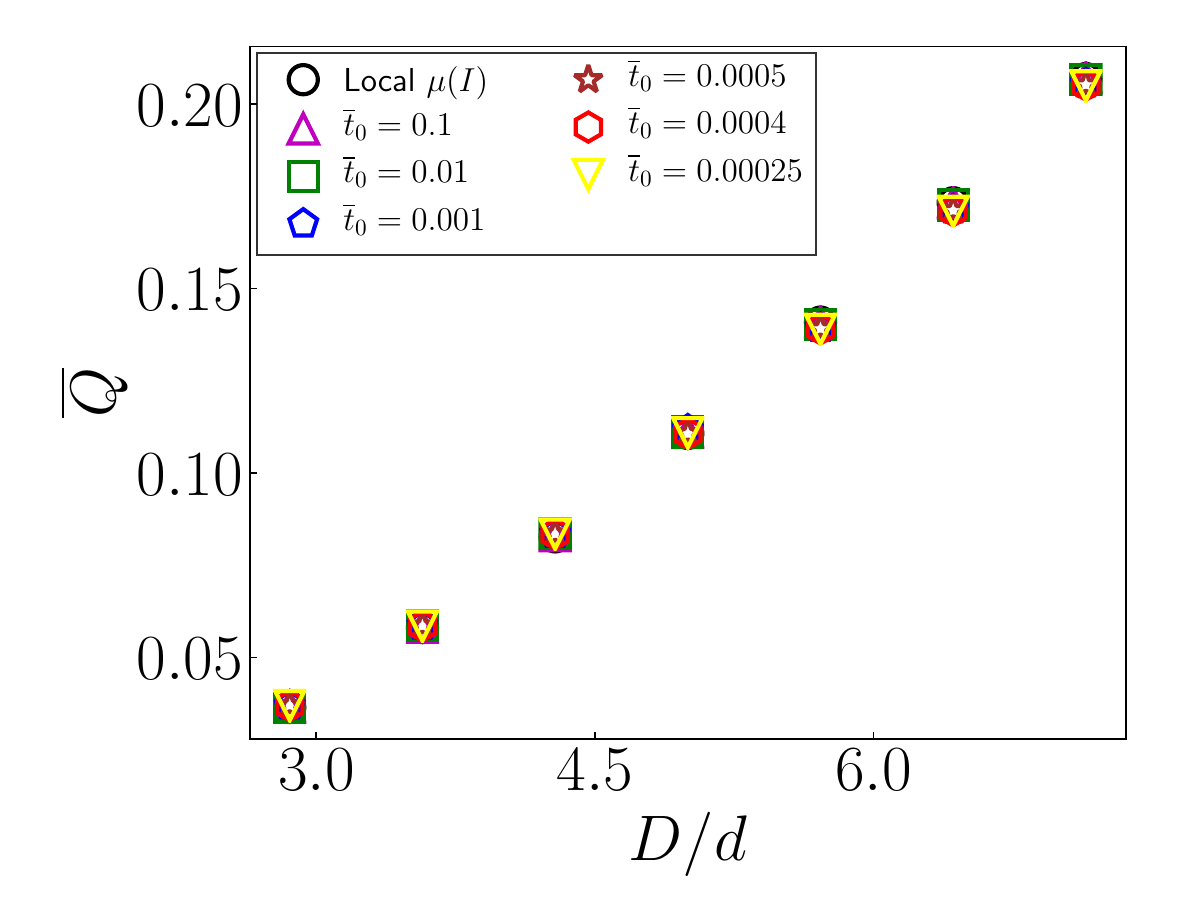}
\includegraphics[height=5cm]
{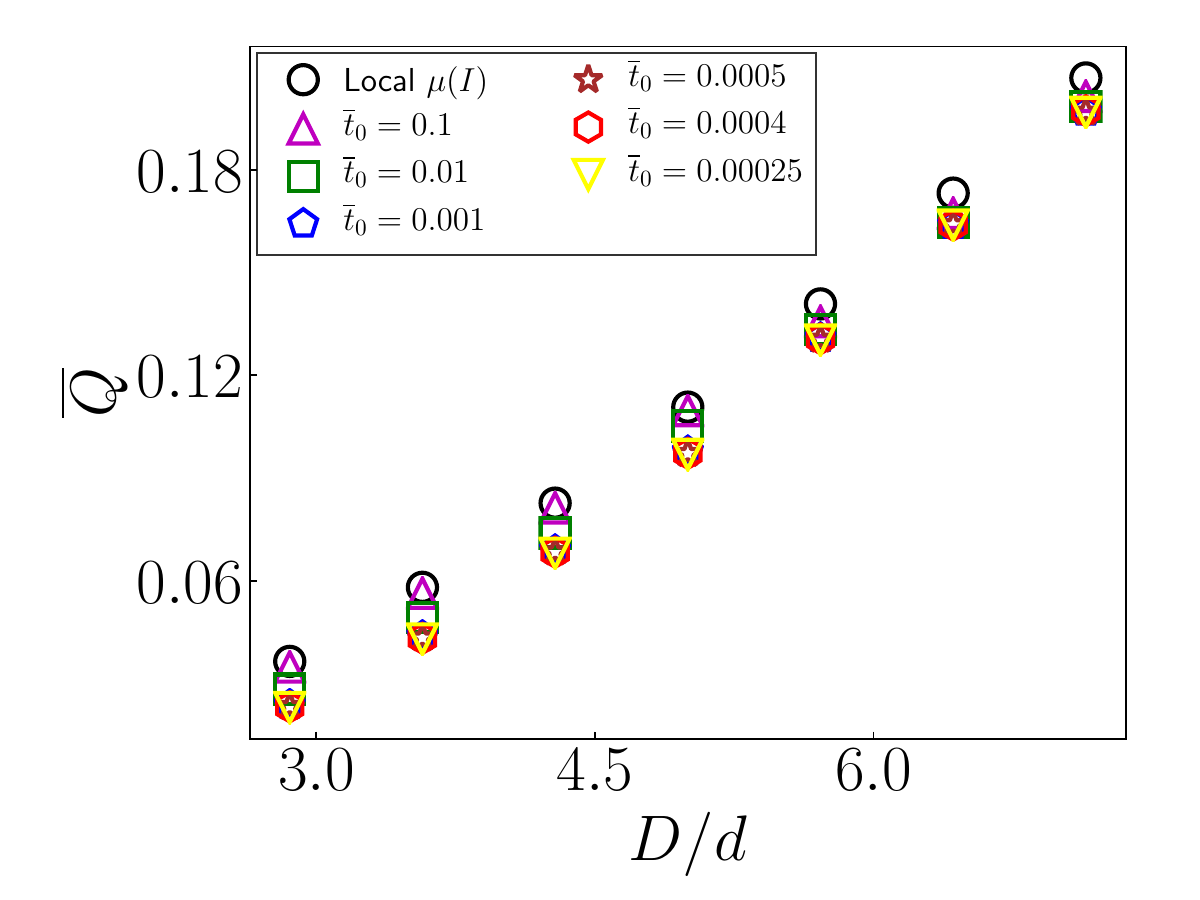}

\caption{\label{T0diffmuItheta}  Mean dimensionless flow rate $\overline{Q}$ as a function of $D/d$ for the local $\mu(I)$ model and the $\mu(I, \Theta)$ model with different values of $\overline{t}_0 = t_{0}/\sqrt{L/G}$  (ranging from $0.00025$ to $0.1$). (a) $A= 0 .5$, (b) $A = 30$.}	
\end{centering}	
\end{figure}

 \begin{figure}
\begin{centering}	
 \hspace{-5.5cm} $(a)$ \hspace{6.cm} $(b)$\\
 \includegraphics[height=5cm]
{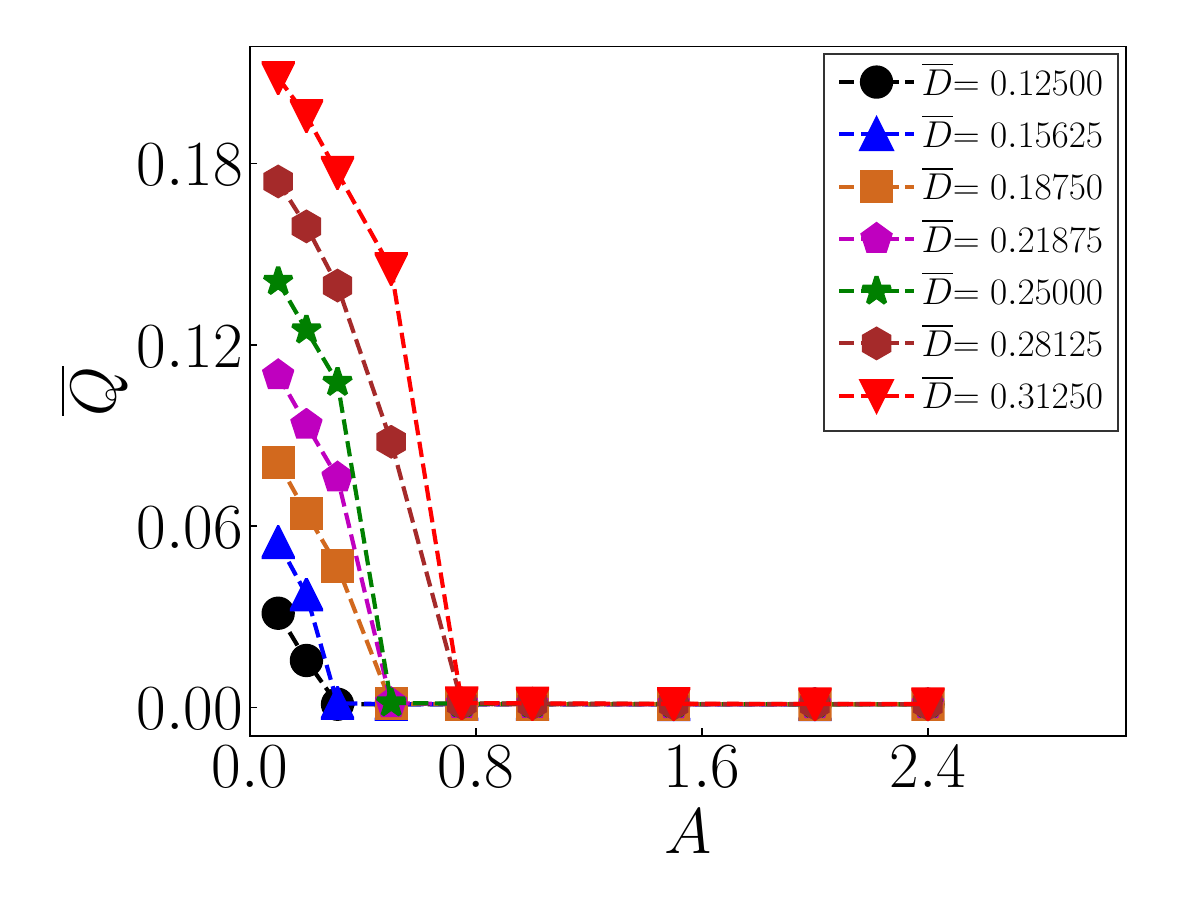}
\includegraphics[height=5cm]
{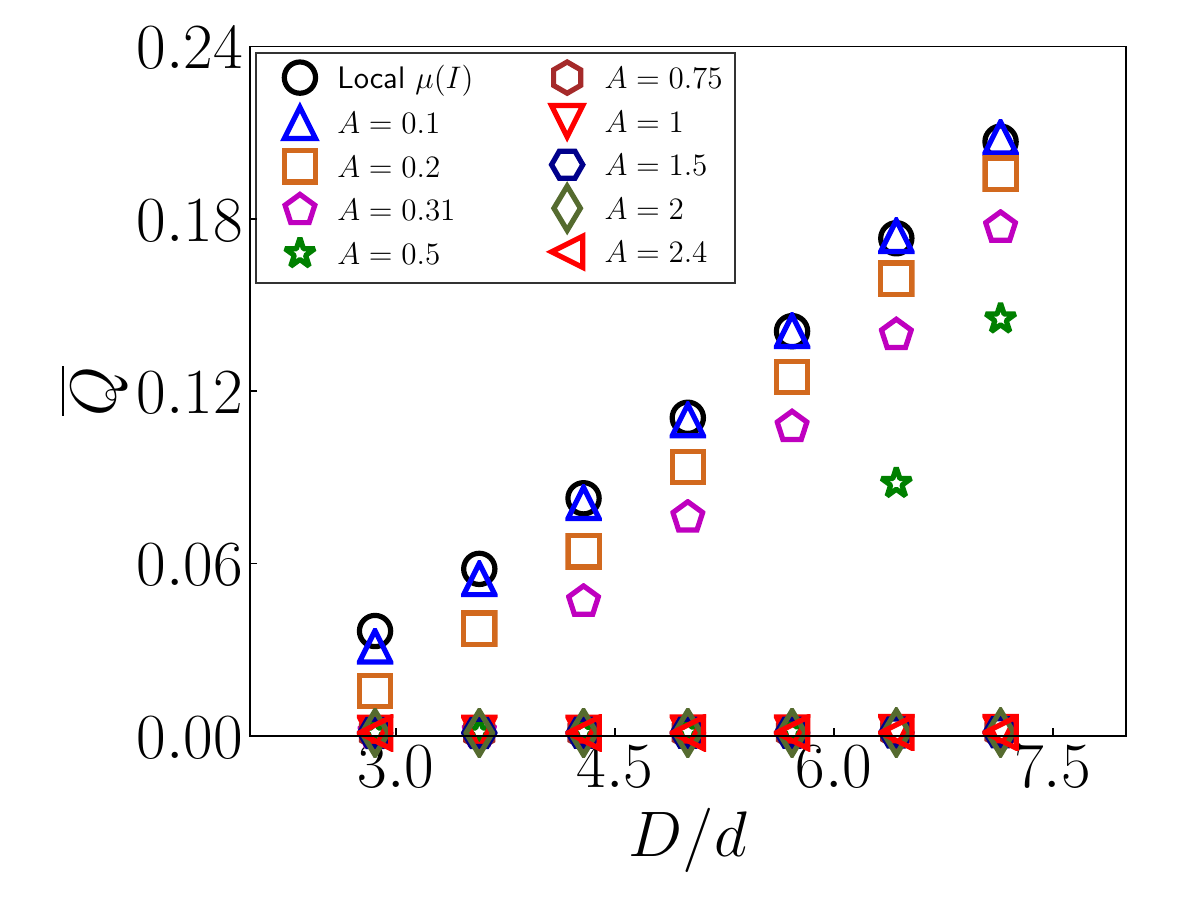}

\caption{\label{NGFflowrate} Results of the mean dimensionless flow rate $\overline{Q}$ obtained with the NGF model for various values of $A$: (a) $\overline{Q}$ versus $A$ for selected values of $D/L$; (b) $\overline{Q}$ versus $D/d$ for selected values of $A$.}	
\end{centering}	
\end{figure}

The implicit treatment eliminates the restrictive stability constraint associated with explicit time stepping, allowing the use of larger time steps and ensuring robust convergence across a wide range of parameters. Motivated by this advantage, we consider the $\mu(I,\Theta)$ model described in \S\ref{sec:muItheta}. In this model, the granular-temperature transport equation is discretised using an unconditionally stable, fully implicit scheme. This allows us to investigate the influence of $\bar{t}_0$ over a wide range, including values much smaller than $\Delta t$, without any numerical stability constraint. The results from the $\mu(I,\Theta)$ model thus serve as a robust validation of the trend observed in the dynamic NGF framework. Figure~\ref{T0diffmuItheta} presents the results for two values of $A$. It shows $\overline{Q}$ as a function of $D/d$ for the local $\mu(I)$ model and for the $\mu(I,\Theta)$ model with $\bar{t}_0$ ranging from $0.00025$ to $0.1$. At $A = 0.5$ (figure~\ref{T0diffmuItheta}(a)), the flow rate is only weakly affected by $\bar{t}_0$, and the data points nearly overlap. At this value of $A$, the $\mu(I,\Theta)$ predictions remain close to the local $\mu(I)$ limit. Consequently, varying $\bar{t}_0$ has a negligible effect on the steady-state discharge rate. At $A = 30$ (figure~\ref{T0diffmuItheta}(b)), however, a clear and systematic trend emerges: as $\bar{t}_0$ decreases, the predicted flow rate decreases monotonically, confirming the behaviour observed in the dynamic NGF model. This $A$-dependence further suggests that $\bar{t}_0$ plays a role only when non-local effects are prominent.

 \begin{figure}
\begin{centering}	
 \hspace{-5.5cm} $(a)$ \hspace{6.cm} $(b)$\\
 \includegraphics[height=5cm]
{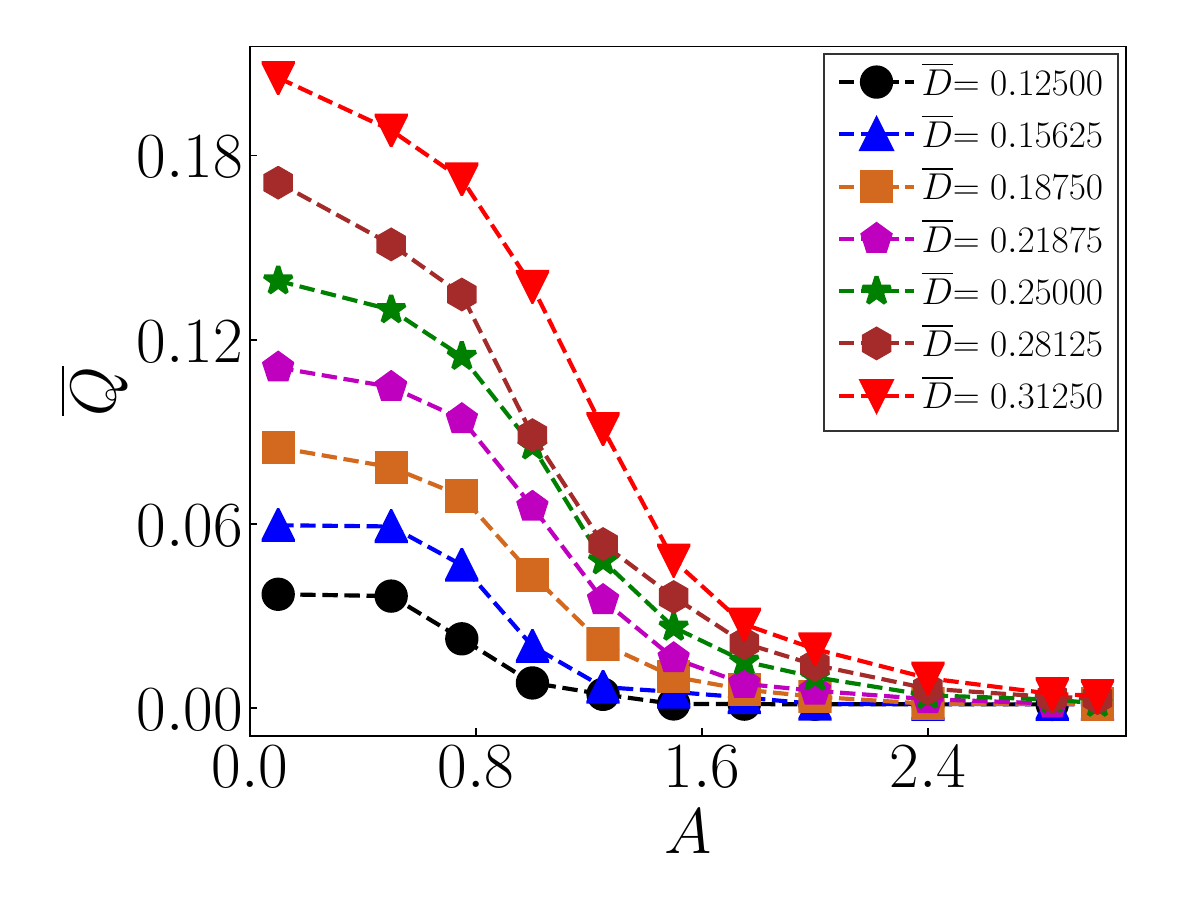}
\includegraphics[height=5cm]
{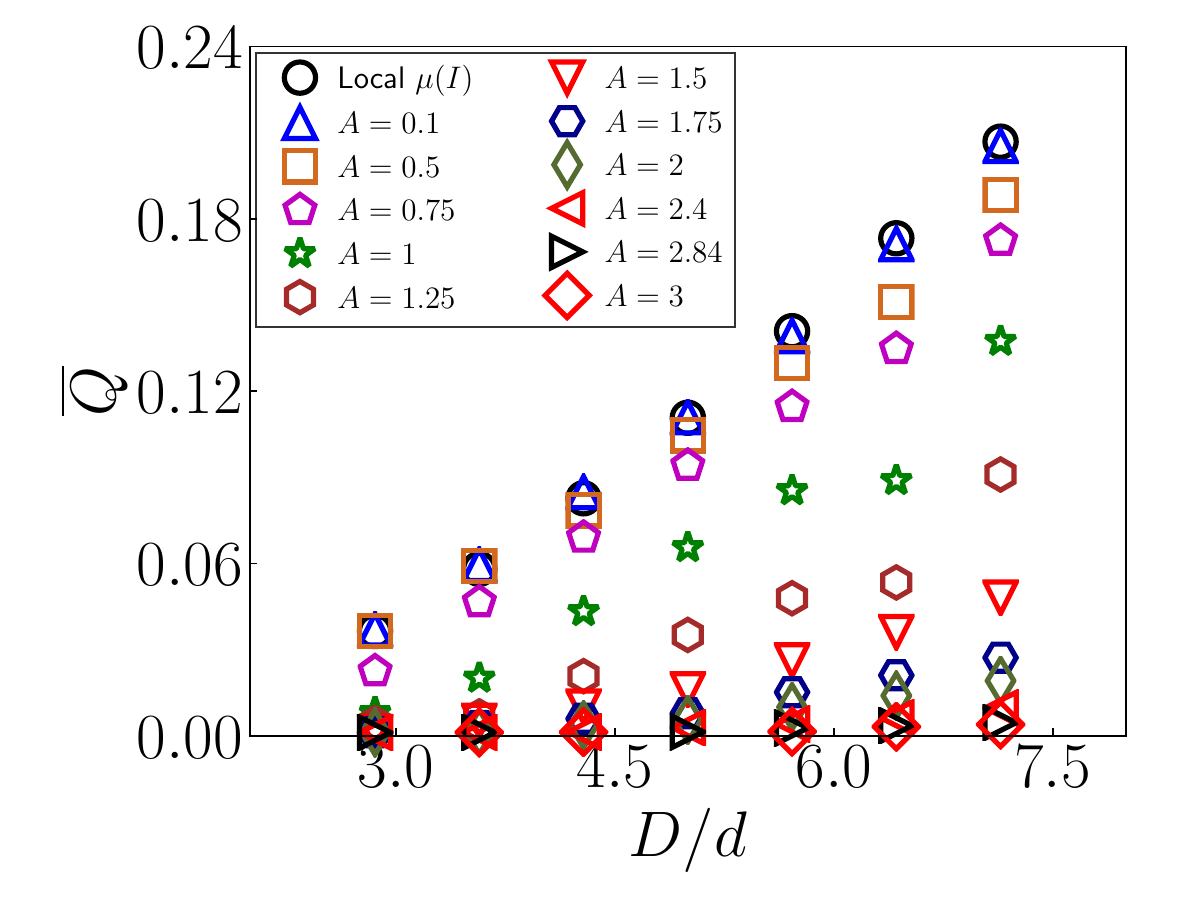}

\caption{\label{Igradientflowrate} Results of the mean dimensionless flow rate $\overline{Q}$ obtained with the $I$-gradient model for various values of $A$: (a) $\overline{Q}$ versus $A$ for selected values of $D/L$; (b) $\overline{Q}$ versus $D/d$ for selected values of $A$.}	
\end{centering}	
\end{figure}

In the previous section (\S\ref{DynamicNGFresults}), we presented the flow rate predictions of the dynamic NGF model and its dependence on $A$ and $D/d$. We now examine the relationship between $\overline{Q}$ and $D/d$ as well as $A$ for three additional non-local models: the NGF model (\S\ref{sec:NGF}), the $I$-gradient model (\S\ref{sec:Igradient}) and the $\mu(I, \Theta)$ model (\S\ref{sec:muItheta}). The NGF model predictions are shown in Figure~\ref{NGFflowrate}. Figure~\ref{NGFflowrate}(a) shows $\overline{Q}$ as a function of $A$ for selected values of $D/L$, while figure~\ref{NGFflowrate}(b) shows $\overline{Q}$ as a function of $D/d$ for selected values of $A$, with the local model results (open black circles) included for reference. The flow rate decreases rapidly as $A$ increases. Notably, at $A = 0.5$, the flow rate becomes zero for most opening sizes, and at $A = 0.75$, it approaches zero across the entire range of $D/d$. This is consistent with the value $A \approx 0.5$ reported in the literature for glass beads (\cite{Henann2013, Dunatunga2022}).

Figure~\ref{Igradientflowrate} displays the corresponding results from the $I$-gradient model, using the same two-panel layout. Our $I$-gradient simulations reveal that the flow rate decreases progressively as $A$ increases, and becomes very close to zero at $A = 2.84$, corresponding to $A^2 = 8.08$. This value is chosen to match the non-local coupling parameter $\nu = 8.08$ reported by \cite{Bouzid2013} for frictional grains. In their shear cell geometry, this divergence reflects the increasing spatial range of non-local correlations near the yield point. In our silo geometry with a free outlet, the same divergence promotes cooperative coupling across the system, leading to a significant reduction in the flow rate.

 \begin{figure}
\begin{centering}	
 \hspace{-5.5cm} $(a)$ \hspace{6.cm} $(b)$\\
 \includegraphics[height=5cm]
{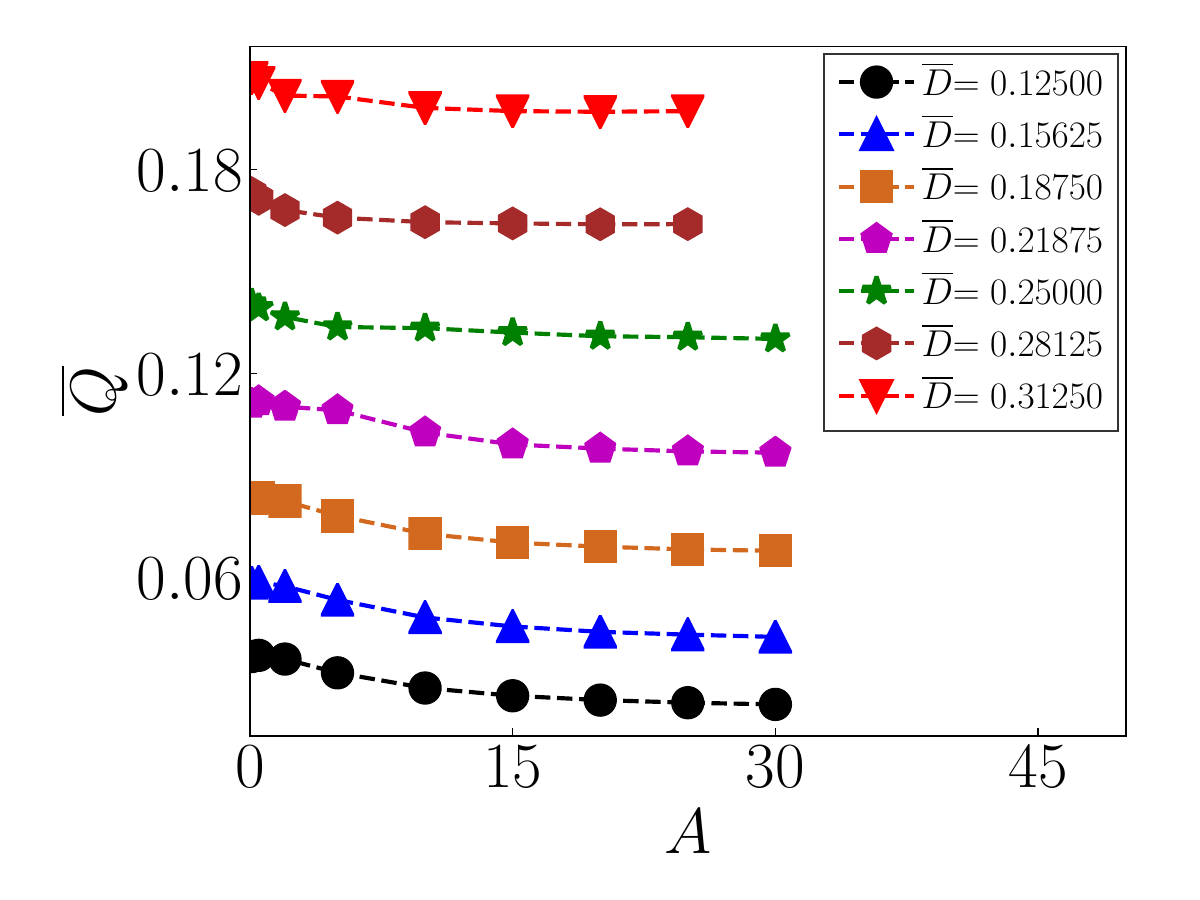}
\includegraphics[height=5cm]
{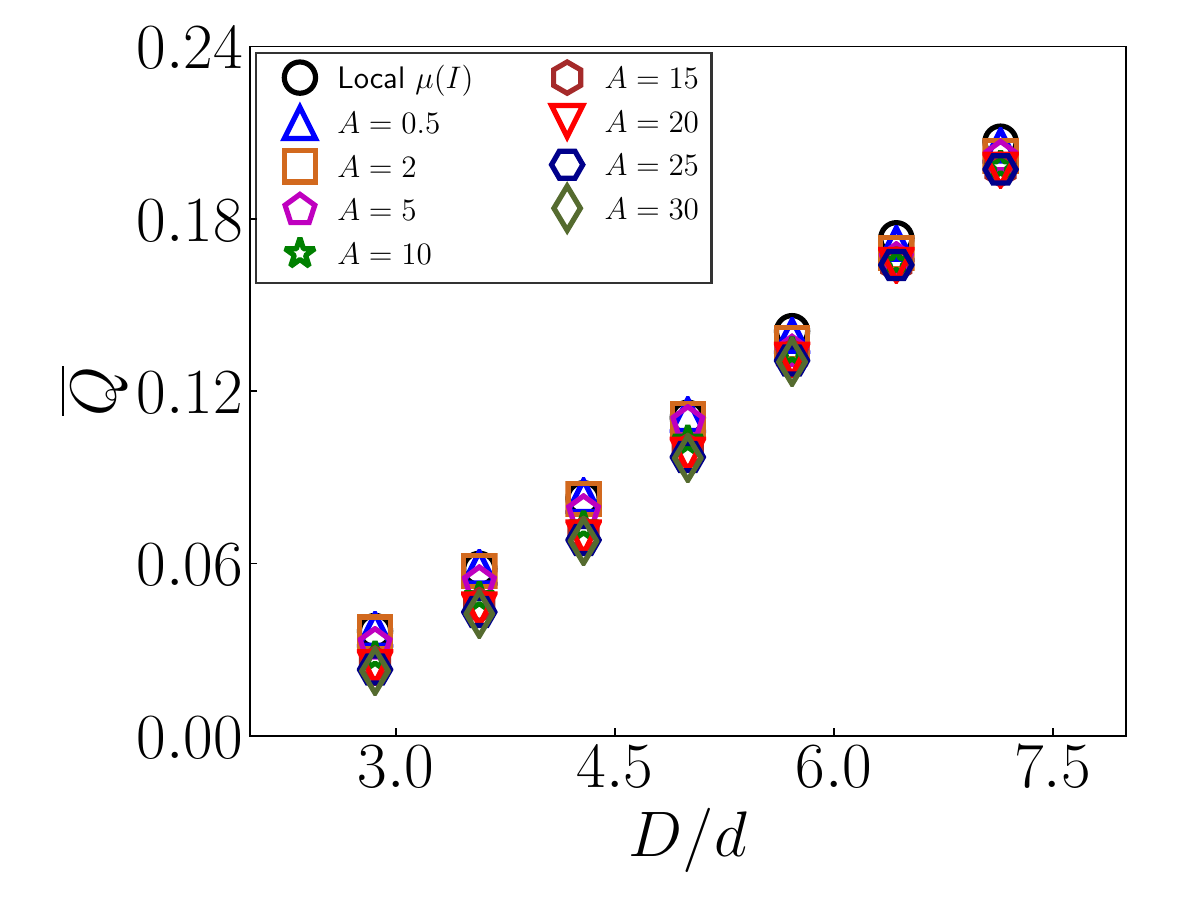}

\caption{\label{GTEflowrate}  Results of the mean dimensionless flow rate $\overline{Q}$ obtained with the $\mu(I, \Theta)$ model for various values of $A$: (a) $\overline{Q}$ versus $A$ for selected values of $D/L$; (b) $\overline{Q}$ versus $D/d$ for selected values of $A$.}	
\end{centering}	
\end{figure}

 \begin{figure}
\begin{centering}	
 \hspace{-5.5cm} $(a)$ \hspace{6.cm} $(b)$\\
 \includegraphics[height=5cm]
{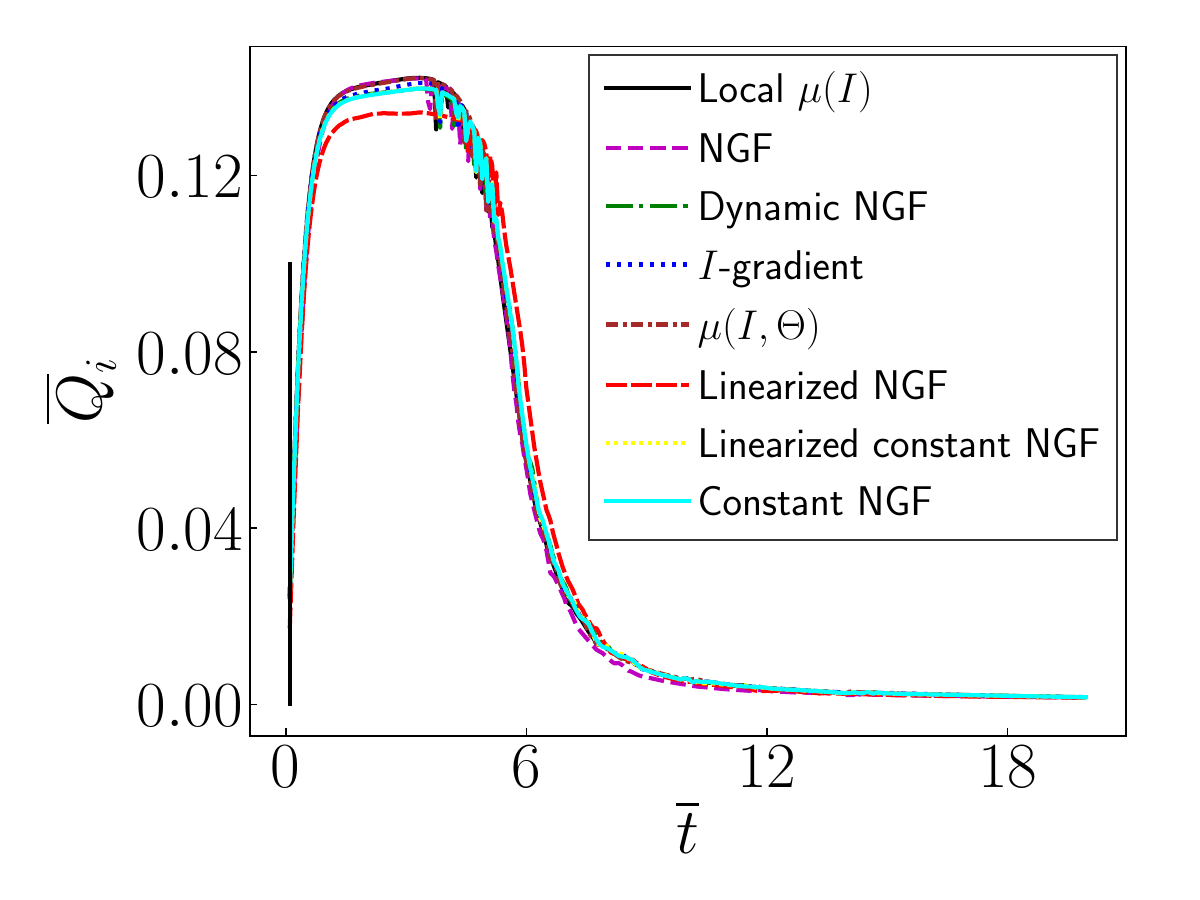}
 \includegraphics[height=5cm]
{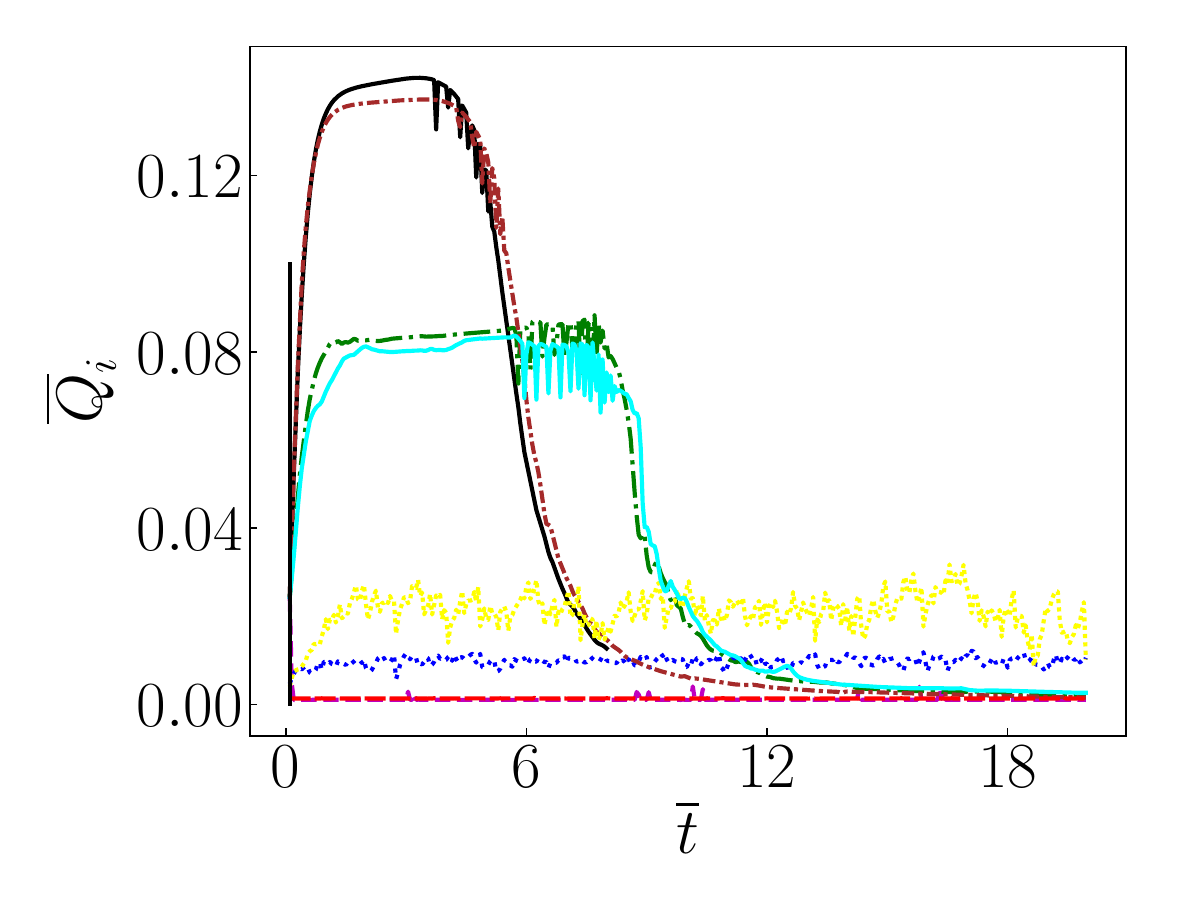}\\
 \hspace{-5.5cm} $(c)$ \hspace{6.cm} $(d)$\\
\includegraphics[height=5cm]
{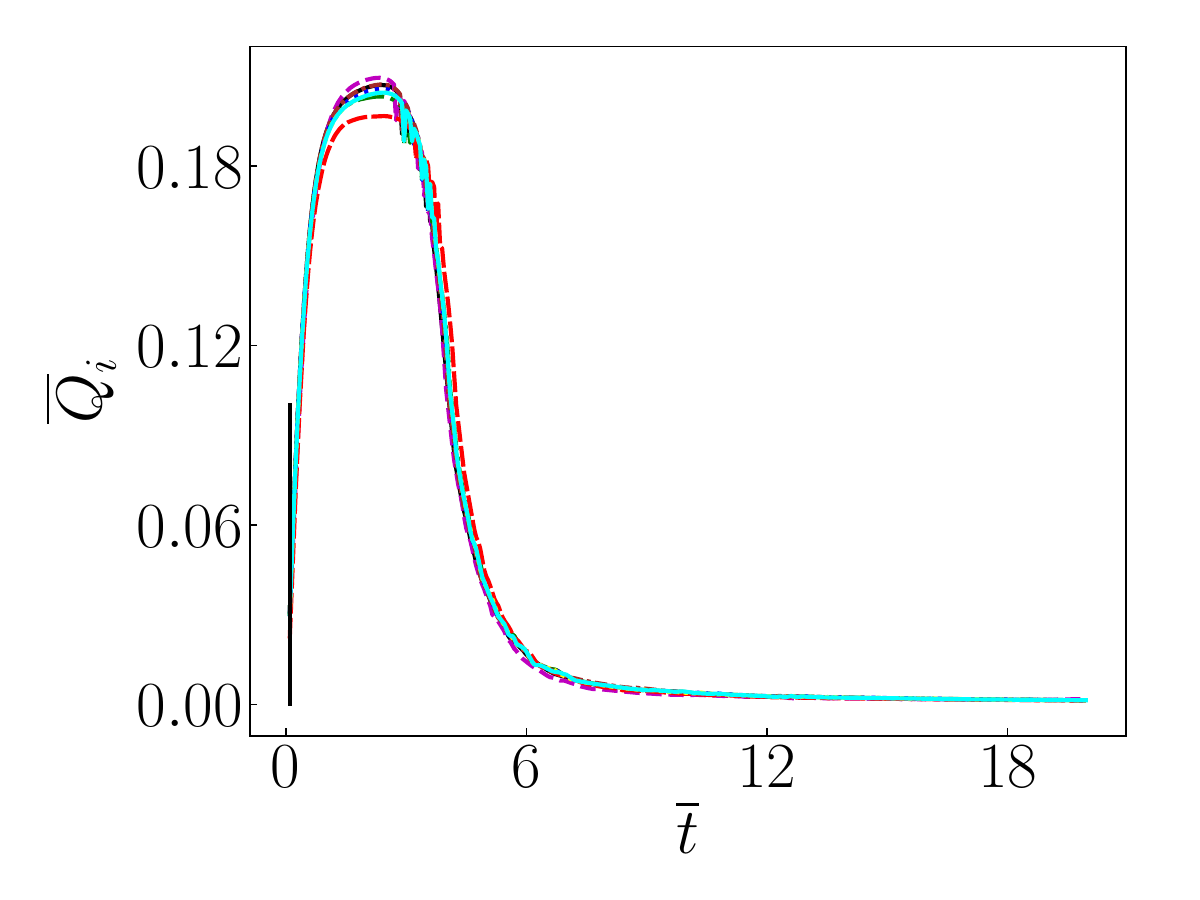}
 \includegraphics[height=5cm]
{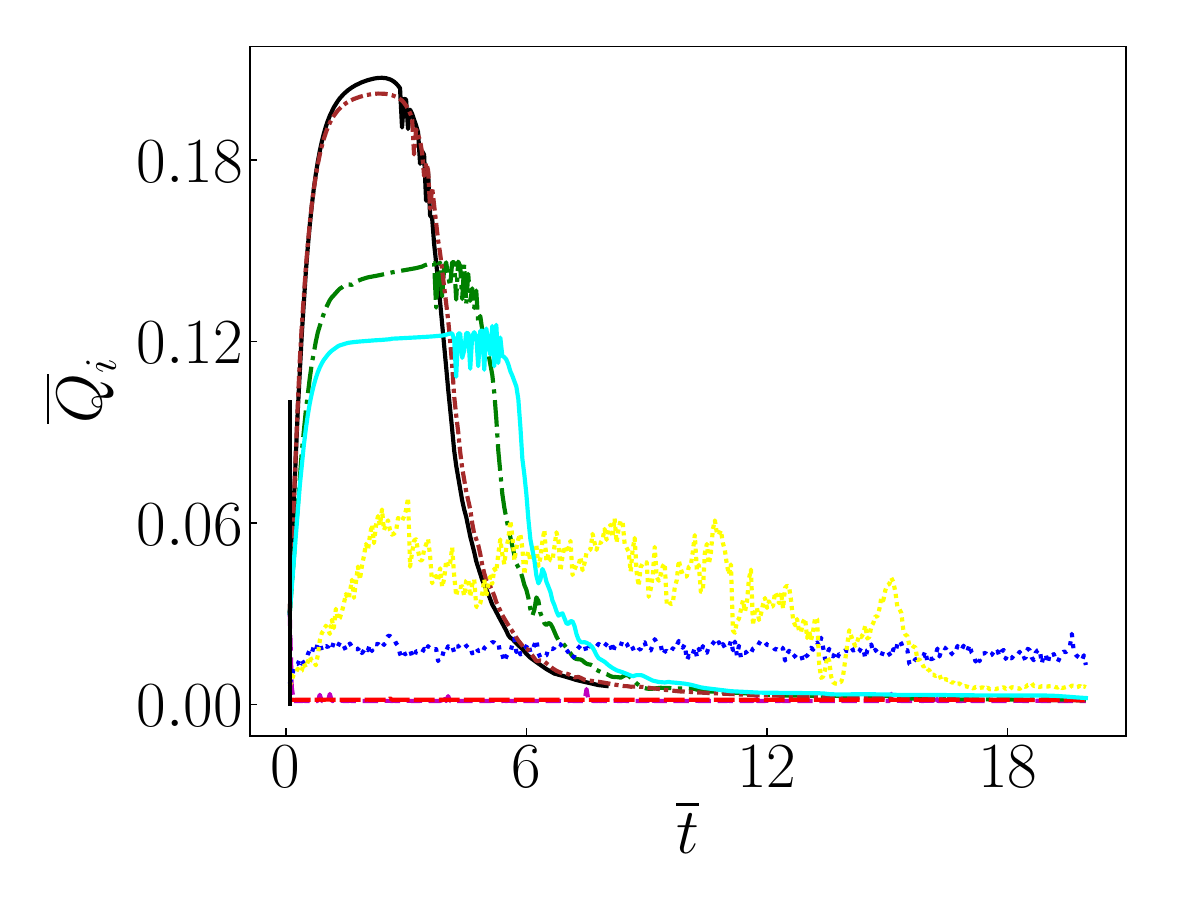}
\caption{\label{fig:flowrate} Dimensionless instantaneous flow rate $\overline{Q}_i$ plotted against $\overline{t}$ for different constitutive models at two $D/L$ ratios: (a, b) $D/L = 0.25$ with (a) $A = 0.1$ and (b) $A = 2$; (c, d) $D/L = 0.3125$ with (c) $A = 0.1$ and (d) $A = 2$. The vertical solid line in each panel indicates the switching time from the local to the non-local simulation, $t_{\text{switch}}/\sqrt{L/G} = 0.1$.
}	
\end{centering}	
\end{figure}

 \begin{figure}
  	\begin{center}
	 \hspace{-12.5cm} $(a)$\\
  \includegraphics[height=7.5cm]{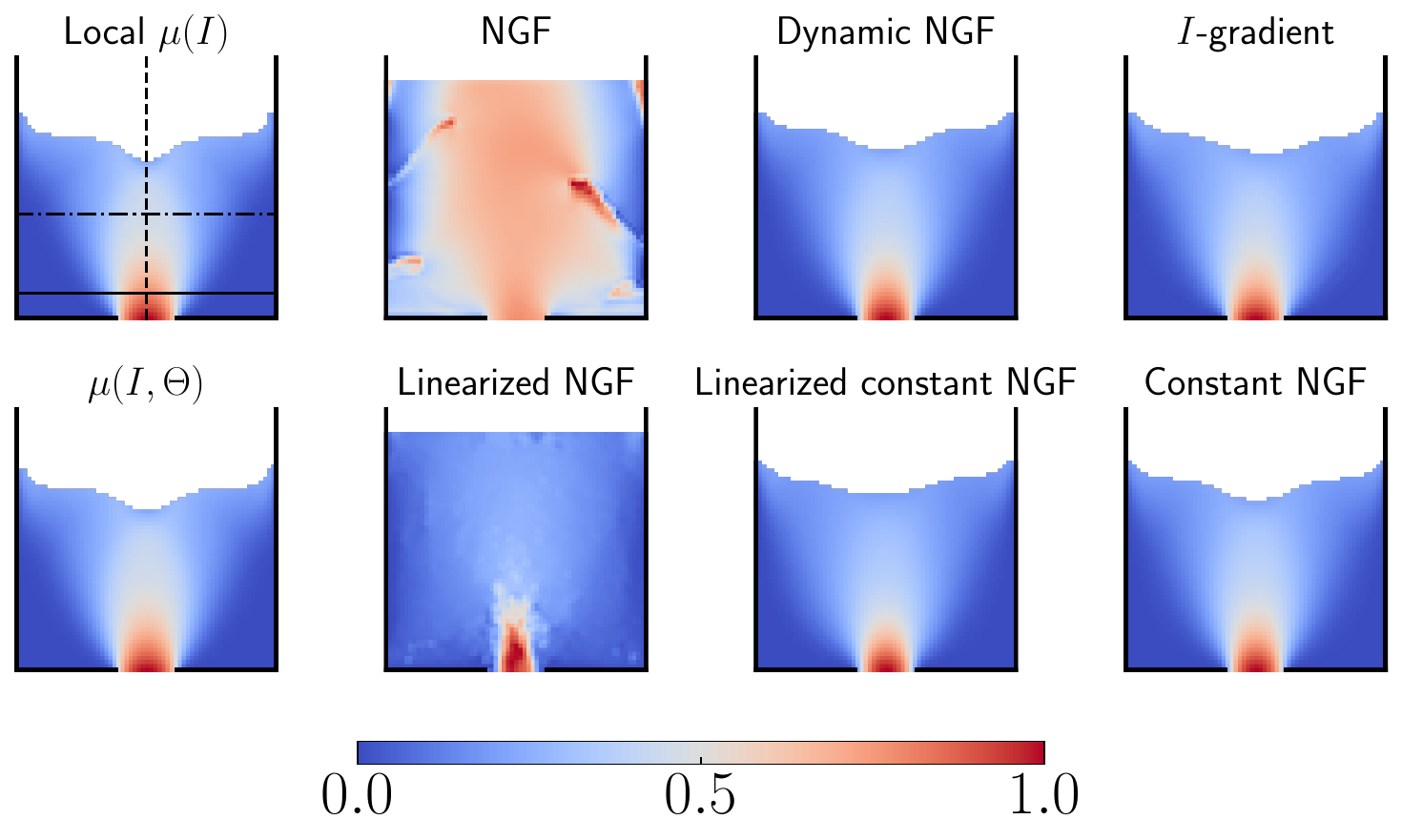}\\
    	 \hspace{-3.5cm} $(b)$ \hspace{4.cm} $(c)$ \hspace{4.cm} $(d)$\\
   \includegraphics[height=3.2cm]
{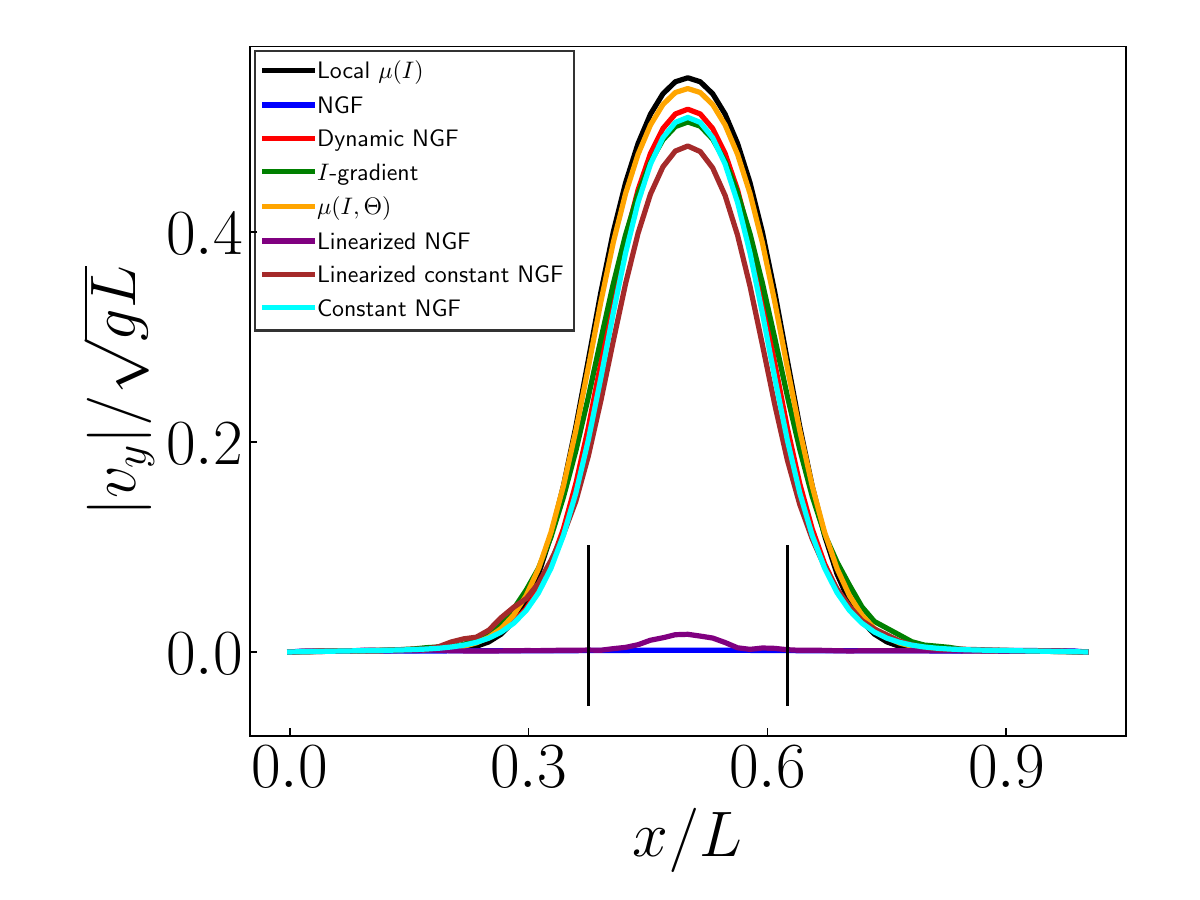}
   \includegraphics[height=3.2cm]
{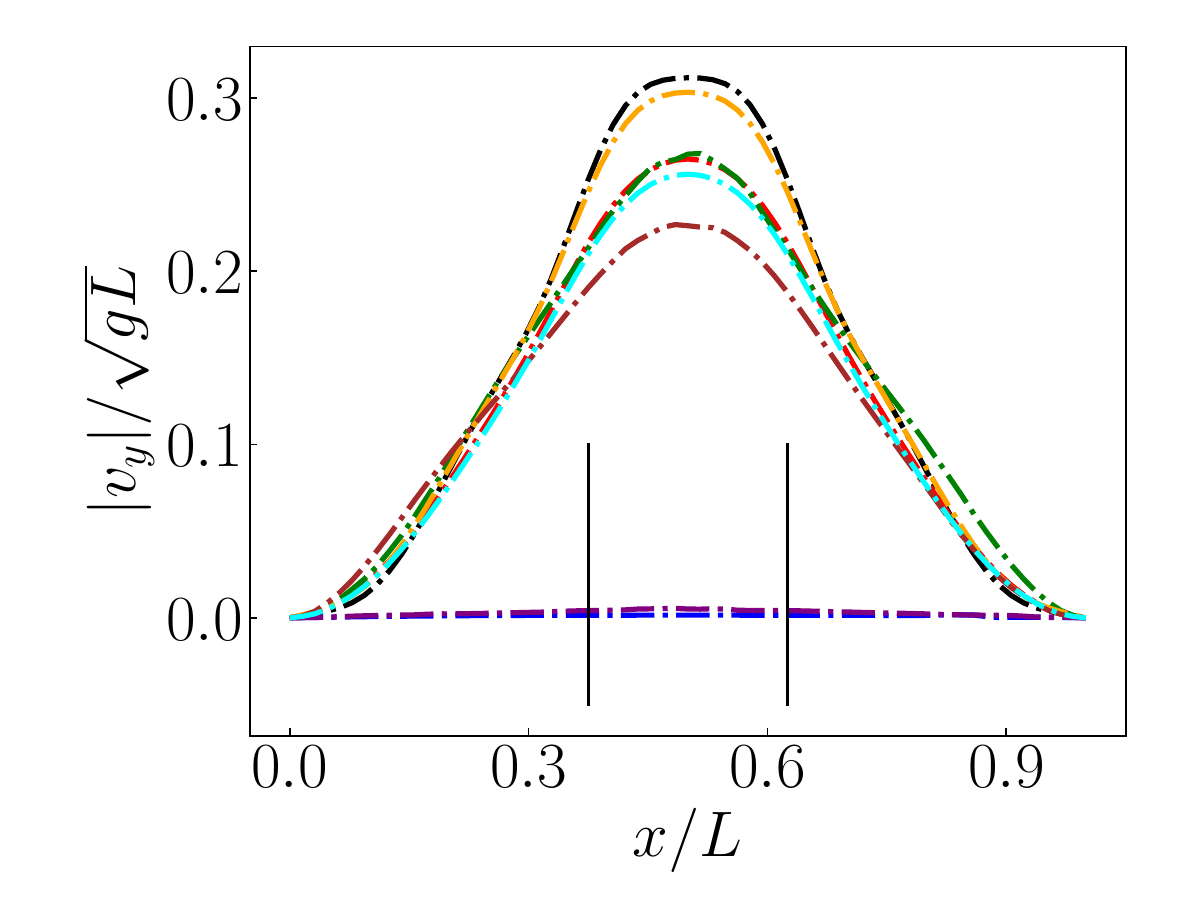}
\includegraphics[height=3.2cm]
{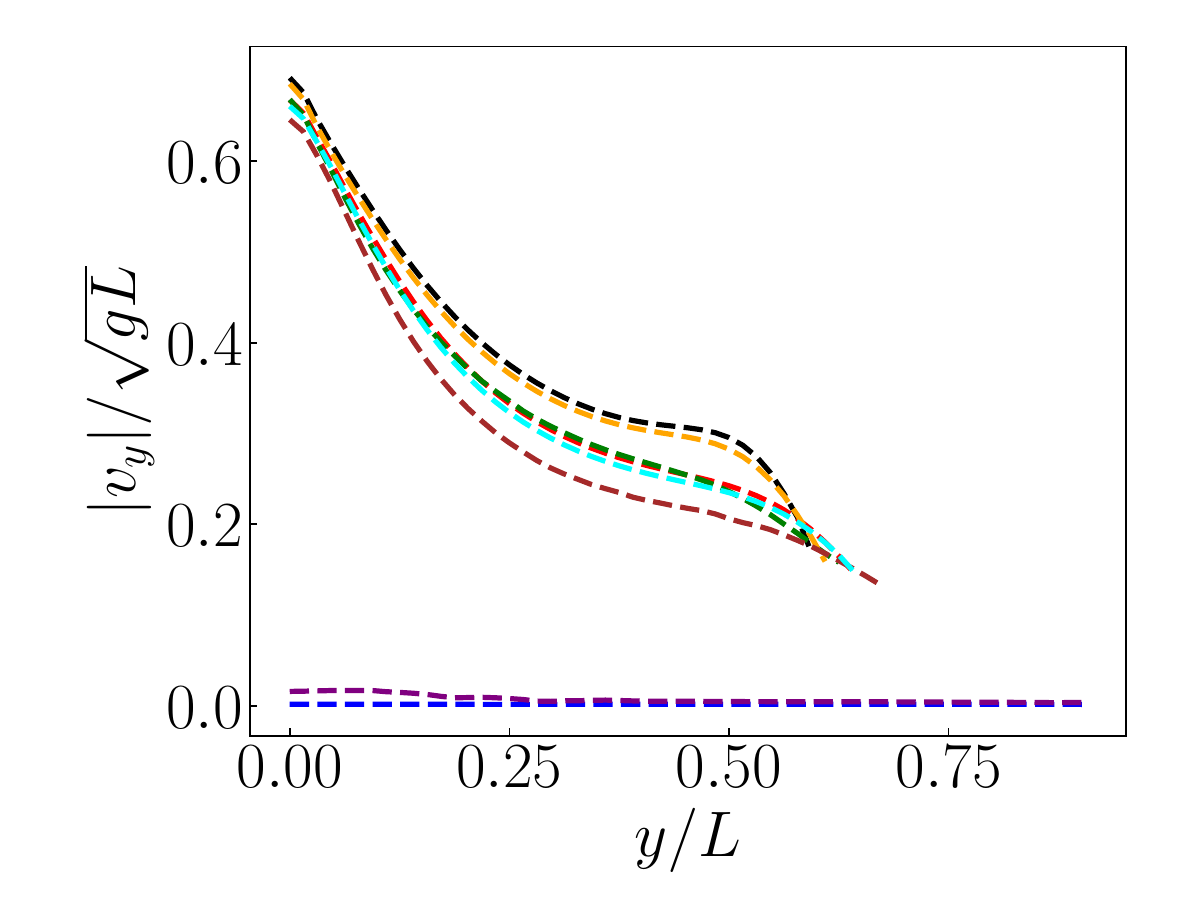}
	\caption{\label{fig:velocity_fieldscompareA05} Normalized vertical velocity fields and profiles at $\overline{t} = 2$, $A = 0.5$, and $\overline{D} = 0.25$ for different constitutive models. (a) Visualization of the normalized vertical velocity field $|v_y|/|v_y|_{\max}$. The solid horizontal line indicates $y/L = 0.1$, the dashed dotted horizontal line indicates $y/L = 0.4$, and the dashed vertical line indicates $x/L = 0.5$. (b) Normalized vertical velocity profiles $|v_y|/\sqrt{GL}$ along the horizontal line at $y/L = 0.1$. (c) Normalized vertical velocity profiles $|v_y|/\sqrt{GL}$ along the horizontal line at $y/L = 0.4$. The vertical solid lines in (b) and (c) indicate the locations of the orifice edges. (d) Normalized vertical velocity profiles $|v_y|/\sqrt{GL}$ along the vertical centerline at $x/L = 0.5$.}
  	\end{center}
\end{figure}

Figure~\ref{GTEflowrate} provides the $\mu(I,\Theta)$ model predictions, again with the identical format. Unlike the other non-local models examined in this work (the $I$-gradient, NGF and dynamic NGF models), the $\mu(I,\Theta)$ model exhibits a distinct behaviour. As $A$ increases, the flow rate decreases and saturates to a finite non-zero value at large $A$. The primary reason lies in the constitutive relation (equation~\eqref{eq:muITheta}), where the exponent $P$ is small. This functional form confines the effective friction coefficient $\mu$ to remain close to the local value $\mu_{\text{loc}}(I)$, preventing it from becoming arbitrarily large even when granular temperature $\Theta$ deviates significantly from its local counterpart.

Having presented the steady-state flow-rate predictions,  we now examine the temporal evolution of the instantaneous dimensionless flow rate $\overline{Q}_i$ for different constitutive models and parameters. This allows us to determine the appropriate time for the velocity and effective viscosity visualisations below. Figure~\ref{fig:flowrate} presents $\overline{Q}_i$ as a function of dimensionless time $\overline{t}$ for two values of $D/L$ ($0.25$ and $0.3125$) and two values of $A$ ($0.1$ and $2$). For $A = 0.1$, all models yield comparable flow rates and exhibit relatively stable behaviour, although mild oscillations appear near the end of the discharge process. For $A = 2$, the differences between models become more pronounced. The dynamic NGF, constant NGF, linearised constant NGF and $I$-gradient models show more pronounced sustained oscillations. The NGF and linearised NGF models predict near-zero discharge rates and exhibit signs of numerical divergence. The $\mu(I,\Theta)$ model also shows some instability, though less severe than the other non-local models and slightly more stable than the local $\mu(I)$ model. Nevertheless, the variations remain bounded, and the overall behaviour is still representative. Based on these results, we choose $\overline{t} = 2$ as a representative time for extracting flow fields, as all models have reached a quasi-steady state by this time. We now examine the spatial distributions of velocity and viscosity predicted by different models at $\overline{t} = 2$ and $D/L = 0.25$. For $A = 0.1$, all models produce nearly identical fields, as non-local effects are weak and the local rheology dominates. We therefore focus on the cases where non-local effects become more pronounced, namely $A = 0.5$ and $A = 2$.

 \begin{figure}
  	\begin{center}
	 \includegraphics[height=7.5cm]{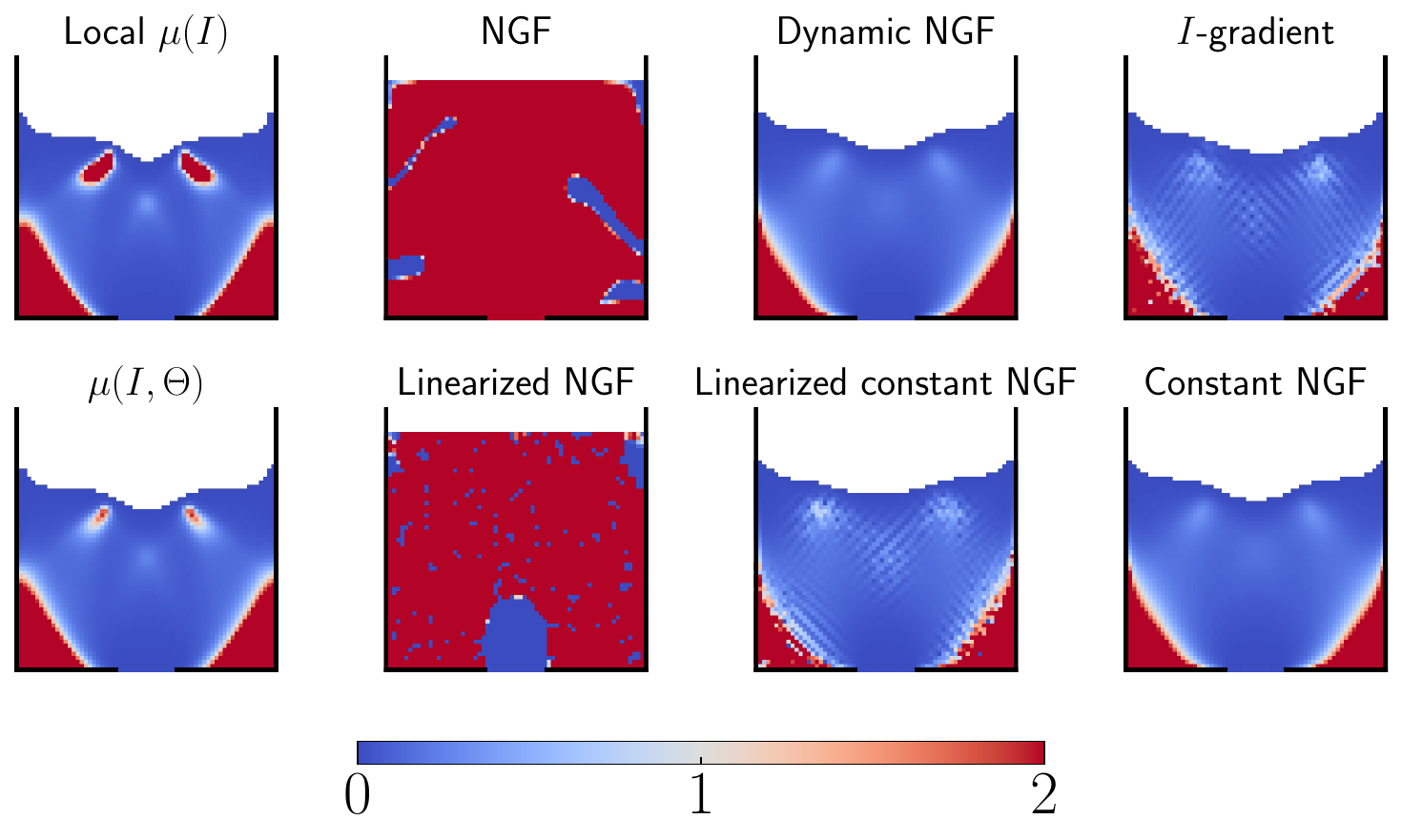}
	  \vspace{-5pt}
	\caption{\label{etadiffA05}Viscosity field $\eta_{eff}/\rho \sqrt{GL^3}$ at $\overline{t} = 2$, $A = 0.5$, and $\overline{D} = 0.25$ for different constitutive models.}
  	\end{center}
\end{figure}

Figure~\ref{fig:velocity_fieldscompareA05} presents the normalized vertical velocity fields and profiles for $A = 0.5$ and $D/L = 0.25$. Panel (a) shows the velocity field contours for different models. Except for the NGF and linearised NGF models, the velocity fields of the other models are qualitatively similar. Panel (b) displays the velocity profiles along the horizontal line at $y/L = 0.1$ (near the outlet). Here, the velocities predicted by most non-local models are slightly lower than those of the local model, while the NGF and linearised NGF models yield velocities very close to zero. Panel (c) shows the velocity profiles at  $y/L = 0.4$. Similar to the local model, the $\mu(I,\Theta)$ model produces a plug-like flow, characterised by a nearly flat velocity profile in the central region. By contrast, the dynamic NGF, $I$-gradient and constant NGF models predict similar but more rounded profiles. These profiles are less plug-like and more consistent with the Gaussian-like distributions observed experimentally. Panel (d) presents the velocity profiles along the vertical centerline ($x/L = 0.5$). Again, only the NGF and linearised NGF models produce velocities close to zero, while the other models show finite flow.

Figure~\ref{etadiffA05} shows the corresponding viscosity fields $\eta_{eff}/\rho \sqrt{GL^3}$. The NGF and linearised NGF models exhibit extremely large viscosity values, consistent with their nearly zero velocities. The remaining models yield comparable viscosity distributions. Notably, the $I$-gradient and linearised constant NGF models display small-scale spatial fluctuations (spots) in the viscosity field, whereas the dynamic NGF, constant NGF and $\mu(I,\Theta)$ models produce smoother fields. This difference arises because both the $I$-gradient model, which includes a correction term $\nu d^{2} \nabla^{2} I / I$, and the linearised constant NGF model, which involves $\xi^{2} \nabla^{2} g / g_{loc}$, require explicit evaluation of second-order derivatives. Such explicit treatment is sensitive to local variations in the solution, leading to small-scale fluctuations. In contrast, the constant NGF, dynamic NGF and $\mu(I,\Theta)$ models are governed by elliptic-type equations, which naturally produce smoother fields.

 \begin{figure}
  	\begin{center}
	 \includegraphics[height=7.5cm]{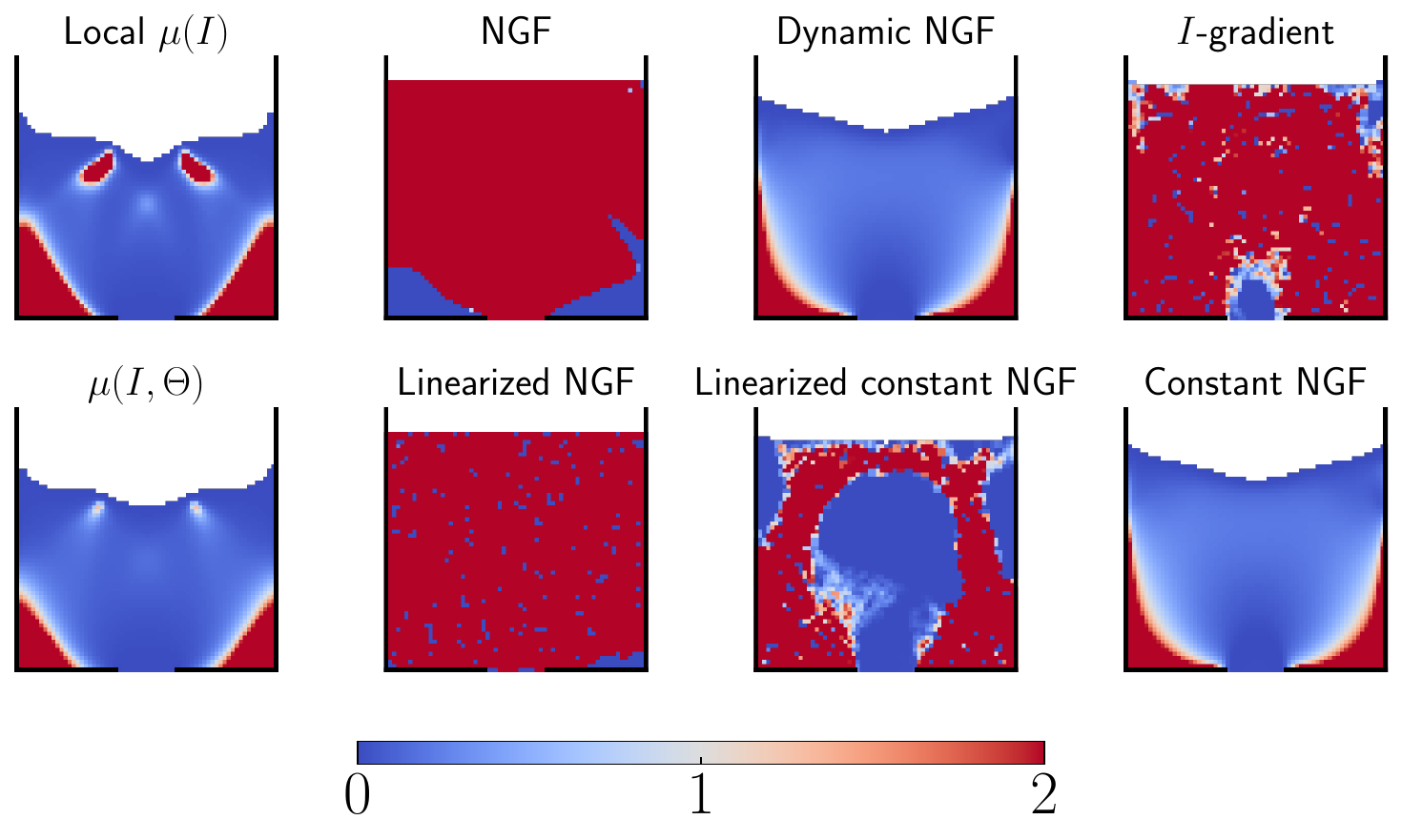}
	  \vspace{-5pt}
	\caption{\label{etadiffA2}Viscosity field $\eta_{eff}/\rho \sqrt{GL^3}$ at $\overline{t} = 2$, $A = 2$, and $\overline{D} = 0.25$ for different constitutive models. }
  	\end{center}
\end{figure}

For $A = 2$, the differences among the models become even more pronounced. Figure~\ref{etadiffA2} shows the corresponding viscosity fields. The NGF, $I$-gradient, linearised NGF and linearised constant NGF models predict extremely large viscosities, indicating flow arrest. By contrast, the local $\mu(I)$, $\mu(I,\Theta)$, constant NGF and dynamic NGF models predict finite viscosities and continued flow. The velocity fields exhibit the same division: the first group predicts nearly zero velocities, whereas the second predicts finite velocities. Since the qualitative features of the velocity fields are similar to those observed at $A = 0.5$, we omit the corresponding velocity figures for brevity.

Before closing this section, we briefly comment on the numerical stability of the simulations presented above. As shown in figure~\ref{fig:flowrate}, the discharge rates predicted by the different non-local models exhibit some scatter. This suggests that the ill-posedness associated with the local $\mu(I)$ rheology may persist, to some extent, in the non-local formulations considered here. The non-local models generally delay the onset of these fluctuations relative to the local model. However, the solutions become less stable and exhibit larger variations at high $A$. Over most of the parameter range, the discharge rates and velocity fields remain relatively stable, and the different models produce consistent overall trends. These observations support the robustness of our main findings.

Beyond the stability issue, the comparison of the eight constitutive models reveals a clear grouping in terms of their predicted flow behaviour. For a given value of $A$, the local and $\mu(I,\Theta)$ models yield very similar results; the dynamic NGF and constant NGF models are closely aligned; the $I$-gradient and linearised constant NGF models form another group; and the NGF and linearised NGF models exhibit comparable behaviour. Moreover, the NGF and linearised NGF models stand out from the other approaches, as they predict a near-zero discharge at $A = 0.5$, consistent with the values reported for glass beads in \cite{Henann2013, Dunatunga2022}. It is also worth noting that the $I$-gradient model yields near-zero discharge at $A = 2.84$ (corresponding to $\nu = 8.08$), consistent with \cite{Bouzid2013}. This grouping reflects the underlying mathematical structure of the models and may provide guidance for the development and selection of constitutive relations in future studies.

Following \cite{Lin2020, Lin2021}, we impose $I = 0$ at the solid walls. However, the inertial number fields obtained from the local and dynamic NGF models (figure \ref{fig:I_fields}) show that $I$ does not always vanish exactly near the walls. This is particularly evident in the upper region of the silo and for small values of $A$. This observation suggests that the commonly used boundary condition $I = 0$ at the wall may not be strictly accurate in all configurations. Nevertheless, we have verified that the discharge rate is only weakly sensitive to this choice of boundary condition, and the main conclusions of this work are independent of it.

\section{Conclusion} \label{conclusion}

In this work, we have simulated silo discharge using various non-local constitutive models implemented in a finite-volume Navier-Stokes solver. Using the FVM with the dynamic NGF model, we obtained results that are quantitatively consistent with those reported by \cite{Dunatunga2022}, despite the fact that our numerical method differs from their MPM-based approach. Our continuum simulations capture the full Gaussian velocity profile, including the $1/\sqrt{y}$ prefactor, which provides a more complete quantitative characterisation of the diffusive spreading observed in their work. However, since we did not introduce a separated phase, we observed that when the outlet size $D/d$ becomes very small, the pressure near the orifice becomes negative. From a continuum simulation perspective, this negative pressure can be interpreted as an indicator of arch formation.

We also proposed an alternative approach to predict clogging that differs from the method of \cite{Dunatunga2022}. Instead of introducing a separated phase to directly capture arch formation, we focused on the discharge rate $Q$ and sought to correlate it with the clogging probability $J$. In our study, we adopted more physically realistic rheological parameters and investigated primarily the effect of the orifice size $D$, whereas \cite{Dunatunga2022} varied the grain size $d$ and employed a large value of $\mu_2$ for numerical convenience. From our continuum simulations, we established two relationships. First, we fitted $Q$ as a function of $A$ and $D/d$ and used this fit to extract the Beverloo cutoff size. The resulting cutoff agrees well with experimental observations (\cite{Benyamine2014, Choi2005}). 
Second, based on the probabilistic framework of \cite{Janda2008}, we proposed a unified, physically based expression for $J(D/d, A)$. The predicted clogging probabilities are qualitatively consistent with the experimental trends. This approach bridges continuum simulations with the intrinsically stochastic nature of clogging by providing a probabilistic description, whereas \cite{Dunatunga2022} sought a deterministic prediction of flow arrest. 

Finally, we systematically compared the predictions of different non-local models for identical values of $A$. For $A > 0$, most models predict a reduction in flow rate relative to the local $\mu(I)$ rheology; for very small $A$, their predictions converge. The models exhibit a clear grouping: the local and $\mu(I,\Theta)$ models yield similar results; the dynamic NGF and constant NGF models form one group; the $I$-gradient and linearised constant NGF models form another; and the NGF and linearised NGF models behave similarly. Notably, the NGF and linearised NGF models predict near-zero discharge at $A = 0.5$, consistent with glass bead parameters reported in \cite{Henann2013, Dunatunga2022}, while the $I$-gradient model yields near-zero discharge at $A = 2.84$ ($\nu = 8.08$), consistent with \cite{Bouzid2013}. As $A$ increases, the differences between groups become more pronounced. This grouping reflects the underlying mathematical structure of the models and may guide the selection of constitutive relations in future studies. We also note that the boundary conditions for $g$ and $I$ at solid walls are not yet firmly established, and future experimental work is needed to better constrain them.

Most simulations presented in this work are computationally efficient: a typical single calculation can be completed within approximately $10$ minutes on a standard Linux computer. The only exception is the finest convergence test case ($\Delta x = L/2^8$, $\Delta t = 0.0000625$ s), which requires approximately 1 hour. The complete source code and simulation scripts for all models presented in this study are available at \url{https://basilisk.fr/sandbox/yixian/}. Future work will extend the present framework by incorporating the $\phi(I)$ rheology to account for volume fraction variations, which are known to play a crucial role in clogging. In addition, more detailed quantitative comparisons with DEM simulations will be performed to further validate the predictive capability of the non-local models across a wider range of flow conditions, including simpler configurations such as inclined plane flows and shear cells.

\clearpage

\appendix

\section{Influence of particle size}\label{app:grain_size}

 \begin{figure}
\begin{centering}	

\includegraphics[height=5cm]
{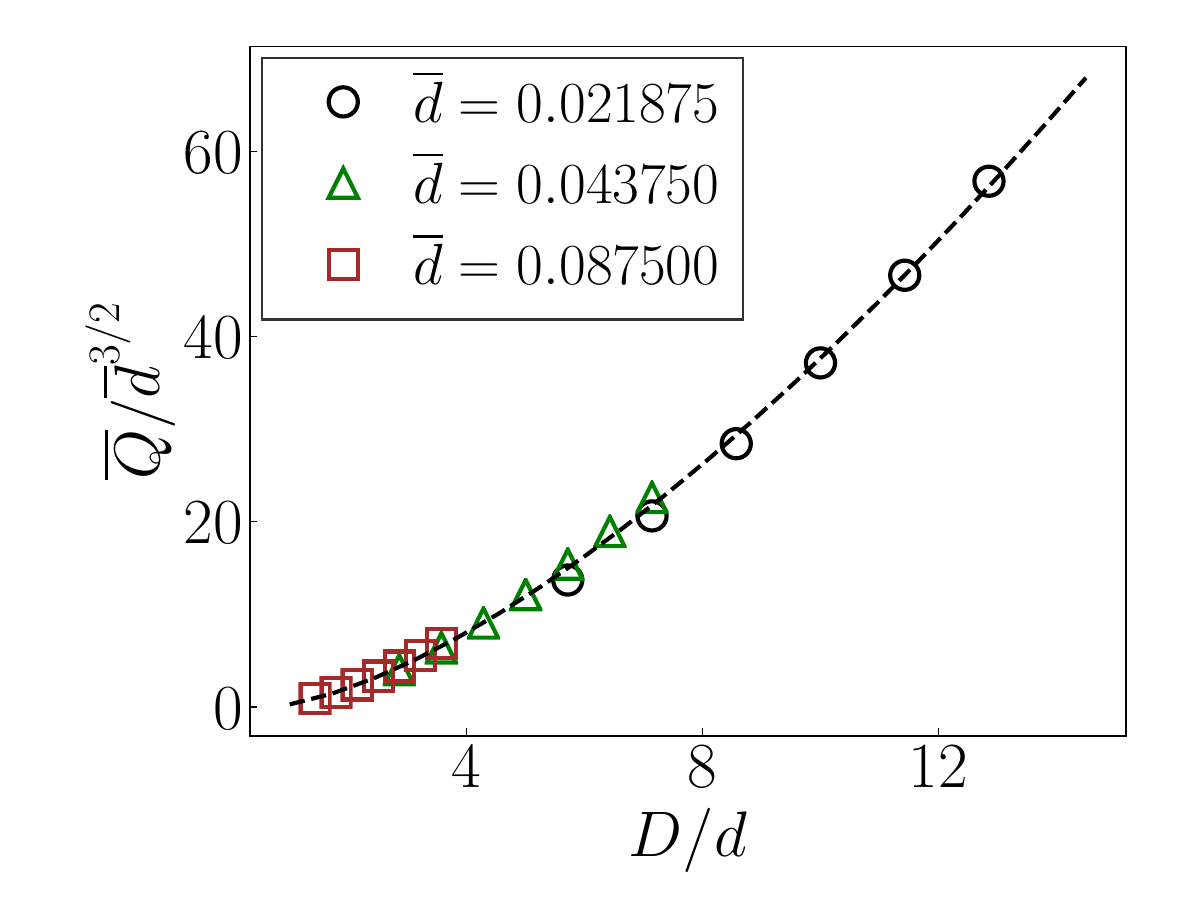}

\caption{\label{QvsDBevddiff} Dimensionless flow rate $\overline{Q}$ normalized by $\overline{d}^{3/2}$ as a function of $D/d$ for the local $\mu(I)$ rheology with different particle sizes, using the modified parameter set (table~\ref{tab:params}). The black dashed line corresponds to equation~\eqref{Qbev} with $C_{Bev} = 1.32$ and $k_{Bev} = 0.68$.}	
\end{centering}	
\end{figure}

\begin{figure}
\begin{centering}
 \hspace{-5.5cm} $(a)$ \hspace{6.cm} $(b)$\\
\includegraphics[height=5cm]
{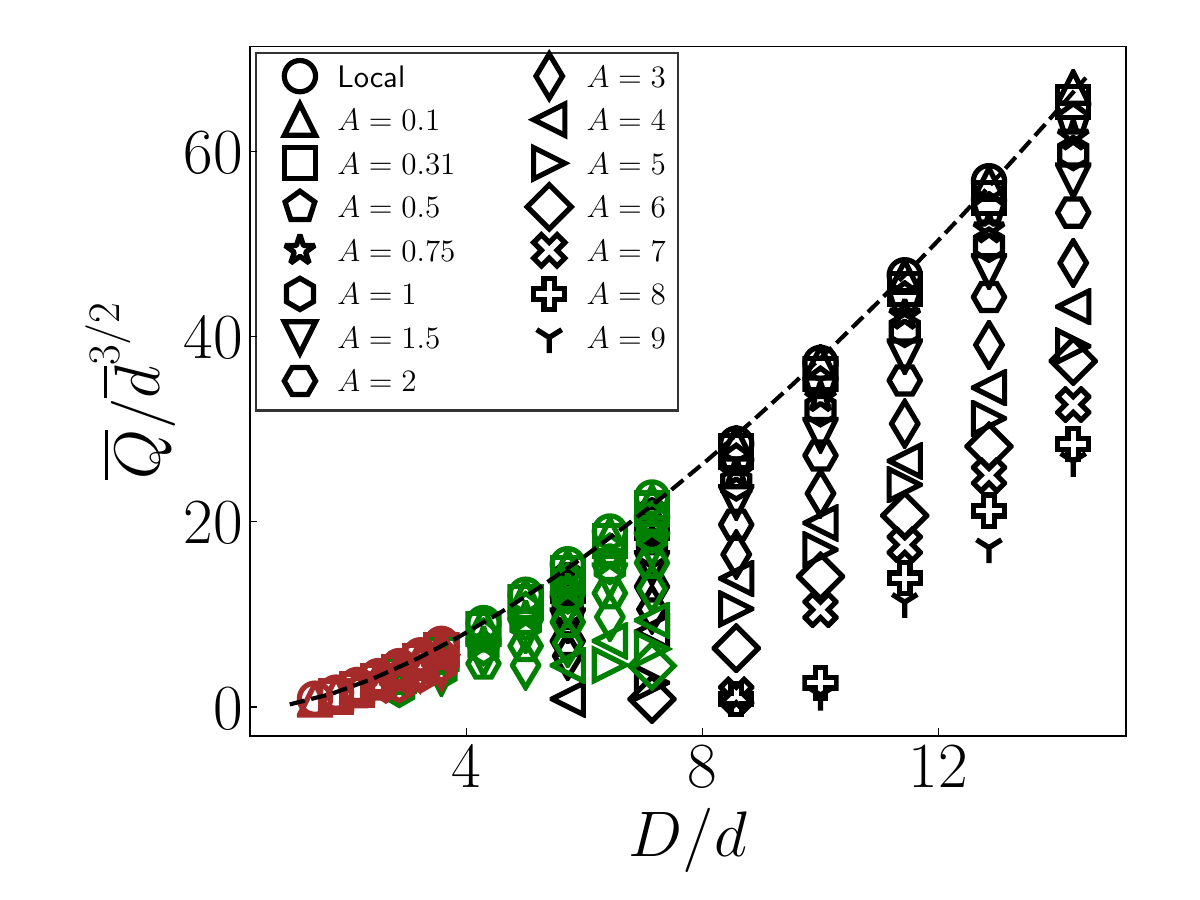}
\includegraphics[height=5cm]
{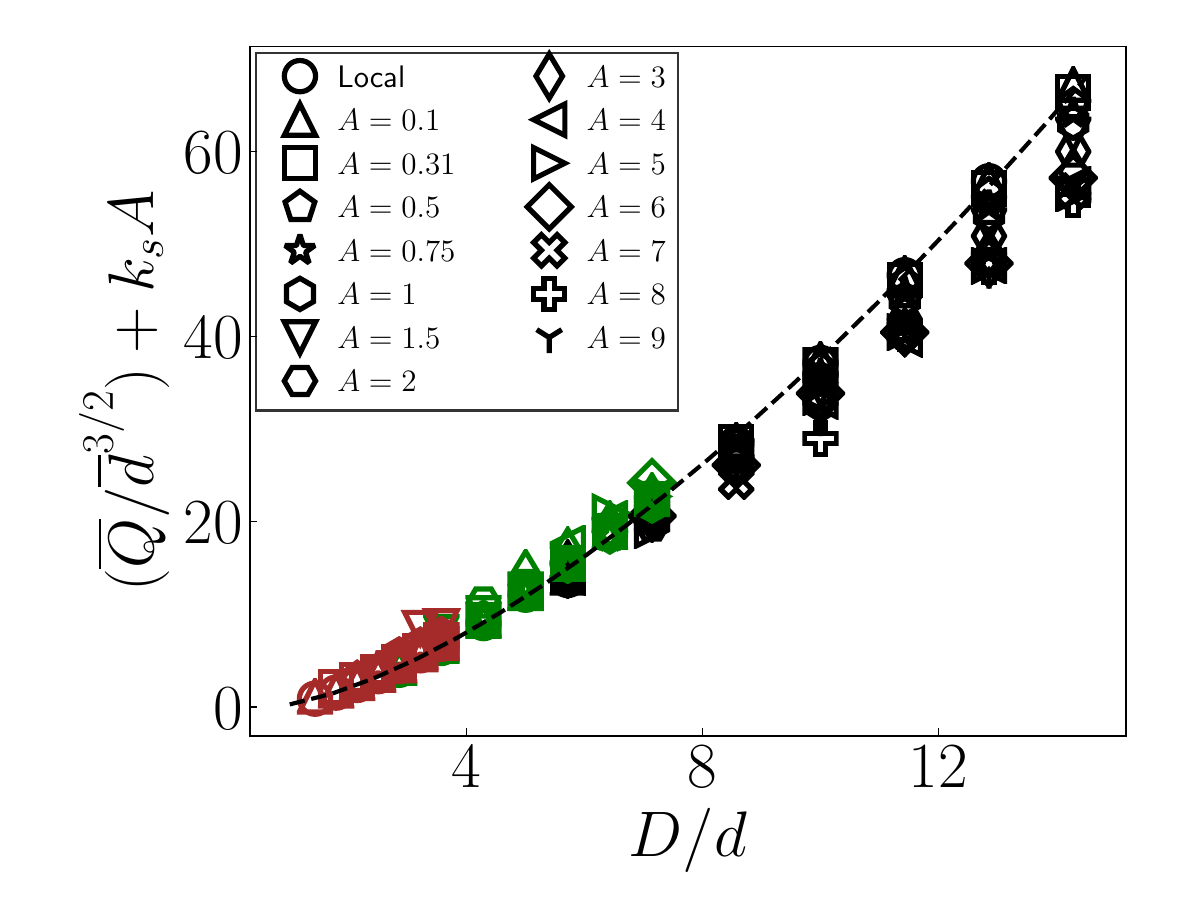}
\caption{\label{beverloo_loiddiffAdiffall} Scaling of the flow rate for different particle sizes and non-local amplitudes, obtained from the dynamic NGF model using the modified parameter set (table~\ref{tab:params}). (a) Dimensionless flow rate $\overline{Q}$ normalized by $\overline{d}^{3/2}$ as a function of $D/d$ for different values of $A$ and particle size $\overline{d}$; (b) $(\overline{Q}/\overline{d}^{3/2}) + k_s A$ as a function of $D/d$ for different values of $A$ and particle size $\overline{d}$, where $k_s = 3.35$. Marker styles distinguish different $A$ values, while colors distinguish different $\overline{d}$ values, as in figure~\ref{QvsDBevddiff}. The black dashed lines correspond to equation~\eqref{Qbev} with $C_{\text{Bev}} = 1.32$ and $k_{\text{Bev}} = 0.68$. Data are shown only for $D/d > (D/d)_c$.}
\end{centering}
\end{figure}

We now investigate the effect of particle size on the flow rate as an additional validation of our continuum numerical framework. Figure~\ref{QvsDBevddiff} shows the dimensionless flow rate $\overline{Q}$ normalized by $\overline{d}^{3/2}$ as a function of $D/d$ for the local model with different particle sizes, using the modified parameter set (table~\ref{tab:params}). The results collapse onto the Beverloo correlation (equation~\eqref{Qbev}), confirming that our simulations capture the classical size-dependent discharge behaviour. 
Figure~\ref{beverloo_loiddiffAdiffall} presents the corresponding results for the dynamic NGF model using the same modified parameter set. The data in figure~\ref{beverloo_loiddiffAdiffall}(a) show the dimensionless flow rate $\overline{Q}$ normalized by $\overline{d}^{3/2}$ as a function of $D/d$ for different values of $A$ and particle size $\overline{d}$; the data do not collapse under this simple scaling. In figure~\ref{beverloo_loiddiffAdiffall}(b), after applying the additional correction $k_s A$ (equation~\eqref{Qalaw}), all data collapse onto a single curve, which follows the Beverloo relation. This demonstrates that the proposed scaling law, which includes the non-local parameter $A$, successfully accounts for the combined effects of particle size and non-locality. Notably, even in continuum simulations, the particle size dependence is well described by the Beverloo equation and its extension, confirming the predictive capability of the present framework.

 \begin{figure}
  	\begin{center}
	 \hspace{-12.5cm} $(a)$\\
  \includegraphics[height=9cm]{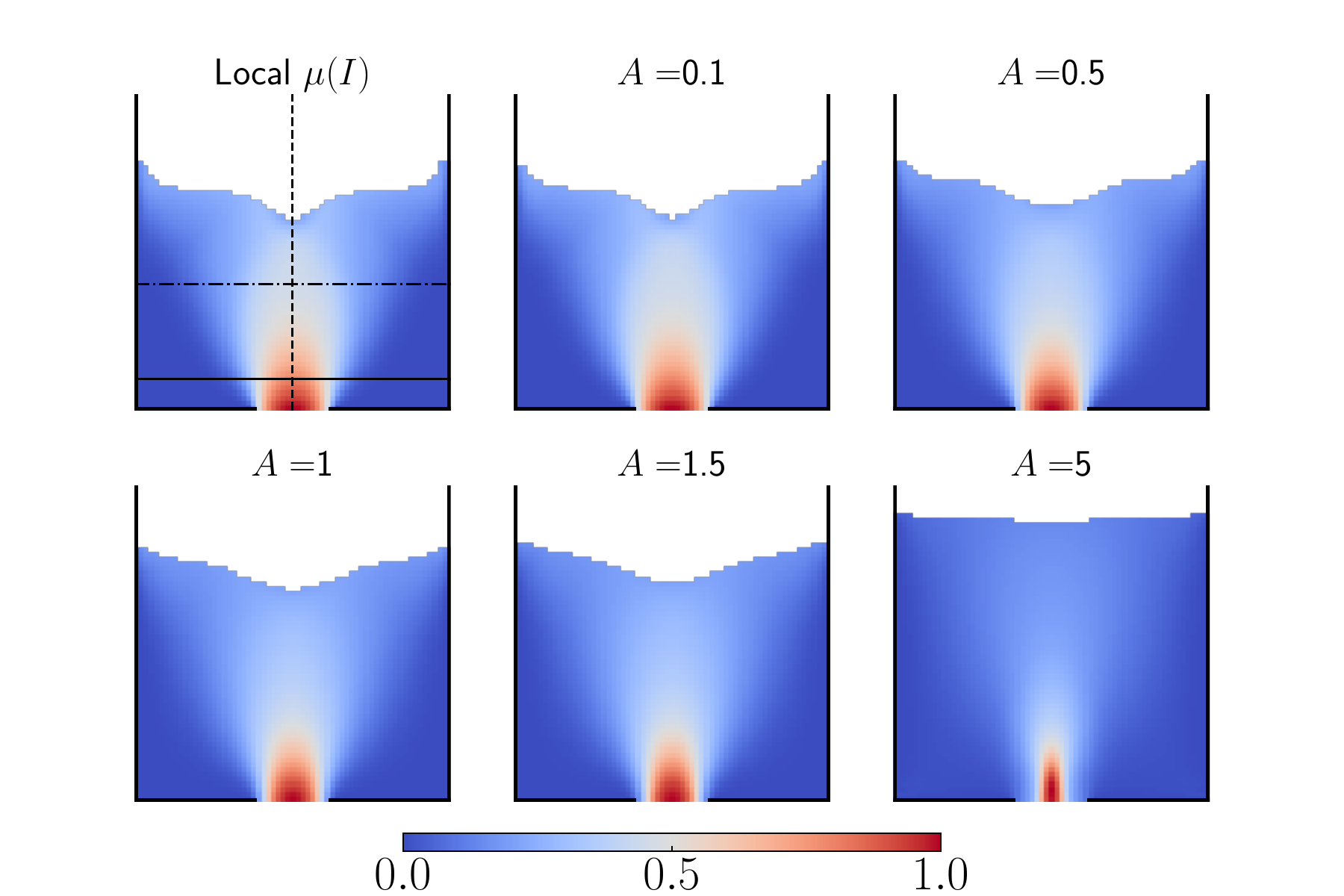}\\
   \hspace{-3.5cm} $(b)$ \hspace{4.cm} $(c)$ \hspace{4.cm} $(d)$\\
   \includegraphics[height=3.2cm]
{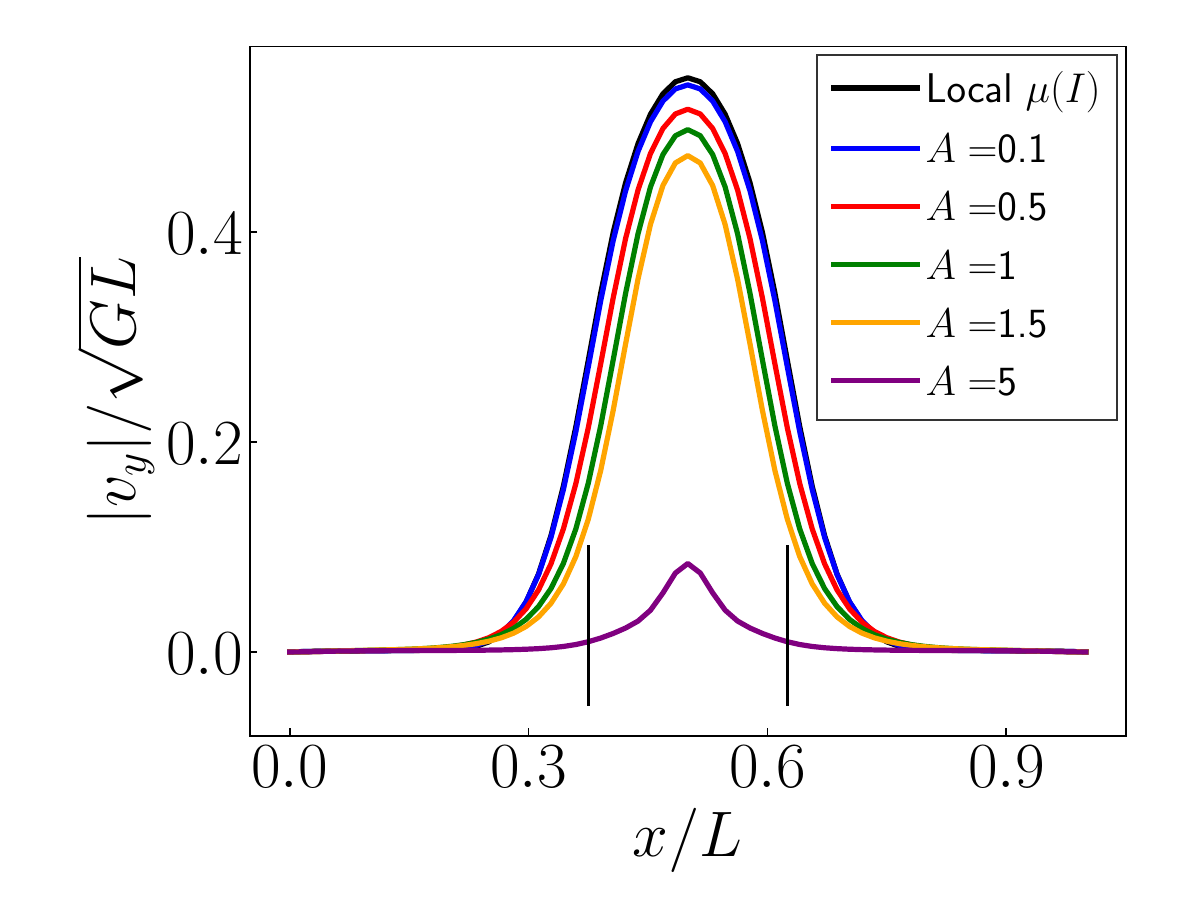}
   \includegraphics[height=3.2cm]
{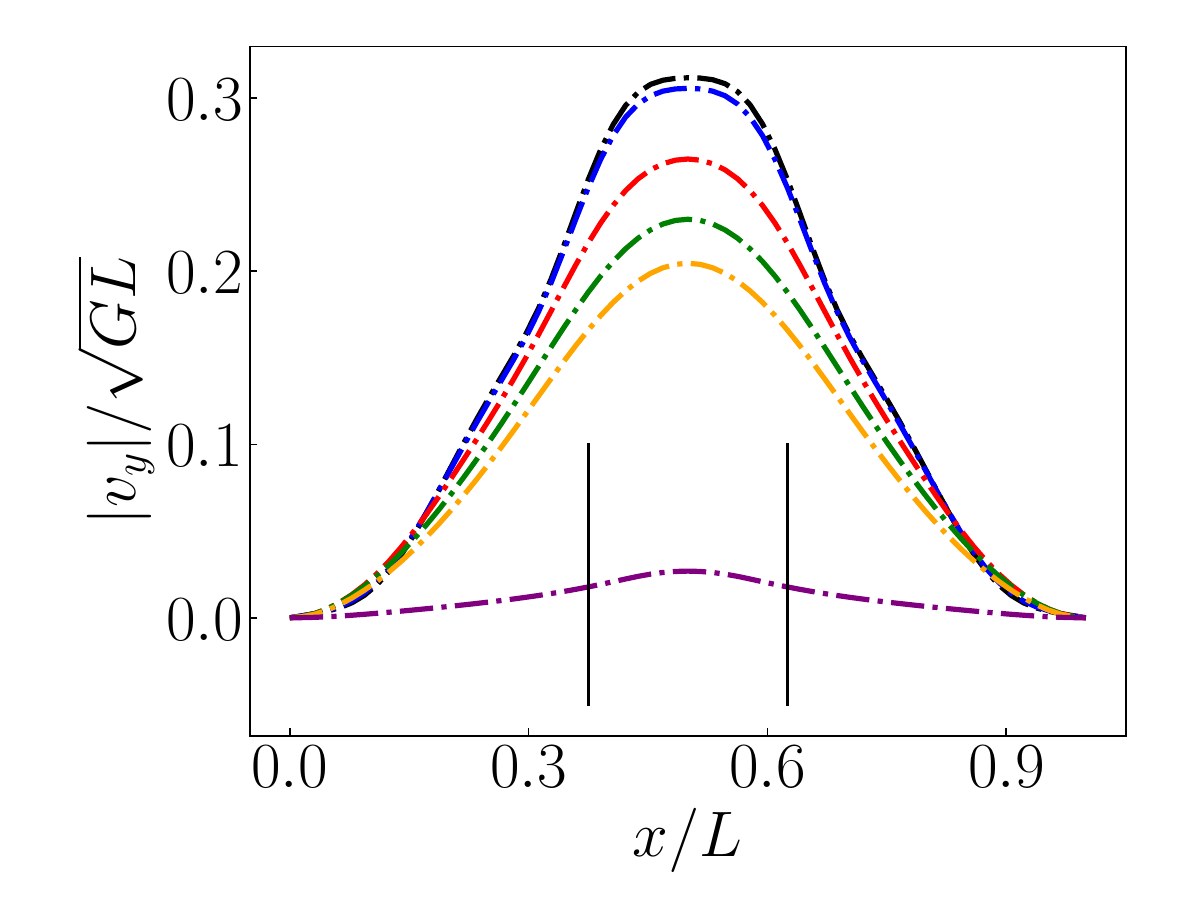}
\includegraphics[height=3.2cm]
{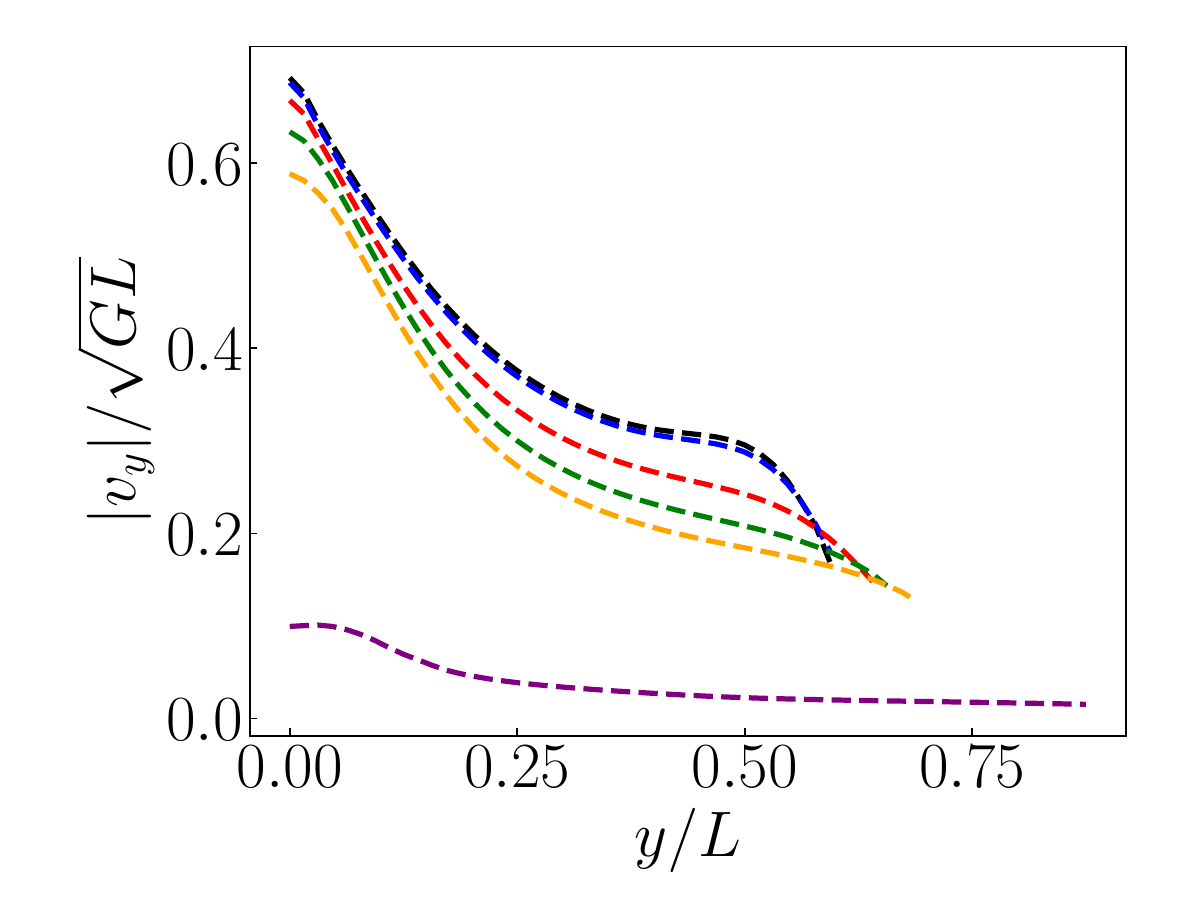}
	\caption{\label{fig:velocity_fields}  Normalized vertical velocity fields and profiles at $\overline{t}=2$ for $\overline{D}=0.25$, ccomparing the local $\mu(I)$ model and the dynamic NGF model with various non-local amplitudes $A$, using the modified parameter set (table~\ref{tab:params}). (a) Normalized vertical velocity $|v_y|/|v_y|_{\max}$ field visualization. The solid horizontal line represents $y/L = 0.1$, the dashed dotted horizontal line indicates $y/L = 0.4$, and the dashed vertical line indicates $x/L = 0.5$. (b) Normalized vertical velocity $|v_y|/\sqrt{GL}$ profiles along the horizontal line at $y/L = 0.1$. (c) Normalized vertical velocity profiles $|v_y|/\sqrt{GL}$ along the horizontal line at $y/L = 0.4$. The vertical solid lines in (b) and (c) indicate the locations of the orifice edges. (d) Normalized vertical velocity $|v_y|/\sqrt{GL}$ profiles along the vertical centerline at $x/L = 0.5$.}
  	\end{center}
\end{figure}
 \begin{figure}
  	\begin{center}
	 \hspace{-12.5cm} $(a)$\\
 \includegraphics[height=9cm]{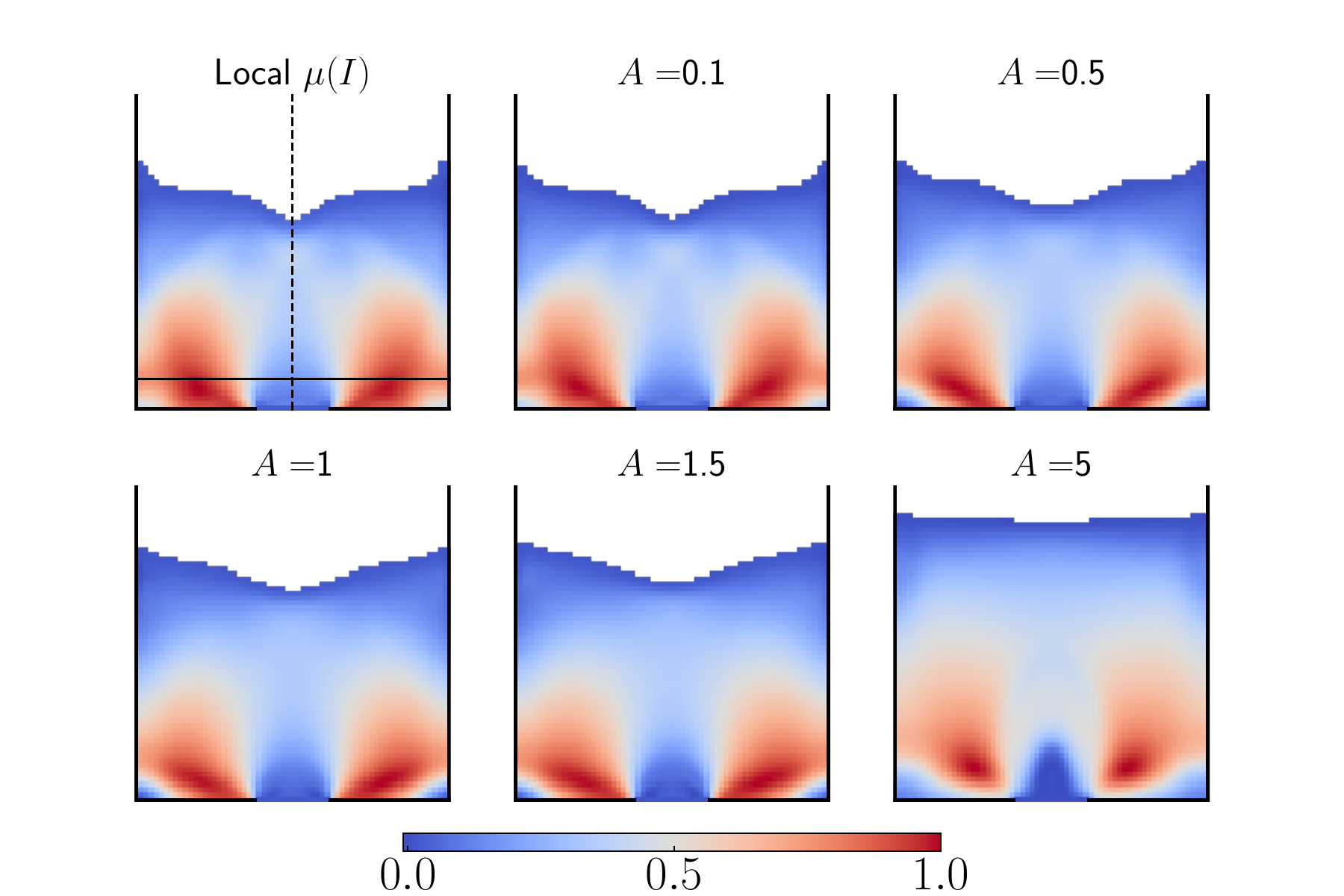}\\
	 \hspace{-5.5cm} $(b)$ \hspace{6.cm} $(c)$\\
   \includegraphics[height=4.5cm]
{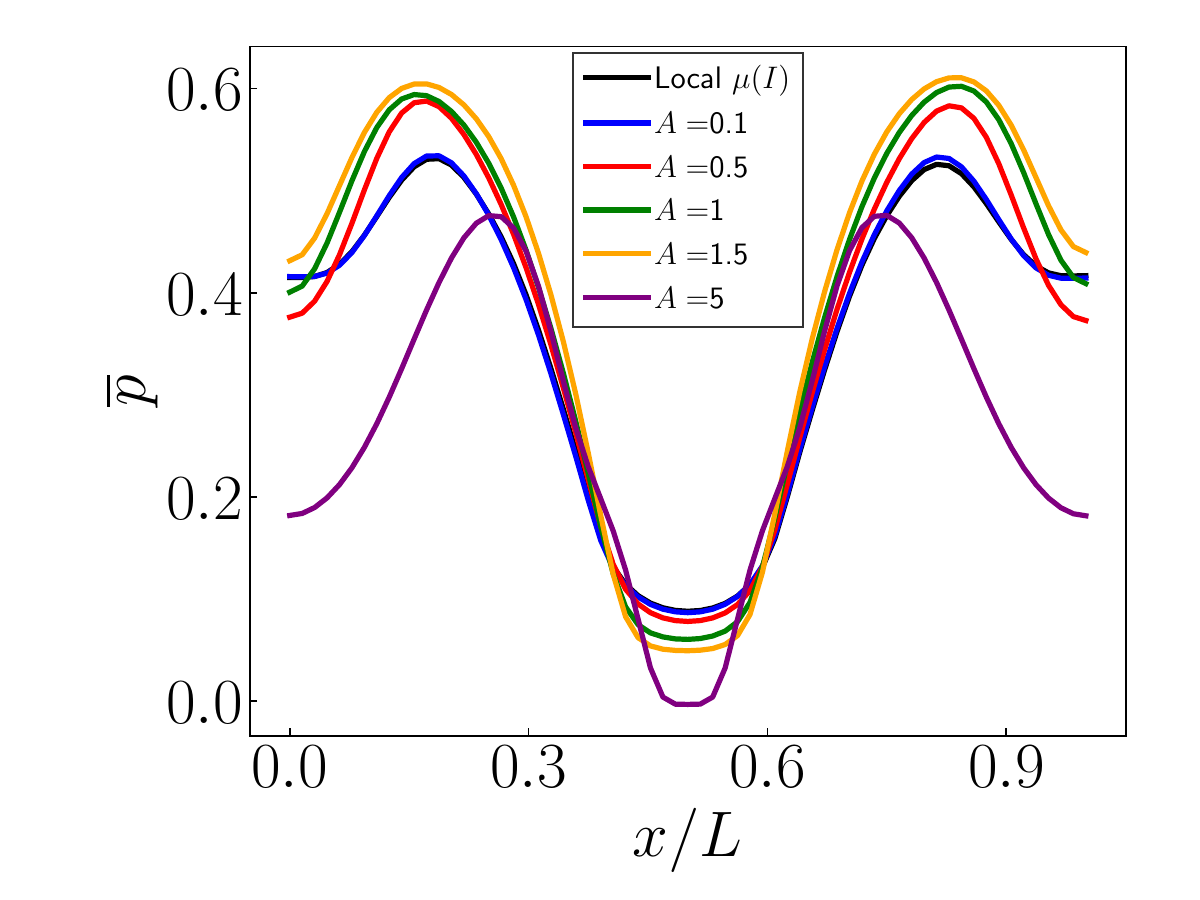}
\includegraphics[height=4.5cm]
{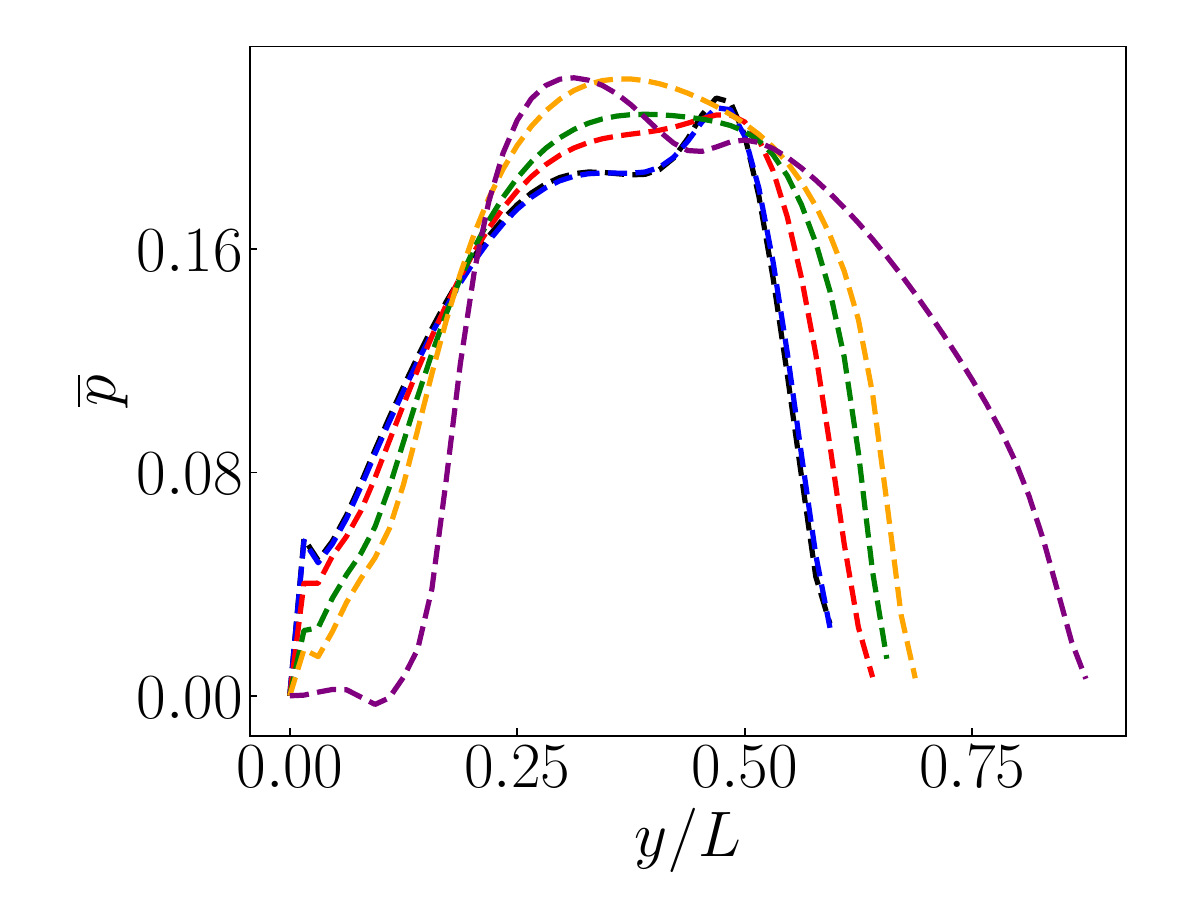}
	\caption{\label{fig:pressure_fields}  Normalized pressure fields and profiles at $\overline{t}=2$ for $\overline{D}=0.25$, comparing the local $\mu(I)$ model and the dynamic NGF model with various non-local amplitudes $A$, using the modified parameter set (table~\ref{tab:params}). (a) Normalized pressure $p/p_{\max}$ field visualization. The solid horizontal line represents $y/L = 0.1$, and the dashed vertical line represents $x/L = 0.5$. (b) Normalized pressure $\overline{p} = p/ (\rho G L)$ profiles along the horizontal line at $y/L = 0.1$. (c) Normalized pressure $\overline{p} = p/ (\rho G L)$ profiles along the vertical centerline at $x/L = 0.5$.}
  	\end{center}
\end{figure}

\section{Supplementary Fields}\label{app:fields}

In this appendix, we present supplementary results for the velocity and pressure fields obtained with the dynamic NGF model at different non-local amplitudes $A$, complementing the figures shown in \S\ref{FielddistributionsDNGF}. Figure~\ref{fig:velocity_fields} shows the normalized vertical velocity fields and profiles at $\overline{t}=2$ for $\overline{D}=0.25$. From the field visualisation in figure~\ref{fig:velocity_fields}(a), we observe that as $A$ increases, the free surface rises, indicating a reduction in flow rate. The velocity profiles in figure~\ref{fig:velocity_fields}(b) along $y/L = 0.1$ show that all non-local models yield lower velocities than the local model. At $y/L = 0.4$ (figure~\ref{fig:velocity_fields}(c)), the local model and non-local models with small $A$ exhibit a plug-like profile (nearly flat in the central region), while those with larger $A$ produce more rounded profiles, indicating enhanced non-local diffusion. The centreline profiles in figure~\ref{fig:velocity_fields}(d) further confirm that increasing $A$ reduces the vertical velocity throughout the domain, consistent with the trend observed in the flow rate.

Figure~\ref{fig:pressure_fields} presents the normalized pressure fields and profiles. The contour plot in figure~\ref{fig:pressure_fields}(a) shows the typical pressure distribution, with negative pressure regions artificially set to zero for visualisation purposes. Figure~\ref{fig:pressure_fields}(b) shows the horizontal pressure profiles at $y/L = 0.1$, where we observe three distinct features: a marked dip of pressure above the outlet, two high-pressure regions on either side of the outlet, and a slight decrease near the sidewalls. These features are consistent with previous discrete and continuum simulations (\cite{Staron2014}). As $A$ increases, the pressure at the centreline decreases further, while the pressure in the two high-pressure regions increases, leading to a more pronounced pressure variation across the width. Figure~\ref{fig:pressure_fields}(c) presents the vertical pressure profiles along the centreline ($x/L = 0.5$). For $A \leq 1$, the pressure starts from nearly zero at the bottom, increases with depth, and gradually saturates, consistent with the Janssen effect (\cite{Sperl2006}). The maximum pressure increases with $A$. Notably, for $A = 5$, the pressure near the outlet approaches zero or becomes negative, as seen in figures~\ref{fig:pressure_fields}(b) and (c). Although negative pressure is physically impossible in granular materials, its emergence in our continuum simulations may be interpreted as an indicator of the potential for arch formation. In regions where $p \le 0$, we set $\eta_{\text{eff}} = 10^{-5}$ as a numerical regularization. This choice does not affect the numerical solution, as the viscosity in these high-shear-rate regions is already small.

\clearpage
\begin{acknowledgments}
We gratefully acknowledge fruitful discussions with S. Popinet, L. Staron and O. Pouliquen. The authors also thank the National Natural Science Foundation of China (Grant No. 12272134) and the China Scholarship Council (CSC) for their financial support. We acknowledge the assistance of DeepSeek in language polishing during manuscript preparation.
\end{acknowledgments}

\bibliographystyle{jfm}
\bibliography{jfm}

@article{Dunatunga2022,
	Author = {Dunatunga, S. and Kamrin, K.},
	C7 = {A14},
	Db = {Cambridge Core},
	Doi = {DOI: 10.1017/jfm.2022.241},
	Dp = {Cambridge University Press},
	Et = {2022/04/06},
	Isbn = {0022-1120},
	Journal = {J. Fluid Mech.},
	Pages = {A14},
	Publisher = {Cambridge University Press},
	Title = {Modelling silo clogging with non-local granular rheology},
	Ty = {JOUR},
	Url = {https://www.cambridge.org/core/product/85A4AAB47282896F1826DDC3447FE10F},
	Volume = {940},
	Year = {2022}}

@article{Janda2008,
	Author = {Janda, A. and Zuriguel, I. and Garcimart{\'\i}n, A. and Pugnaloni, L. A. and Maza, D.},
	Doi = {10.1209/0295-5075/84/44002},
	Journal = {Europhys. Lett.},
	Month = {nov},
	Number = {4},
	Pages = {44002},
	Title = {Jamming and critical outlet size in the discharge of a two-dimensional silo},
	Url = {https://dx.doi.org/10.1209/0295-5075/84/44002},
	Volume = {84},
	Year = {2008}}

@article{Bouzid2013,
	Author = {Bouzid, M. and Trulsson, M. and Claudin, P. and Cl\'ement, E. and Andreotti, B.},
	Doi = {10.1103/PhysRevLett.111.238301},
	Issue = {23},
	Journal = {Phys. Rev. Lett.},
	Month = {Dec},
	Numpages = {5},
	Pages = {238301},
	Publisher = {American Physical Society},
	Title = {Nonlocal Rheology of Granular Flows across Yield Conditions},
	Url = {https://link.aps.org/doi/10.1103/PhysRevLett.111.238301},
	Volume = {111},
	Year = {2013}}

@article{Henann2013,
	Author = {Henann, D. L. and Kamrin, K.},
	Doi = {10.1073/pnas.1219153110},
	Journal = {Proc. Natl Acad. Sci. USA},
	Number = {17},
	Pages = {6730-6735},
	Title = {A predictive, size-dependent continuum model for dense granular flows},
	Url = {https://www.pnas.org/doi/abs/10.1073/pnas.1219153110},
	Volume = {110},
	Year = {2013},
	}

@article{BEV61,
Author = {W. A. Beverloo and H. A. Leniger and J. Van de Velde},
Journal = {Chem. Eng. Sci.},
Pages = {260-269},
Title = {The flow of granular solids through orifices},
Volume = {15},
Year = {1961}}

@book{NED92,
Author = {R. M. Nedderman},
Publisher = {Cambridge University Press},
Title = {Statics and kinematics of granular materials},
Year = {1992}}

@article{To2001,
Author = {K. To and P.-Y. Lai and H.K. Pak},
Doi = {10.1103/PhysRevLett.86.71},
Journal = {Phys. Rev. Lett.},
Pages = {71},
Title = {Jamming of Granular Flow in a Two-Dimensional Hopper},
Url = {http://dx.doi.org/10.1103/PhysRevLett.86.71},
Volume = {86},
Year = {2001}}

@article{Zuriguel2005,
Author = {Zuriguel, I. and Garcimart\'{\i}n, A. and Maza, D. and Pugnaloni, L. A. and Pastor, J. M.},
Doi = {10.1103/PhysRevE.71.051303},
Issue = {5},
Journal = {Phys. Rev. E},
Month = {May},
Numpages = {9},
Pages = {051303},
Publisher = {American Physical Society},
Title = {Jamming during the discharge of granular matter from a silo},
Url = {https://link.aps.org/doi/10.1103/PhysRevE.71.051303},
Volume = {71},
Year = {2005}}

@article{Zhou2026,
Author = {Zhou, Y. and Li, M. and Guan, Y. and Wang, Y. and Liu, Y. and Zou, Z.},
Doi = {10.1103/qs87-8yts},
Issue = {2},
Journal = {Phys. Rev. E},
Month = {Feb},
Numpages = {6},
Pages = {L023401},
Publisher = {American Physical Society},
Title = {Clogging-flowing transition of granular media in a two-dimensional vertical pipe},
Url = {https://link.aps.org/doi/10.1103/qs87-8yts},
Volume = {113},
Year = {2026}}

@article{Zuriguel2003,
Author = {Zuriguel, I. and Pugnaloni, L. A. and Garcimart\'{\i}n, A. and Maza, D.},
Doi = {10.1103/PhysRevE.68.030301},
Issue = {3},
Journal = {Phys. Rev. E},
Month = {Sep},
Numpages = {4},
Pages = {030301},
Publisher = {American Physical Society},
Title = {Jamming during the discharge of grains from a silo described as a percolating transition},
Url = {https://link.aps.org/doi/10.1103/PhysRevE.68.030301},
Volume = {68},
Year = {2003}}

@article{Sheldon2010,
Author = {Sheldon, H. G. and Durian, D. J.},
Doi = {10.1007/s10035-010-0198-3},
Issn = {1434-7636},
Journal = {Granul. Matter},
Number = {6},
Pages = {579--585},
Title = {Granular discharge and clogging for tilted hoppers},
Url = {http://dx.doi.org/10.1007/s10035-010-0198-3},
Volume = {12},
Year = {2010}}

@article{Zuriguel2014,
Author = {Zuriguel, I. and Parisi, D. R. and Hidalgo, R. C. and Lozano, C. and Janda, A. and Gago, Paula A. and Peralta, J. P. and Ferrer, L. M. and Pugnaloni, L. A. and Cl{\'e}ment, E. and Maza, D. and Pagonabarraga, I. and Garcimart{\'\i}n, A.},
Da = {2014/12/04},
Doi = {10.1038/srep07324},
Id = {Zuriguel2014},
Isbn = {2045-2322},
Journal = {Sci. Rep.},
Number = {1},
Pages = {7324},
Title = {Clogging transition of many-particle systems flowing through bottlenecks},
Ty = {JOUR},
Url = {https://doi.org/10.1038/srep07324},
Volume = {4},
Year = {2014}}

@article{Zuriguel2011,
Author = {Zuriguel, I. and Janda, A. and Garcimart\'{\i}n, A. and Lozano, C. and Ar\'evalo, R. and Maza, D.},
Doi = {10.1103/PhysRevLett.107.278001},
Issue = {27},
Journal = {Phys. Rev. Lett.},
Month = {Dec},
Numpages = {5},
Pages = {278001},
Publisher = {American Physical Society},
Title = {Silo Clogging Reduction by the Presence of an Obstacle},
Url = {https://link.aps.org/doi/10.1103/PhysRevLett.107.278001},
Volume = {107},
Year = {2011}}

@article{Gella2018,
Author = {Gella, D. and Zuriguel, I. and Maza, D.},
Doi = {10.1103/PhysRevLett.121.138001},
Issue = {13},
Journal = {Phys. Rev. Lett.},
Month = {Sep},
Numpages = {5},
Pages = {138001},
Publisher = {American Physical Society},
Title = {Decoupling Geometrical and Kinematic Contributions to the Silo Clogging Process},
Url = {https://link.aps.org/doi/10.1103/PhysRevLett.121.138001},
Volume = {121},
Year = {2018}}

@article{To2005,
Author = {To, K.},
Doi = {10.1103/PhysRevE.71.060301},
Issue = {6},
Journal = {Phys. Rev. E},
Month = {Jun},
Numpages = {4},
Pages = {060301(R)},
Publisher = {American Physical Society},
Title = {Jamming transition in two-dimensional hoppers and silos},
Url = {https://link.aps.org/doi/10.1103/PhysRevE.71.060301},
Volume = {71},
Year = {2005}}

@article{Thomas2013,
Author = {Thomas, C. C. and Durian, D. J.},
Doi = {10.1103/PhysRevE.87.052201},
Issue = {5},
Journal = {Phys. Rev. E},
Month = {May},
Numpages = {8},
Pages = {052201},
Publisher = {American Physical Society},
Title = {Geometry dependence of the clogging transition in tilted hoppers},
Url = {https://link.aps.org/doi/10.1103/PhysRevE.87.052201},
Volume = {87},
Year = {2013}}

@article{Arevalo2014,
Author = {Ar\'evalo, R. and Zuriguel, I. and Maza, D. and Garcimart\'{\i}n, A.},
Doi = {10.1103/PhysRevE.89.042205},
Issue = {4},
Journal = {Phys. Rev. E},
Month = {Apr},
Numpages = {5},
Pages = {042205},
Publisher = {American Physical Society},
Title = {Role of driving force on the clogging of inert particles in a bottleneck},
Url = {https://link.aps.org/doi/10.1103/PhysRevE.89.042205},
Volume = {89},
Year = {2014}}

@article{Zhou_15,
Author = {Zhou, Y. and Ruyer, P. and Aussillous, P.},
Doi = {10.1103/PhysRevE.92.062204},
Journal = {Phys. Rev. E},
Pages = {062204},
Title = {Discharge flow of a bidisperse granular media from a silo: discrete particle simulations.},
Url = {http://link.aps.org/doi/10.1103/PhysRevE.92.062204},
Volume = {92},
Year = {2015}}

@article{Arevalo2016,
Author = {Ar{\'e}valo, R. and Zuriguel, I.},
Doi = {10.1039/C5SM01599E},
Issue = {1},
Journal = {Soft Matter},
Pages = {123-130},
Publisher = {The Royal Society of Chemistry},
Title = {Clogging of granular materials in silos: effect of gravity and outlet size},
Url = {http://dx.doi.org/10.1039/C5SM01599E},
Volume = {12},
Year = {2016}}

@article{STA12,
Author = {L. Staron and P.-Y. Lagr\'ee and S. Popinet},
Doi = {10.1063/1.4757390},
Journal = {Phys. Fluids},
Pages = {103301},
Title = {The granular silo as a continuum plastic flow: The hour-glass vs the clepsydra},
Url = {http://dx.doi.org/10.1063/1.4757390},
Volume = {24},
Year = {2012}}

@article{Staron2014,
Author = {Staron, L. and Lagr\'ee, P.-Y. and Popinet, S.},
Doi = {10.1140/epje/i2014-14005-6},
Issn = {1292-895X},
Journal = {Eur. Phys. J. E},
Number = {1},
Pages = {1--12},
Title = {Continuum simulation of the discharge of the granular silo},
Url = {http://dx.doi.org/10.1140/epje/i2014-14005-6},
Volume = {37},
Year = {2014}}

@article{Dunatunga2015,
Author = {Dunatunga, S. and Kamrin, K.},
Doi = {10.1017/jfm.2015.383},
Journal = {J. Fluid Mech.},
Pages = {483--513},
Title = {Continuum modelling and simulation of granular flows through their many phases},
Volume = {779},
Year = {2015}}

@article{Zhou2017,
Author = {Zhou, Y. and Lagr{\'e}e, P. -Y. and Popinet, S. and Ruyer, P. and Aussillous, P.},
Db = {Cambridge Core},
Doi = {DOI: 10.1017/jfm.2017.543},
Dp = {Cambridge University Press},
Et = {2017/09/21},
Isbn = {0022-1120},
Journal = {J. Fluid Mech.},
Pages = {459-485},
Publisher = {Cambridge University Press},
Title = {Experiments on, and discrete and continuum simulations of, the discharge of granular media from silos with a lateral orifice},
Ty = {JOUR},
Url = {https://www.cambridge.org/core/product/0B924467BC7DF119B7E03B190EDDA999},
Volume = {829},
Year = {2017}}

@article{Fullard2019,
Author = {Fullard, L. and Holland, D. J. and Galvosas, P. and Davies, C. and Lagr\'ee, P.-Y. and Popinet, S.},
Doi = {10.1103/PhysRevFluids.4.074302},
Issue = {7},
Journal = {Phys. Rev. Fluids},
Month = {Jul},
Numpages = {20},
Pages = {074302},
Publisher = {American Physical Society},
Title = {Quantifying silo flow using MRI velocimetry for testing granular flow models},
Url = {https://link.aps.org/doi/10.1103/PhysRevFluids.4.074302},
Volume = {4},
Year = {2019}}

@article{GDR2004,
Author = {{G.D.R.} MIDI},
Journal = {Eur. Phys. J. E},
Pages = {341-365},
Title = {On dense granular flows},
Volume = {14},
Year = {2004}}

@article{daCruz2005,
Author = {Da Cruz, F. and Emam, S. and Prochnow, M. and Roux, J.-N. and Chevoir, F.},
Doi = {10.1103/PhysRevE.72.021309},
Issue = {2},
Journal = {Phys. Rev. E},
Month = {Aug},
Numpages = {17},
Pages = {021309},
Publisher = {American Physical Society},
Title = {Rheophysics of dense granular materials: Discrete simulation of plane shear flows},
Url = {https://link.aps.org/doi/10.1103/PhysRevE.72.021309},
Volume = {72},
Year = {2005}}

@article{Jop06,
Author = {P. Jop and Y. Forterre and O. Pouliquen},
Doi = {10.1038/nature04801},
Journal = {Nature},
Pages = {727-730},
Title = {A constitutive law for dense granular flows},
Url = {http://dx.doi.org/10.1038/nature04801},
Volume = {441},
Year = {2006}}

@article{Lagree2011,
Author = {Lagr\'ee, P.-Y. and Staron, L. and Popinet, S.},
Journal = {J. Fluid Mech.},
Pages = {378-408},
Title = {The granular column collapse as a continuum: validity of a two-dimensional {N}avier--{S}tokes model with a $\mu$({I})-rheology},
Volume = {686},
Year = {2011}}

@article{Pouliquen1999,
Author = {Pouliquen, O.},
Journal = {Phys. Fluids},
Number = {3},
Pages = {542-548},
Title = {Scaling laws in granular flows down rough inclined planes},
Volume = {11},
Year = {1999}}

@article{Nichol2010,
Author = {Nichol, K. and Zanin, A. and Bastien, R. and Wandersman, E. and van Hecke, M.},
Doi = {10.1103/PhysRevLett.104.078302},
Issue = {7},
Journal = {Phys. Rev. Lett.},
Month = {Feb},
Numpages = {4},
Pages = {078302},
Publisher = {American Physical Society},
Title = {Flow-Induced Agitations Create a Granular Fluid},
Url = {https://link.aps.org/doi/10.1103/PhysRevLett.104.078302},
Volume = {104},
Year = {2010}}

@article{Reddy2011,
Author = {Reddy, K. A. and Forterre, Y. and Pouliquen, O.},
Doi = {10.1103/PhysRevLett.106.108301},
Issue = {10},
Journal = {Phys. Rev. Lett.},
Month = {Mar},
Numpages = {4},
Pages = {108301},
Publisher = {American Physical Society},
Title = {Evidence of Mechanically Activated Processes in Slow Granular Flows},
Url = {https://link.aps.org/doi/10.1103/PhysRevLett.106.108301},
Volume = {106},
Year = {2011}}

@book{Andreotti2013,
Author = {Andreotti, B. and Forterre, Y. and Pouliquen, O.},
Place = {Cambridge},
Publisher = {Cambridge University Press},
Title = {Granular Media: Between Fluid and Solid},
Year = {2013}}

@article{Kamrin2012,
Author = {Kamrin, K. and Koval, G.},
Doi = {10.1103/PhysRevLett.108.178301},
Issue = {17},
Journal = {Phys. Rev. Lett.},
Month = {Apr},
Numpages = {5},
Pages = {178301},
Publisher = {American Physical Society},
Title = {Nonlocal Constitutive Relation for Steady Granular Flow},
Url = {https://link.aps.org/doi/10.1103/PhysRevLett.108.178301},
Volume = {108},
Year = {2012}}

@article{Kamrin2015SM,
Author = {Kamrin, K. and Henann, D. L.},
Doi = {10.1039/C4SM01838A},
Issue = {1},
Journal = {Soft Matter},
Pages = {179-185},
Publisher = {The Royal Society of Chemistry},
Title = {Nonlocal modeling of granular flows down inclines},
Url = {http://dx.doi.org/10.1039/C4SM01838A},
Volume = {11},
Year = {2015}}

@article{Kamrin2024,
Author = {Kamrin, K. and Hill, K. M. and Goldman, D. I. and Andrade, J. E.},
Doi = {https://doi.org/10.1146/annurev-fluid-121021-022045},
Issn = {1545-4479},
Journal = {Annu. Rev. Fluid Mech.},
Number = {Volume 56, 2024},
Pages = {215-240},
Publisher = {Annual Reviews},
Title = {Advances in Modeling Dense Granular Media},
Type = {Journal Article},
Url = {https://www.annualreviews.org/content/journals/10.1146/annurev-fluid-121021-022045},
Volume = {56},
Year = {2024}}

@article{Pouliquen2026,
Author = {Pouliquen, O.},
Da = {2026/04/07},
Doi = {10.1140/epje/s10189-026-00579-7},
Id = {Pouliquen2026},
Isbn = {1292-895X},
Journal = {Eur. Phys. J. E},
Number = {4},
Pages = {32},
Title = {Non-local rheology in granular media: a perspective on the 2015 EPJE Paper by Bouzid et al.},
Ty = {JOUR},
Url = {https://doi.org/10.1140/epje/s10189-026-00579-7},
Volume = {49},
Year = {2026}}

@article{Pouliquen2009,
Author = {Pouliquen, O. and Forterre, Y.},
Doi = {10.1098/rsta.2009.0171},
Issn = {1364-503X},
Journal = {Philos. Trans. R. Soc. A},
Month = {12},
Number = {1909},
Pages = {5091-5107},
Title = {A non-local rheology for dense granular flows},
Url = {https://doi.org/10.1098/rsta.2009.0171},
Volume = {367},
Year = {2009}}

@article{Dsouza2020,
Author = {Dsouza, P. V. and Nott, P. R.},
Doi = {10.1017/jfm.2020.62},
Journal = {J. Fluid Mech.},
Pages = {R3},
Title = {A non-local constitutive model for slow granular flow that incorporates dilatancy},
Volume = {888},
Year = {2020}}

@article{Berzi2024,
Author = {Berzi, D.},
Doi = {10.1103/PhysRevFluids.9.034304},
Issue = {3},
Journal = {Phys. Rev. Fluids},
Month = {Mar},
Numpages = {26},
Pages = {034304},
Publisher = {American Physical Society},
Title = {On granular flows: From kinetic theory to inertial rheology and nonlocal constitutive models},
Url = {https://link.aps.org/doi/10.1103/PhysRevFluids.9.034304},
Volume = {9},
Year = {2024}}

@article{Alessio2026,
Author = {Alessio, B. M. and Edwards, M. R and Lai, C.-Y.},
Journal = {Arkiv},
Pages = {01907},
Title = {Dense granular rheology from fluctuations},
Url = {https://arxiv.org/html/2601.01907v1},
Volume = {2501},
Year = {2026}}

@article{Goyon2008,
Author = {Goyon, J. and Colin, A. and Ovarlez, G. and Ajdari, A. and Bocquet, L.},
Da = {2008/07/01},
Doi = {10.1038/nature07026},
Id = {Goyon2008},
Isbn = {1476-4687},
Journal = {Nature},
Number = {7200},
Pages = {84--87},
Title = {Spatial cooperativity in soft glassy flows},
Ty = {JOUR},
Url = {https://doi.org/10.1038/nature07026},
Volume = {454},
Year = {2008}}

@article{Henann2014,
Author = {Henann, D. L. and Kamrin, K.},
Doi = {https://doi.org/10.1016/j.ijplas.2014.05.002},
Issn = {0749-6419},
Journal = {Int. J. Plast.},
Pages = {145-162},
Title = {Continuum thermomechanics of the nonlocal granular rheology},
Url = {https://www.sciencedirect.com/science/article/pii/S0749641914000989},
Volume = {60},
Year = {2014}}

@article{Zhang2017,
Author = {Zhang, Q. and Kamrin, K.},
Doi = {10.1103/PhysRevLett.118.058001},
Issue = {5},
Journal = {Phys. Rev. Lett.},
Month = {Jan},
Numpages = {6},
Pages = {058001},
Publisher = {American Physical Society},
Title = {Microscopic Description of the Granular Fluidity Field in Nonlocal Flow Modeling},
Url = {https://link.aps.org/doi/10.1103/PhysRevLett.118.058001},
Volume = {118},
Year = {2017}}

@article{Bouzid2015,
Author = {Bouzid, M. and Izzet, A. and Trulsson, M. and Cl{\'e}ment, E. and Claudin, P. and Andreotti, B.},
Da = {2015/11/30},
Doi = {10.1140/epje/i2015-15125-1},
Id = {Bouzid2015},
Isbn = {1292-895X},
Journal = {Eur. Phys. J. E},
Number = {11},
Pages = {125},
Title = {Non-local rheology in dense granular flows},
Ty = {JOUR},
Url = {https://doi.org/10.1140/epje/i2015-15125-1},
Volume = {38},
Year = {2015}}

@article{Kim2020,
Author = {Kim, S. and Kamrin, K.},
Doi = {10.1103/PhysRevLett.125.088002},
Issue = {8},
Journal = {Phys. Rev. Lett.},
Month = {Aug},
Numpages = {6},
Pages = {088002},
Publisher = {American Physical Society},
Title = {Power-Law Scaling in Granular Rheology across Flow Geometries},
Url = {https://link.aps.org/doi/10.1103/PhysRevLett.125.088002},
Volume = {125},
Year = {2020}}

@article{Irmer2025,
Author = {Irmer, M. G. and Brodsky, E. E. and Clark, A. H.},
Doi = {10.1103/PhysRevLett.134.048202},
Issue = {4},
Journal = {Phys. Rev. Lett.},
Month = {Jan},
Numpages = {6},
Pages = {048202},
Publisher = {American Physical Society},
Title = {Granular Temperature Controls Local Rheology of Vibrated Granular Flows},
Url = {https://link.aps.org/doi/10.1103/PhysRevLett.134.048202},
Volume = {134},
Year = {2025}}

@article{Poon2023,
Author = {Poon, R. N. and Thomas, A. L. and Vriend, N. M.},
Doi = {10.1103/PhysRevE.108.064902},
Issue = {6},
Journal = {Phys. Rev. E},
Month = {Dec},
Numpages = {6},
Pages = {064902},
Publisher = {American Physical Society},
Title = {Microscopic origin of granular fluidity: An experimental investigation},
Url = {https://link.aps.org/doi/10.1103/PhysRevE.108.064902},
Volume = {108},
Year = {2023}}

@article{Clarke2024,
Author = {Clarke, D. A. and Poata, J. and Galvosas, P. and Holland, D. J.},
Journal = {Phys. Fluids},
Number = {5},
Pages = {053317},
Title = {Investigation of nonlocal granular fluidity models using nuclear magnetic resonance},
Volume = {36},
Year = {2024}}

@article{Hao2023,
Author = {Hao, J. and Guo, Y.},
Journal = {Phys. Fluids},
Number = {10},
Pages = {103310},
Title = {Rheology of sheared polyhedral granular materials in inclined flows},
Volume = {35},
Year = {2023}}

@article{Breard2024,
Author = {Breard, E. C. P. and Fullard, L. and Dufek, J.},
Doi = {10.1103/PhysRevFluids.9.054303},
Issue = {5},
Journal = {Phys. Rev. Fluids},
Month = {May},
Numpages = {18},
Pages = {054303},
Publisher = {American Physical Society},
Title = {Rheology of granular mixtures with varying size, density, particle friction, and flow geometry},
Url = {https://link.aps.org/doi/10.1103/PhysRevFluids.9.054303},
Volume = {9},
Year = {2024}}

@article{Yuan2026,
Author = {Yuan, Z. and Zhao, H. and Wang, D.},
Journal = {J. Fluid Mech.},
Number = {A24},
Title = {Toward a unified continuum framework for dense granular flows: $\mu$({I}) rheology extended with granular temperature},
Volume = {1030},
Year = {2026}}

@article{Kamrin2019,
Author = {Kamrin, K.},
Doi = {10.3389/fphy.2019.00116},
Issn = {2296-424X},
Journal = {Front. Phys.},
Title = {Non-locality in Granular Flow: Phenomenology and Modeling Approaches},
Url = {https://www.frontiersin.org/journals/physics/articles/10.3389/fphy.2019.00116},
Volume = {Volume 7 - 2019},
Year = {2019}}

@article{Lin2021,
Author = {Lin, C. and Yang, F.},
Doi = {10.1063/5.0057598},
Issn = {1070-6631},
Journal = {Phys. Fluids},
Month = {09},
Number = {9},
Pages = {093302},
Title = {Continuum simulation of non-local effects in a granular silo discharge flow using a regularized $\mu$({I}) rheology model},
Url = {https://doi.org/10.1063/5.0057598},
Volume = {33},
Year = {2021}}

@article{Schaeffer1987,
Author = {D. G. Schaeffer},
Doi = {https://doi.org/10.1016/0022-0396(87)90038-6},
Issn = {0022-0396},
Journal = {J. Differ. Equ.},
Number = {1},
Pages = {19-50},
Title = {Instability in the evolution equations describing incompressible granular flow},
Url = {https://www.sciencedirect.com/science/article/pii/0022039687900386},
Volume = {66},
Year = {1987}}

@article{Barker2015,
Author = {Barker, T. and Schaeffer, D. G. and Bohorquez, P. and Gray, J. M. N. T.},
Doi = {10.1017/jfm.2015.412},
Journal = {J. Fluid Mech.},
Pages = {794--818},
Title = {Well-posed and ill-posed behaviour of the $\mu$({I})-rheology for granular flow},
Volume = {779},
Year = {2015}}

@article{Barker2017b,
Author = {Barker, T. and Gray, J. M. N. T.},
Journal = {J. Fluid Mech.},
Pages = {5--32},
Title = {Partial regularisation of the incompressible $\mu$({I})-rheology for granular flow},
Volume = {828},
Year = {2017}}

@article{Franci2019,
Author = {Franci, A. and Cremonesi, M.},
Doi = {https://doi.org/10.1016/j.jcp.2018.11.011},
Issn = {0021-9991},
Journal = {J. Comput. Phys.},
Pages = {257-277},
Title = {3D regularized $\mu$({I})-rheology for granular flows simulation},
Url = {https://www.sciencedirect.com/science/article/pii/S0021999118307290},
Volume = {378},
Year = {2019}}

@article{Barker2017,
Author = {Barker, T. and Schaeffer, D. G. and Shearer, M. and Gray, J. M. N. T.},
Doi = {10.1098/rspa.2016.0846},
Issn = {1364-5021},
Journal = {Proc. R. Soc. A},
Month = {05},
Number = {2201},
Pages = {20160846},
Title = {Well-posed continuum equations for granular flow with compressibility and $\mu$({I})-rheology},
Url = {https://doi.org/10.1098/rspa.2016.0846},
Volume = {473},
Year = {2017}}

@article{Heyman2017,
Author = {Heyman, J. and Delannay, R. and Tabuteau, H. and Valance, A.},
Doi = {10.1017/jfm.2017.612},
Journal = {J. Fluid Mech.},
Pages = {553--568},
Title = {Compressibility regularizes the $\mu$({I})-rheology for dense granular flows},
Volume = {830},
Year = {2017}}

@article{Schaeffer2019,
Author = {Schaeffer, D. G. and Barker, T. and Tsuji, D. and Gremaud, P. and Shearer, M. and Gray, J. M. N. T.},
Doi = {10.1017/jfm.2019.476},
Journal = {J. Fluid Mech.},
Pages = {926--951},
Title = {Constitutive relations for compressible granular flow in the inertial regime},
Volume = {874},
Year = {2019}}

@article{Li2019,
Author = {Li, S. and Henann, D. L.},
Doi = {10.1017/jfm.2019.311},
Journal = {J. Fluid Mech.},
Pages = {799--830},
Title = {Material stability and instability in non-local continuum models for dense granular materials},
Volume = {871},
Year = {2019}}

@article{Zhu2022,
Author = {Zhu, C. and Peng, C. and Wu, W.},
Doi = {https://doi.org/10.1016/j.powtec.2022.117699},
Issn = {0032-5910},
Journal = {Powder Technol.},
Pages = {117699},
Title = {Lagrangian meshfree particle method ({SPH}) based simulation for granular flow in a rotating drum with regularized $\mu$({I}) elastoplastic model},
Url = {https://www.sciencedirect.com/science/article/pii/S0032591022005927},
Volume = {408},
Year = {2022}}

@article{Yang2023,
Author = {Yang, G.C. and Huang, Y.J. and Lu, Y. and Kwok, C.Y. and Sobral, Y.D. and Yao, Q.H.},
Doi = {10.1017/jfm.2023.782},
Journal = {J. Fluid Mech.},
Pages = {A21},
Title = {Frictional boundary condition for lattice {B}oltzmann modelling of dense granular flows},
Volume = {973},
Year = {2023}}

@article{Yang2023b,
Author = {G. C. Yang and S. C. Yang and L. Jing and C. Y. Kwok and Y. D. Sobral},
Doi = {https://doi.org/10.1016/j.jcp.2023.111956},
Issn = {0021-9991},
Journal = {J. Comput. Phys.},
Pages = {111956},
Title = {Efficient lattice {B}oltzmann simulation of free-surface granular flows with $\mu$({I})-rheology},
Url = {https://www.sciencedirect.com/science/article/pii/S0021999123000517},
Volume = {479},
Year = {2023}}

@article{Shen2026,
Article-Number = {036605},
Author = {Shen, Y. and Qiu, L.},
Doi = {10.1063/5.0318595},
Eissn = {1089-7666},
Issn = {1070-6631},
Journal = {Phys. Fluids},
Month = {MAR},
Number = {3},
Title = {A three-dimensional lattice Boltzmann method for simulating the run-out of debris flows},
Unique-Id = {WOS:001711154800001},
Volume = {38},
Year = {2026}}

@article{Zhou2019,
Author = {Zhou, Y. and Lagr\'ee, P.-Y. and Popinet, S. and Ruyer, P. and Aussillous, P.},
Doi = {10.1103/PhysRevFluids.4.124305},
Issue = {12},
Journal = {Phys. Rev. Fluids},
Month = {Dec},
Numpages = {15},
Pages = {124305},
Publisher = {American Physical Society},
Title = {Gas-assisted discharge flow of granular media from silos},
Url = {https://link.aps.org/doi/10.1103/PhysRevFluids.4.124305},
Volume = {4},
Year = {2019}}

@article{Zou2022,
Author = {Zou, Z. and Ruyer, P. and Lagr\'ee, P.-Y. and Aussillous, P.},
Doi = {10.1103/PhysRevFluids.7.064306},
Issue = {6},
Journal = {Phys. Rev. Fluids},
Month = {Jun},
Numpages = {22},
Pages = {064306},
Publisher = {American Physical Society},
Title = {Nonsteady discharge of granular media from a silo driven by a pressurized gas},
Url = {https://link.aps.org/doi/10.1103/PhysRevFluids.7.064306},
Volume = {7},
Year = {2022}}

@article{Jop2005,
Author = {Jop, P. and Forterre, Y. and Pouliquen, O.},
Doi = {10.1017/S0022112005005987},
Journal = {J. Fluid Mech.},
Pages = {167--192},
Title = {Crucial role of sidewalls in granular surface flows: consequences for the rheology},
Volume = {541},
Year = {2005}}

@article{Bocquet2009,
Author = {Bocquet, L. and Colin, A. and Ajdari, A.},
Doi = {10.1103/PhysRevLett.103.036001},
Issue = {3},
Journal = {Phys. Rev. Lett.},
Month = {Jul},
Numpages = {4},
Pages = {036001},
Publisher = {American Physical Society},
Title = {Kinetic Theory of Plastic Flow in Soft Glassy Materials},
Url = {https://link.aps.org/doi/10.1103/PhysRevLett.103.036001},
Volume = {103},
Year = {2009}}

@article{SalvadorVieira2017,
Author = {Salvador-Vieira, G. and Staron, L. and Popinet, S. and Deboeuf, S. and Lagr{\'e}e, P.-Y.},
Doi = {10.1051/epjconf/201714003045},
Journal = {EPJ Web Conf.},
Pages = {03045},
Title = {Modeling flow arrest using a non-local rheology?},
Url = {https://doi.org/10.1051/epjconf/201714003045},
Volume = 140,
Year = 2017}

@article{Tuzun1979,
Author = {U. T{\"u}z{\"u}n and R.M. Nedderman},
Doi = {https://doi.org/10.1016/0032-5910(79)87044-8},
Issn = {0032-5910},
Journal = {Powder Technol.},
Number = {2},
Pages = {257-266},
Title = {Experimental evidence supporting kinematic modelling of the flow of granular media in the absence of air drag},
Url = {https://www.sciencedirect.com/science/article/pii/0032591079870448},
Volume = {24},
Year = {1979}}

@article{Nedderman1979,
Author = {R.M. Nedderman and U. T{\"u}z{\"u}n},
Doi = {https://doi.org/10.1016/0032-5910(79)80030-3},
Issn = {0032-5910},
Journal = {Powder Technol.},
Number = {2},
Pages = {243-253},
Title = {A kinematic model for the flow of granular materials},
Url = {https://www.sciencedirect.com/science/article/pii/0032591079800303},
Volume = {22},
Year = {1979}}

@article{Choi2005,
Author = {Choi, J. and Kudrolli, A. and Bazant, M. Z},
Doi = {10.1088/0953-8984/17/24/011},
Journal = {J. Phys.: Condens. Matter},
Month = {jun},
Number = {24},
Pages = {S2533},
Title = {Velocity profile of granular flows inside silos and hoppers},
Url = {https://doi.org/10.1088/0953-8984/17/24/011},
Volume = {17},
Year = {2005}}

@article{Irvine2023,
Author = {Irvine, S. K. and Fullard L. A. and Holland D. J. and Clarke D. A. and Lynch T. A. and Lagr\'ee, P.-Y.},
Journal = {Adv. Powder Technol.},
Number = {7},
Pages = {104044},
Title = {Capturing the dynamics of a two orifice silo with the $\mu$({I}) model and extensions},
Volume = {34},
Year = {2023}}

@article{Medina1998,
Author = {A. Medina and J.A. C{\'o}rdova and E. L. and C. Trevi{\~n}o},
Doi = {https://doi.org/10.1016/S0375-9601(98)00795-6},
Issn = {0375-9601},
Journal = {Phys. Lett. A},
Number = {1},
Pages = {111-116},
Title = {Velocity field measurements in granular gravity flow in a near 2D silo},
Url = {https://www.sciencedirect.com/science/article/pii/S0375960198007956},
Volume = {250},
Year = {1998}}

@article{Samadani1999,
Author = {Samadani, A. and Pradhan, A. and Kudrolli, A.},
Doi = {10.1103/PhysRevE.60.7203},
Issue = {6},
Journal = {Phys. Rev. E},
Month = {Dec},
Numpages = {0},
Pages = {7203--7209},
Publisher = {American Physical Society},
Title = {Size segregation of granular matter in silo discharges},
Url = {https://link.aps.org/doi/10.1103/PhysRevE.60.7203},
Volume = {60},
Year = {1999}}

@article{Kamrin2007,
Author = {Kamrin, K. and Bazant, M. Z.},
Doi = {10.1103/PhysRevE.75.041301},
Issue = {4},
Journal = {Phys. Rev. E},
Month = {Apr},
Numpages = {28},
Pages = {041301},
Publisher = {American Physical Society},
Title = {Stochastic flow rule for granular materials},
Url = {https://link.aps.org/doi/10.1103/PhysRevE.75.041301},
Volume = {75},
Year = {2007}}

@article{Garcimartin2011,
Author = {Garcimart\'{\i}n, A. and Zuriguel, I. and Janda, A. and Maza, D.},
Doi = {10.1103/PhysRevE.84.031309},
Issue = {3},
Journal = {Phys. Rev. E},
Month = {Sep},
Numpages = {8},
Pages = {031309},
Publisher = {American Physical Society},
Title = {Fluctuations of grains inside a discharging two-dimensional silo},
Url = {https://link.aps.org/doi/10.1103/PhysRevE.84.031309},
Volume = {84},
Year = {2011}}

@article{Zuriguel2019,
Author = {Zuriguel, I. and Maza, D. and Janda, A. and Hidalgo, R. C. Cruz and Garcimart{\'\i}n, A.},
Da = {2019/06/03},
Doi = {10.1007/s10035-019-0903-9},
Id = {Zuriguel2019},
Isbn = {1434-7636},
Journal = {Granul. Matter},
Number = {3},
Pages = {47},
Title = {Velocity fluctuations inside two and three dimensional silos},
Ty = {JOUR},
Url = {https://doi.org/10.1007/s10035-019-0903-9},
Volume = {21},
Year = {2019}}

@article{Popinet2003,
Author = {S. Popinet},
Doi = {https://doi.org/10.1016/S0021-9991(03)00298-5},
Issn = {0021-9991},
Journal = {J. Comput. Phys.},
Number = {2},
Pages = {572-600},
Title = {Gerris: a tree-based adaptive solver for the incompressible Euler equations in complex geometries},
Url = {https://www.sciencedirect.com/science/article/pii/S0021999103002985},
Volume = {190},
Year = {2003}}

@article{Popinet2009,
Author = {S. Popinet},
Journal = {J. Comput. Phys.},
Pages = {5838--5866},
Title = {An accurate adaptive solver for surface-tension-driven interfacial flows},
Volume = {228},
Year = {2009}}

@article{Popinet2015,
Author = {Popinet, S.},
Journal = {J. Comput. Phys.},
Pages = {336-358},
Title = {A quadtree-adaptive multigrid solver for the {S}erre--{G}reen--{N}aghdi equations},
Volume = {302},
Year = {2015}}

@article{Sperl2006,
Author = {M. Sperl},
Doi = {10.1007/s10035-005-0224-z},
Journal = {Granul. Matter},
Number = {2},
Pages = {59-65},
Title = {Experiments on corn pressure in silo cells. {T}ranslation and comment of {J}anssen's paper from 1895},
Url = {https://doi.org/10.1007/s10035-005-0224-z},
Volume = {8},
Year = {2006}}

@article{Benyamine2014,
Author = {Benyamine, M. and Djermane, M. and Dalloz-Dubrujeaud, B. and Aussillous, P.},
Doi = {10.1103/PhysRevE.90.032201},
Issue = {3},
Journal = {Phys. Rev. E},
Month = {Sep},
Numpages = {8},
Pages = {032201},
Publisher = {American Physical Society},
Title = {Discharge flow of a bidisperse granular media from a silo},
Url = {http://link.aps.org/doi/10.1103/PhysRevE.90.032201},
Volume = {90},
Year = {2014}}

@article{Kamrin2010,
Author = {K. Kamrin},
Doi = {https://doi.org/10.1016/j.ijplas.2009.06.007},
Issn = {0749-6419},
Journal = {Int. J. Plasticity},
Number = {2},
Pages = {167-188},
Title = {Nonlinear elasto-plastic model for dense granular flow},
Url = {https://www.sciencedirect.com/science/article/pii/S0749641909000898},
Volume = {26},
Year = {2010}}

@article{Lin2020,
	Author = {Lin, C. and Yang, F.},
	Doi = {https://doi.org/10.1016/j.jcp.2020.109708},
	Issn = {0021-9991},
	Journal = {J. Comput. Phys.},
	Pages = {109708},
	Title = {Continuum simulation for regularized non-local $\mu$({I}) model of dense granular flows},
	Url = {https://www.sciencedirect.com/science/article/pii/S0021999120304824},
	Volume = {420},
	Year = {2020}}

@phdthesis{Zhou2016,
	Author = {Zhou, Y.},
	Doi = {https://doi.org/10.70675/866f2dd2z225az41c6z98f3zc6abbf538a5a},
	Note = {2016AIXM4731},
	School = {l'Universit{\'e} d'Aix-Marseille},
	Title = {Ejection de gaz et de grains suite {\`a} la rupture d'un crayon de combustible nucl{\'e}aire : mod{\'e}lisation de la dynamique},
	Url = {https://theses.fr/2016AIXM4731/document},
	Year = {2016}}
\end{document}